\documentclass[12pt]{article}

\usepackage{graphicx}
\usepackage[tight]{subfigure}	% subfigure package
\usepackage{pslatex}
\newcommand{\goodgap}{%
	\hspace{\subfigtopskip}%
	\hspace{\subfigbottomskip}}

\title{Nobility and Stupidity:\\
Modeling the Evolution of Class Endogamy}

\author{Theodore C. Belding\\
        Center for the Study of Complex Systems\\ 
	University of Michigan\\ 
	Ann Arbor, MI 48109-1120, USA\\
	\texttt{Ted.Belding@umich.edu}\\
	\texttt{http://www-personal.umich.edu/{\textasciitilde}streak/}}

\date{June 3, 2004}

\begin{document}

\maketitle

%\begin{center}
%\vspace{0.25in}
%\Huge
%DRAFT
%\normalsize
%\vspace{0.25in}
%\end{center}

\begin{abstract}
Class endogamy is a phenomenon in which nobles only marry other nobles
and commoners only marry other commoners. The origin of class
endogamy, and of social stratification in general, is a major open
question in archaeology. This paper implements a verbal model proposed
by Marcus and Flannery as a class of agent-based computer models by
generalizing and simplifying a mathematical model of marriage markets
developed by Burdett and Coles. One force that can produce class
endogamy occurs if agents are only willing to marry suitors having
status no less than some fixed value below the status of their
highest-status suitor, which they can learn. Another such force
results if children inherit the average of their parents' statuses. In
contrast, status achieved over an agent's lifetime can be viewed as
noise, analogous to mutation in biological evolution. I propose that
class endogamy may have resulted from the interaction of forces such
as these, along with other factors such as ideology. Simulation
results are presented, and potential areas for future research are
sketched out. The validity of these models for any particular culture
depends, of course, on whether these forces were actually operating in
that society.
\end{abstract}

%\doublespacing % begin double spacing after title

% main body

\section{Introduction}

The origin of social inequality is one of the core questions in the
social sciences. In archaeology, a related question is the origin of
{\em class endogamy,} where men and women marry only within their own
class. At first glance, this might not seem to need any explanation:
Social stratification and class endogamy were widespread in most
Western societies until relatively recently. However, many other
societies had neither.  The question thus arises as to why class
endogamy should be present in one society and not in another, and what
factors cause it to evolve.

The vast array of human societies have often been classified by
anthropologists and archaeologists into several broad
categories~\cite{steward:1955,service:1962,fried:1967,flannery:1972,flannery:1995}.
While these categories are generalizations that do not necessarily fit
all societies well, they are still useful theoretical
constructs~\cite{flannery:1995}. In {\em hunter-gatherer
bands}~\cite{ingold:etal:1988}, no one individual is allowed to
accumulate significantly more status or wealth than any other. The
decisions of a band are collective, with even the most able or
experienced members bowing to the group consensus.  In {\em
tribes}~\cite{service:1962,sahlins:1963} or {\em autonomous village
societies}, an individual is allowed to accumulate wealth and status
but must continually expend this wealth in feasts or as gifts in order
to maintain that status. A {\em big man} is a member of a tribe who
has accumulated enough status to be able to mobilize others to action;
however, a big man may be deposed if he demands too much or fails to
provide enough gifts.  In these societies, an individual's possessions
are often destroyed after death, and children do not inherit their
parents' status.  In {\em chiefdoms}~\cite{carneiro:1981,kirch:1984}
or {\em rank societies}, wealth and rank can be inherited at birth,
along with whatever individuals accumulate in their own
lifetimes. However, all members of the society are considered to be at
least distantly related to one another, and rank is continuous: There
is no chiefly or noble class. These societies are ruled by a {\em
chief} who can command others by virtue of his office alone, and the
office may be hereditary. While the chief may have a group of
lieutenants, they have no specialized functions. In contrast, in a
{\em complex chiefdom}~\cite{earle:1978,wright:1994} or {\em
stratified society}, there is a ``discontinuity in rank between chiefs
and commoners''~\cite[p. 12]{earle:1978}: The chiefs are no longer
related to the lowest-ranking members of society, and their close
relatives constitute a chiefly or noble class, who practice class
endogamy.  In addition, the chief's lieutenants now have specialized
roles in the government. Finally, in a {\em
state}~\cite{service:1962,service:1975,flannery:1972,wright:1977,carneiro:1981},
features such as a specialized bureaucracy and a standing professional
army arise.

Assuming that the first human societies were small, egalitarian bands
of hunter-gatherers~\cite{flannery:1995}, a major question in
archaeology is why sedentary agricultural societies emerged, with rank
differences or even stratification into classes. Hunter-gatherers
generally appear to have more leisure time and less disease than
sedentary farmers,
and it seems against human nature to voluntarily give a portion of
one's production to a ``macroparasite''~\cite{mcneill:1976}
such as a chief or king. Even if sedentism, agriculture, and rank
differences between individuals are explained, there remains the
question of how stratification into classes of nobles and commoners
arose.  These questions are important if we want to understand
contemporary human societies, as well as how they came to be.

This paper attempts to address the question: Given a chiefdom where
individuals both inherit status at birth and also achieve or lose
status over their lifetimes, what are the necessary and sufficient
conditions for a stratified society with class endogamy to arise? More
specifically, what is the simplest class of agent-based computer
models~\cite{schelling:1978, reynolds:1986, resnick:1994,
kohler:etal:1999, lansing:1999, kohler:gumerman:1999,
epstein:axtell:1996, axelrod:1997a, holland:1995} that produce a
result qualitatively identical to class endogamy, given a plausible
set of assumptions?  How rational must the modeled individuals be to
form classes of nobles and commoners? In other words, how stupid can
the agents be and still produce class endogamy?
 
Following Wright~\cite[p. 68]{wright:1994}, I define a {\em
noble class} as ``a ranked group whose members [compete] with each
other for access to controlling positions and stand together in
opposition to other people''. For the purposes of these computer
models, however, a {\em class} will simply be a set of men and a set
of women, each from contiguous status intervals, who practice {\em
endogamy}: Most of the time, they marry only each other.

I will show that one force that can produce class endogamy occurs if
agents are only willing to marry suitors having status no less than
some fixed value below the status of their highest-status
suitor. Furthermore, agents can easily learn what the status of their
highest-status suitor is. Another such force results if children
inherit the average of their parents' statuses. In contrast, status
achieved over an agent's lifetime can be viewed as noise, analogous to
mutation in biological evolution. I propose that class endogamy may
have been produced by the interaction of forces such as these, along
with other factors such as ideology~\cite{wright:1994}. Whether these
models have any bearing on how class endogamy actually evolved in any
particular human society must of course be tested against the
ethnographic, ethnohistorical, archaeological, and psychological
data. However, such testing is beyond the scope of this paper.

The paper is organized as follows: This section has laid out the
background and the question. The next section,
Section~\ref{sec:arch-model}, introduces a verbal model from
archaeology that has been proposed to explain the evolution of class
endogamy. Section~\ref{sec:econ-model} describes a mathematical model
from economics that will be generalized and simplified to implement
this verbal model as a class of agent-based computer models, which are
presented in Section~\ref{sec:computer-models}, along with their
results. Section~\ref{sec:conclusion} summarizes the paper, and
Section~\ref{sec:future-work} outlines some avenues for future
research.

\section{A Verbal Model from Archaeology} \label{sec:arch-model}

My starting point for this paper is a verbal model of the emergence of
class endogamy proposed by Marcus and Flannery:
\begin{quote}
In most chiefdoms, there is a continuum of differences in rank from
top to bottom. People are ranked in terms of their genealogical
distance from the paramount chief, with the lowliest persons being
very distantly related indeed. In most archaic states, there was an
actual genealogical gap between the stratum of nobles and the stratum
of commoners. Lesser nobles knew they were at least distantly related
to the king. Commoners were not considered related to him at all. As
we have seen, the two strata were kept separate by class endogamy, the
practice of marrying within one's class.

There are many possible scenarios for the evolution of stratified
society out of chiefly society. An actor-centered explanation might
begin with a chief's need to ensure that his offspring would succeed
him. The only way he could ensure that goal was by marrying the
highest-ranking woman available. The genealogical gap mentioned earlier
could arise through intense competition for the most advantageous
marriages. Eventually the more genealogically distant members of
society --- marriage to whom would only condemn one's offspring to
lower rank --- might have their kin ties to the elite severed. This
apparently happened to low-ranking families in Hawaii just prior to
state formation.\cite[p. 181]{marcus:flannery:1996}
\end{quote}
The main goals of this paper are to explore how to implement Marcus
and Flannery's verbal model as a class of agent-based computer models
and to examine whether they in fact produce class endogamy as
expected. Computer models have the virtue that the outcomes follow
rigorously from the model's assumptions. Furthermore, the model's
assumptions and mechanisms must be specified explicitly.

Specifically, my goal is to find the {\em simplest} class of
agent-based computer models that produce class endogamy, starting with
a simple ranked society. In contrast to many computer simulations in
the archaeological and anthropological literature, I do not attempt to
include every feature that appears to be relevant; rather, my goal is
to include the minimal set of features necessary and sufficient to
produce class endogamy. I do this for several reasons: First,
following Occam's razor, it makes little sense to design a complicated
model when a simpler model suffices. Second, a simple model is easier
to implement and debug than a complicated one. Finally, a simple model
produces results that are easier to understand. If the model were as
complicated as the real system, the results would be much more
realistic than a simple model, but just as hard to understand as the
real world.

In this paper, I also make no attempt to incorporate empirical data
such as status distributions from real societies. While such data
would be valuable for the validation of the model for any particular
society, in this paper I have adopted a data-independent approach, in
order to construct a general class of models. Any empirical data that
might be incorporated would be at least partially incorrect,
especially for societies that no longer exist. A model that held for
{\em all} possible status distributions would obviously be inherently
% troy
more general and reliable than one constructed under the assumption of
any specific ``real'' status distribution. Any illusion of support
given by such data would be outweighed by the risk that the data were
in fact wrong, since in that case, a model built around that data
could be fatally flawed. Of course, once a class of general,
data-independent models has been developed, any empirical data that
are available can and should be used to test whether the models are
applicable to a particular society. Such testing is beyond the scope
of this paper, however.

As Renfrew writes concerning modeling in archaeology,
\begin{quote}
\dots [A]ny models which we may set up are likely to be over
simplified, but I believe the effort is worthwhile. For it then allows
us, when considering any specific case of change, to refer it back to
the general models which we have, and to see if they are of help in
explaining what has been observed. In many cases I believe that they
are. This undertaking of attempting some sort of explanation through
generalization is what is termed in contemporary archaeology the
processual approach. It has the merit of making our explanations
explicit, which is a very effective way of bringing their weaknesses
to light, and hence also of investigating their
strengths.~\cite[pp. 120--121]{renfrew:1987}
\end{quote}

\section{A Mathematical Model from Economics} \label{sec:econ-model}

How might Marcus and Flannery's model be implemented on a computer,
and how exactly might class endogamy emerge from agents competing for
high-status spouses? One possible source of answers for these
questions is economics, where much work has been done to model
marriage markets and the related domain of labor markets. In
particular, Burdett and Coles~\cite{burdett:coles:1997} have shown
mathematically that classes emerge endogenously in a marriage market
given certain assumptions. They considered a population of agents that
decide whether to marry one another based on their respective
``pizazz'', or desirability, which is modeled as a real number. The
utility an agent receives from a marriage is equal to its spouse's
pizazz, discounted according to how long it had to wait before
marrying.  Burdett and Coles assume that the agents try to maximize
their utility, and that they know the rate at which they meet agents
of the opposite sex, as well as the distribution of pizazz for those
agents. Under these assumptions, they show that an equilibrium exists,
in which each agent will use a strategy to maximize its expected
discounted lifetime utility by only proposing to agents who have
pizazz equal to or greater than this expected utility, which can be
calculated.  Furthermore, endogamy emerges: The individuals partition
themselves into discrete groups, where the males and females are each
divided into contiguous intervals of pizazz, and females will only
marry males from the same group, and vice versa.  In a second
paper~\cite{burdett:coles:2001}, they extended their model to consider
the effect of self-improvement, i.e., pizazz that an agent can achieve
at some cost during its lifetime. Endogamy also emerges in this second
model, and they prove results concerning the amount of
self-improvement a given agent will choose to invest in.

If status is substituted for pizazz, and achieved status for
self-improvement, then these results can be transfered directly to the
question of how class endogamy emerged in past societies.
Furthermore, this mathematical model could be directly implemented on
a computer, since each agent's optimal strategy can be
calculated. Hence, Burdett and Coles have provided one possible
operationalization of Marcus and Flannery's verbal model and verified
it mathematically, within the context of their assumptions.

This is a significant and interesting result. However, Burdett and
Coles implicitly assume that, in order to determine who they should
propose to, agents are able to calculate which agents are willing to
marry them, the rate at which they encounter such agents, and the
pizazz distribution of those agents.  Anecdotal evidence from
contemporary American society gives us some reason to doubt that men
and women actually know this information in general (though it might
in fact be true for the relatively simple, low-population societies we
are considering in this paper). Furthermore, their model presumes that
agents calculate an optimal strategy; while this allows them to prove
that an equilibrium exists, it begs the question of whether such
optimal strategies are actually necessary to produce endogamy. Are
there other strategies that produce the same effect? {\em Why} exactly
does this type of strategy produce endogamy? And in particular, are
there simple heuristics that also result in endogamy?  Can the agents
be made less rational, i.e., more stupid? The next section generalizes
and simplifies Burdett and Coles's proof in order to answer these
questions.

\subsection{Burdett and Coles's Proof, Generalized and Simplified} \label{sec:proof}

Suppose that agents with status $s$ will only marry suitors with
status $s' \geq s'_{\mathrm{max}}(s) - f(H(s)),$ where
$s'_{\mathrm{max}}(s)$ is the status of the highest-status agent
willing to marry an agent with status $s,$ and $f(H(s)) \geq 0$ is
some function of the distribution of status $H(s)$ among those willing
to marry an agent with status $s.$ This section will prove that, under
those circumstances, class endogamy results.

This proof relies on the following assumptions regarding status:
First, females all use a single, uniform status metric to evaluate
male suitors, and vice versa. Second, the status of any two agents can
be compared, and these comparisons are transitive, i.e., $s(a_{1})
\geq s(a_{2}) \wedge s(a_{2}) \geq s(a_{3}) \Rightarrow s(a_{1}) \geq
s(a_{3}),$ where $s(a_{i})$ is the status of agent $a_{i},$
represented by a real number. Third, agents are always willing to
marry the highest-status agent willing to marry them.  Fourth, all
males will propose to the highest-status female, with status
$s_{\mathrm{max}}^{\mathrm{f}},$ and all females will propose to the
highest-status male, with status $s_{\mathrm{max}}^{\mathrm{m}}.$
Fifth, if an agent is willing to marry a suitor with a given status,
then it is willing to marry all higher-status suitors: $M(a_{1},
a_{2}) \wedge s(a_{2}) \leq s(a_{3}) \Rightarrow M(a_{1}, a_{3}),$
where $M(a_{i}, a_{j})$ is true if and only if an agent $a_{i}$ is
willing to marry suitor $a_{j}.$ Finally, the threshold for accepting
a suitor is nondecreasing with increasing status: $M(a_{1}, a_{3}) \wedge
s(a_{1}) \geq s(a_{2}) \Rightarrow M(a_{2}, a_{3}).$ (This may seem to
be a large number of assumptions to make; however, they are all
plausible statements about the way status operates.)

\begin{figure}
  \begin{center}
    \includegraphics[width=1.5in]{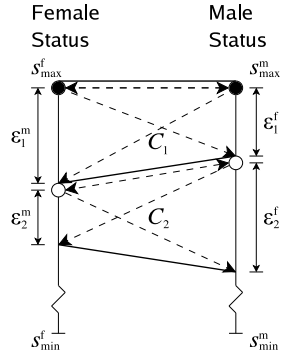}%
  \end{center}
  \caption{Class formation under the proof in
  Section~\ref{sec:proof}. The highest-status female and male, with
  selectivity constants $\epsilon^{\mathrm{f}}_{1} =
  f^{\mathrm{f}}(H(s_{\mathrm{max}}^{\mathrm{f}}))$ and
  $\epsilon^{\mathrm{m}}_{1} =
  f^{\mathrm{m}}(H(s_{\mathrm{max}}^{\mathrm{m}})),$ respectively, are
  represented by solid black circles. The dashed lines indicate the
  agents that the highest-status female and male are willing to marry,
  respectively. All males and females between the first two horizontal
  solid black lines form the first class $C_{1}$ and only marry each
  other. The highest-status female and male outside of $C_{1}$ are
  represented by white circles and have selectivity constants
  $\epsilon^{\mathrm{f}}_{2}$ and $\epsilon^{\mathrm{m}}_{2}.$ All
  agents between the second set of horizontal solid lines form class
  $C_{2}.$ The remaining classes are not shown.}
  \label{fig:classes}%
\end{figure}

The proof is as follows: All males will propose to the highest-status
female, with status $s_{\mathrm{max}}^{\mathrm{f}}.$ This female, in
turn, will propose to the highest-status male, with status
$s_{\mathrm{max}}^{\mathrm{m}},$ as well as to all other males who
have status $s' \geq s_{\mathrm{max}}^{\mathrm{m}} -
f(H(s_{\mathrm{max}}^{\mathrm{f}})).$ Now, since the highest-status
female will propose to these males, all females will propose to
them. Hence, these males all face the same situation as the
highest-status male and will all follow the same strategy. Call this
set of males $C_{1}^{\mathrm{m}}.$ They are all willing to accept the
highest-status female, along with all other females with status $s'
\geq s_{\mathrm{max}}^{\mathrm{f}} -
f(H(s_{\mathrm{max}}^{\mathrm{m}})).$ Call that set of females
$C_{1}^{\mathrm{f}}.$ Those females, in turn, all face the same
situation as the highest-status female and thus use the same strategy
as she does: Propose only to the males in $C_{1}^{\mathrm{m}}.$ Thus,
the females in $C_{1}^{\mathrm{f}}$ will marry only the males in
$C_{1}^{\mathrm{m}},$ and vice versa. $C_{1}^{\mathrm{f}}$ and
$C_{1}^{\mathrm{m}}$ form a class $C_{1},$ which constitutes the
highest social stratum. Figure~\ref{fig:classes} shows the argument
graphically.

To form the next highest class $C_{2},$ remove the individuals in
$C_{1}$ from the population and apply the same reasoning, starting
with the new highest-status female and male, producing
$C_{2}^{\mathrm{f}}$ and $C_{2}^{\mathrm{m}},$ which together
constitute $C_{2}.$ Repeat this procedure, forming classes until there
are either no females or no males remaining. Note that it is possible
for one of the sexes to have one more class than the other, in which
case the members of that lowest class will not be able to find anyone
willing to marry them.

This generalization of Burdett and Coles's proof shows that the
fundamental mechanisms underlying this form of class endogamy are
first, that the agents all decide whether or not to marry a suitor
based on its status relative to the status of the highest-status agent
willing to marry them, and secondly, that there is an upper bound on
status. The function $f(H(s))$ determines how selective an agent
is. The rational strategy derived by Burdett and Coles, where agents
maximize their expected discounted lifetime utility, is thus a special
case of this more general result.  (However, their result is more
general in the sense that it considers discounting.) In fact, for
class endogamy to emerge, it suffices if $f(H(s)) = \epsilon,$ where
$\epsilon$ is a non-negative integer constant; thus it is not
necessary for the agents to know anything about $H(s)$ other than
$s'_{\mathrm{max}}(s).$ The proof still holds if males and females
have different selectivity functions $f^{\mathrm{m}}(H(s))$ and
$f^{\mathrm{f}}(H(s)).$ Note also that it does not depend on the
particular status distribution in the society. (Furthermore, it is
interesting to note that it actually does not depend on the fact that
there are sexes at all: The results would still hold if the agents
were hermaphroditic.) The proof also provides a procedure for
calculating the class boundaries, as well as the highest-status suitor
for any given agent.

This generalized strategy is not necessarily optimal; hence, unlike
Burdett and Coles, I have not shown that an equilibrium exists in
which rational agents will all converge to using this
strategy. However, if a population of agents do all use this strategy,
then class endogamy will result. In an unstratified simple chiefdom,
such as is assumed in these models, it seems plausible to assume that
the agents are undifferentiated and will all use the same
strategy. The next section introduces a class of agent-based computer
models to verify these results.

\section{A Class of Computer Models} \label{sec:computer-models}

The basic modeling framework, adapted from Burdett and
Coles~\cite{burdett:coles:1997, burdett:coles:2001}, consists of a
population of agents, representing a chiefdom or village. Each agent
can be male or female and has an integer status that determines its
rank within the society. Part of this status is inherited at birth,
and part may be achieved (or lost) over the agent's lifetime. The
population is initialized with some number of agents, with each
agent's inherited status drawn randomly from a uniform
distribution. At each time step thereafter, one randomly-chosen female
encounters one randomly-chosen male. If both agents find the other
acceptable, based on their respective statuses, they marry and produce
a pair of children. The parents are removed from the population, and
the children are added to it. However, if at least one of the agents
rejects the other, they do not marry but instead remain in the
population. This cycle continues until some predetermined number of
marriages has occurred.  The agents live forever, until they marry,
and they do not become less selective the longer they remain single
--- there is no discounting.

In these models, the population was initialized with $N = 10,000$
agents, where each agent had a 50\% chance of being male or female,
and each one had an inherited status $s_{\mathrm{i}}$ drawn from a
uniform probability distribution between $s_{\mathrm{min}} = 0$ and
$s_{\mathrm{max}} = 99.$ The models ran until $t_{\mathrm{max}} =
100,000$ marriages had occurred. Fifty runs of each model were
performed, each run initialized with a different random number seed.

\subsection{Cloned Offspring} \label{sec:cloned}

To keep things simple, agents in the first set of models did not have
any achieved status. If a female agent encountered a male agent and
both agreed to marry, each parent produced a clone with the same sex
and status, and the parents were removed from the population. The
distributions of sex and status in the population thus remained
constant over time.

\subsubsection{Strategy $\mathbf{S}_{1}$: Rationality} \label{sec:rational}

I have already shown mathematically in Section~\ref{sec:proof} that
class endogamy results if agents all follow a strategy
$\mathbf{S}_{1}$ of only accepting suitors who have a status $s' \geq
s'_{\mathrm{max}}(s) - \epsilon,$ where $s'_{\mathrm{max}}(s)$ is the
status of the highest-status agent willing to marry an agent with
status $s,$ and $\epsilon$ is some non-negative integer. In this
section, I verify this result using an agent-based computer model,
model $\mathbf{M}_{1}.$ When a female agent encountered a male agent
in this model, both used strategy $\mathbf{S}_{1},$ with $\epsilon =
9.$

\begin{figure}
  \begin{center}
    \subfigure[Female status histo\-gram, for one run]{%
      \includegraphics[width=1.5in]{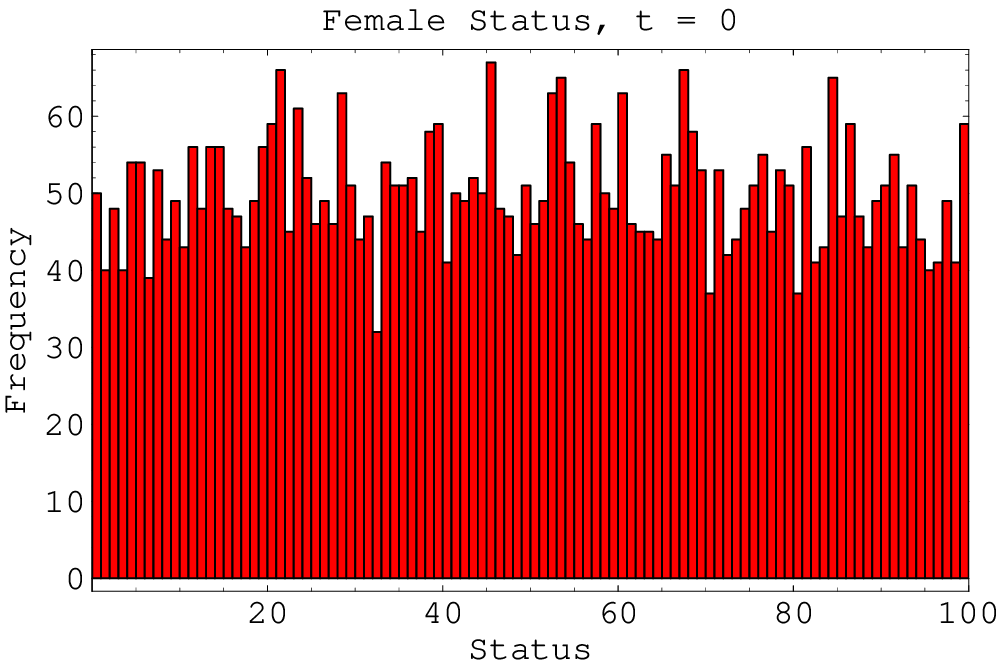}%
      \label{fig:rational-cloning-noachieved.1.fHistogram.0}%
    }
    \goodgap
    \subfigure[Male status histo\-gram, for one run]{%
      \includegraphics[width=1.5in]{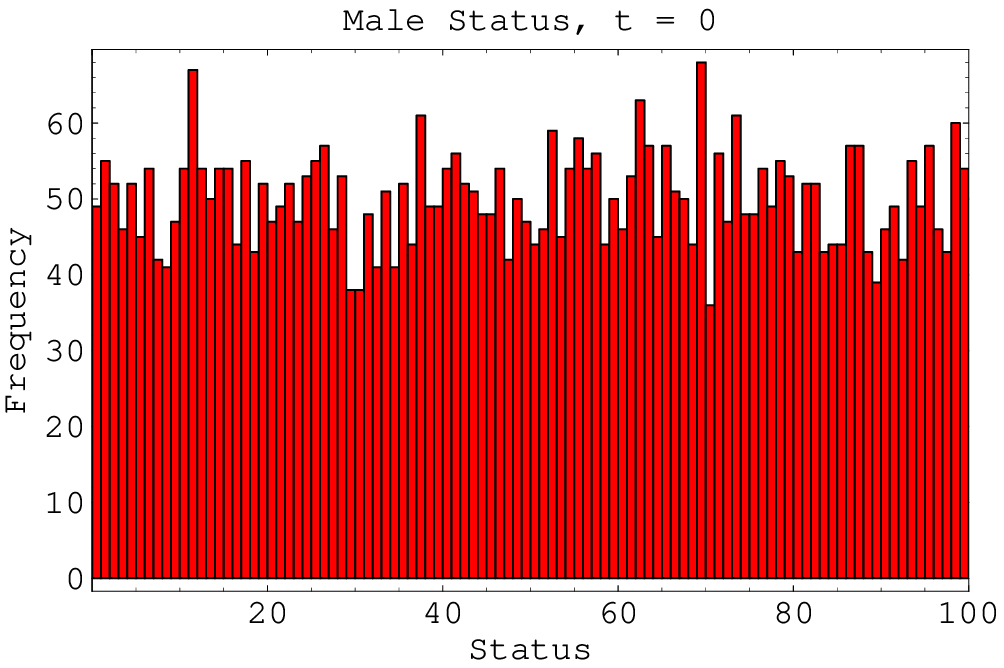}%
      \label{fig:rational-cloning-noachieved.1.mHistogram.0}%
    }
  \end{center}
  \caption{Female and male status histograms for one typical run at $t
  = 0$ marriages, for the model $\mathbf{M}_{1}$ from
  Section~\ref{sec:rational}, using the rational strategy
  $\mathbf{S}_{1}.$ Children were clones of their parents, and agents
  did not achieve any status over their lifetimes. Thus, the histogram
  remained constant over the entire run. The status histograms for the
  other models at $t = 0$ were very similar.}
  \label{fig:rational-cloning-noachieved.1.histograms.0}%
\end{figure}

Every $10,000$ marriages, the number of males and females at each
status level were recorded. The female and male status histograms from
a single typical run are plotted at $t = 0$ marriages in
Figure~\ref{fig:rational-cloning-noachieved.1.histograms.0}; the
distributions remain constant over the entire run.

\begin{figure}
\begin{center}
\subfigure[Marriage frequency, \newline for one run]{%
	\includegraphics[width=1.5in]{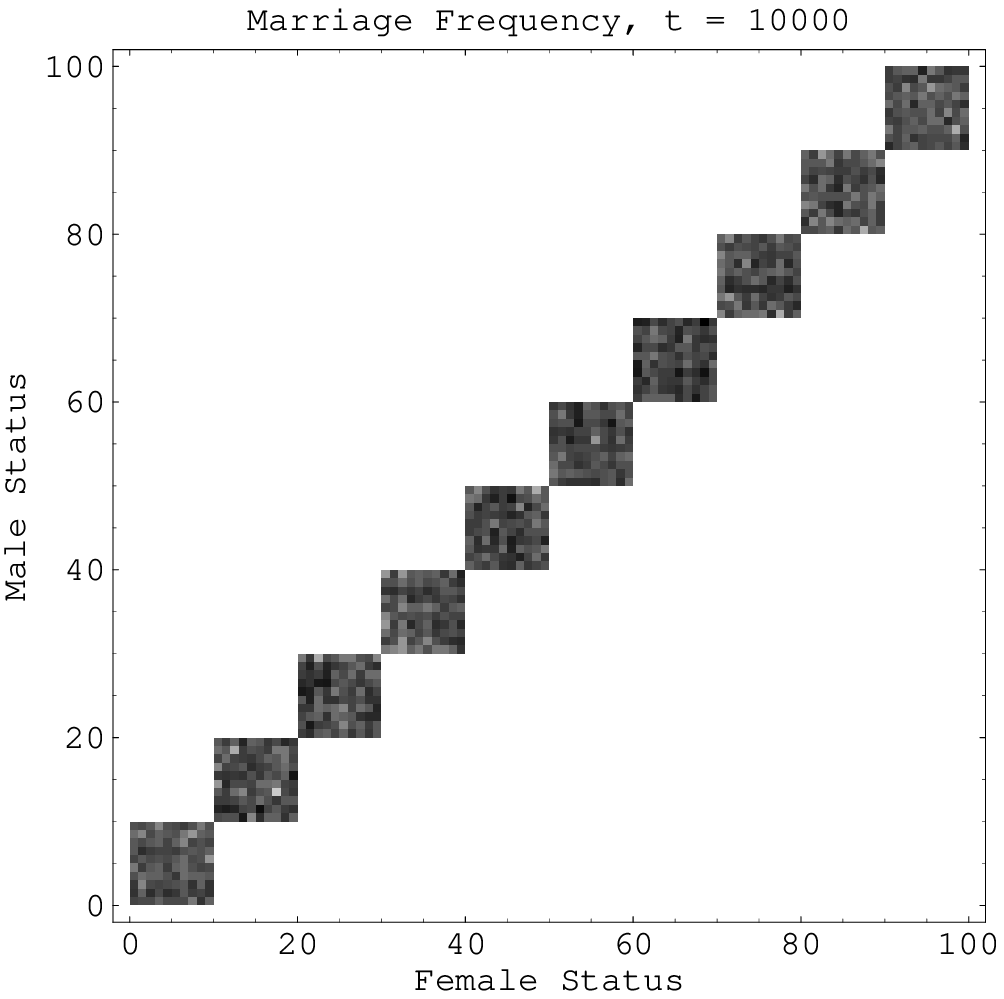}%
	\label{fig:rational-cloning-noachieved.1.marriages.10000}%
}%
\goodgap
\parbox{1.5in}{%
\subfigure[Hypergamy $h,$ for one run]{%
	\includegraphics[width=1.5in]{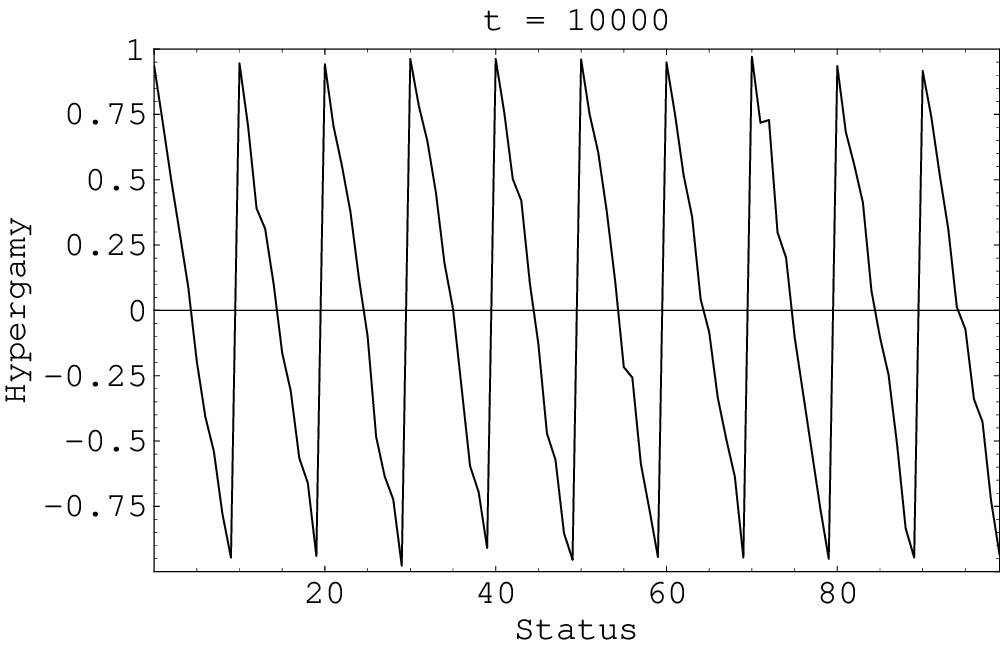}%
	\label{fig:rational-cloning-noachieved.1.hypergamy.10000}%
}%
\\
\subfigure[Hypergamy $h,$ for 50 runs]{%
	\includegraphics[width=1.5in]{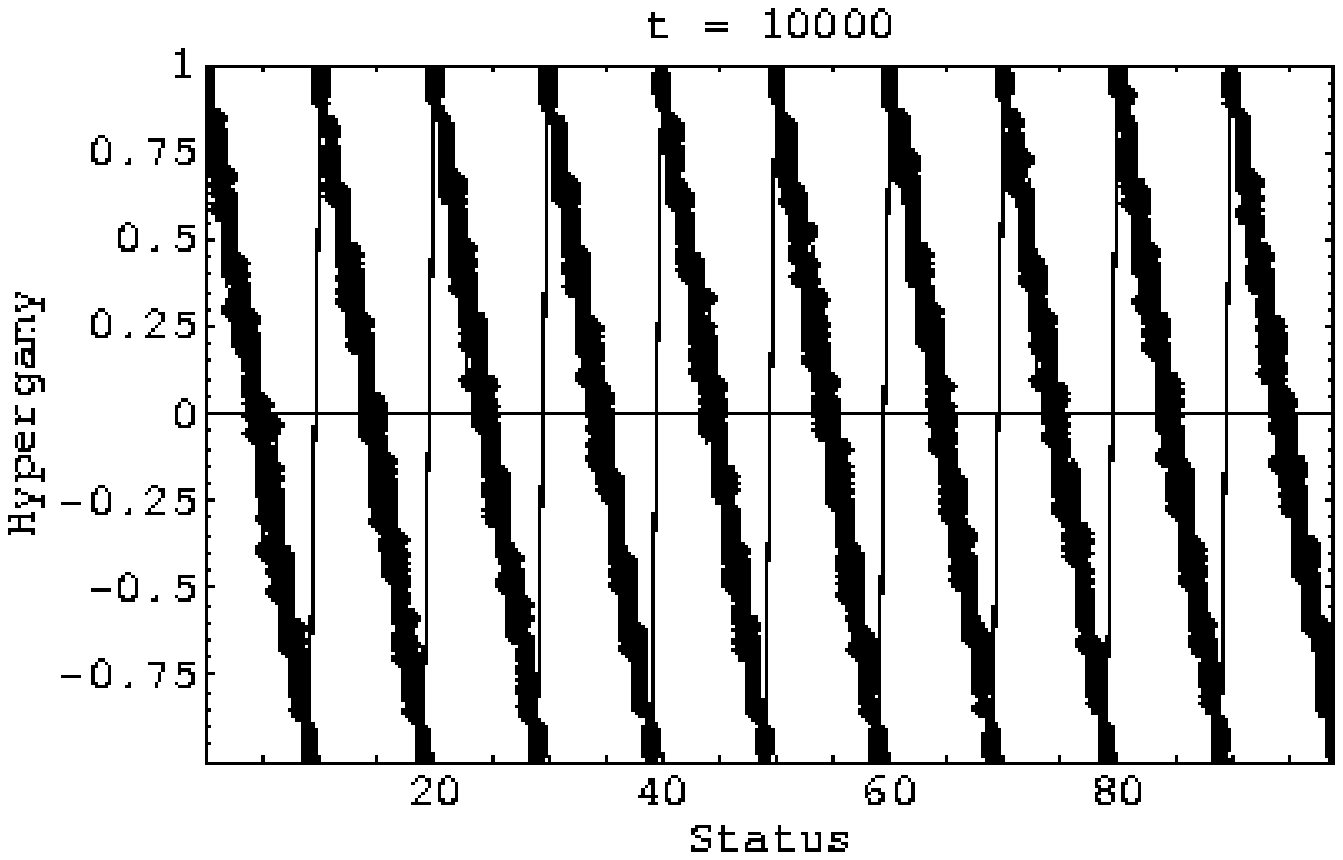}%
	\label{fig:rational-cloning-noachieved.multiHypergamy.10000}%
}%
}%
\goodgap
\parbox{1.5in}{%
\subfigure[Female status histo\-gram, for one run]{%
	\includegraphics[width=1.5in]{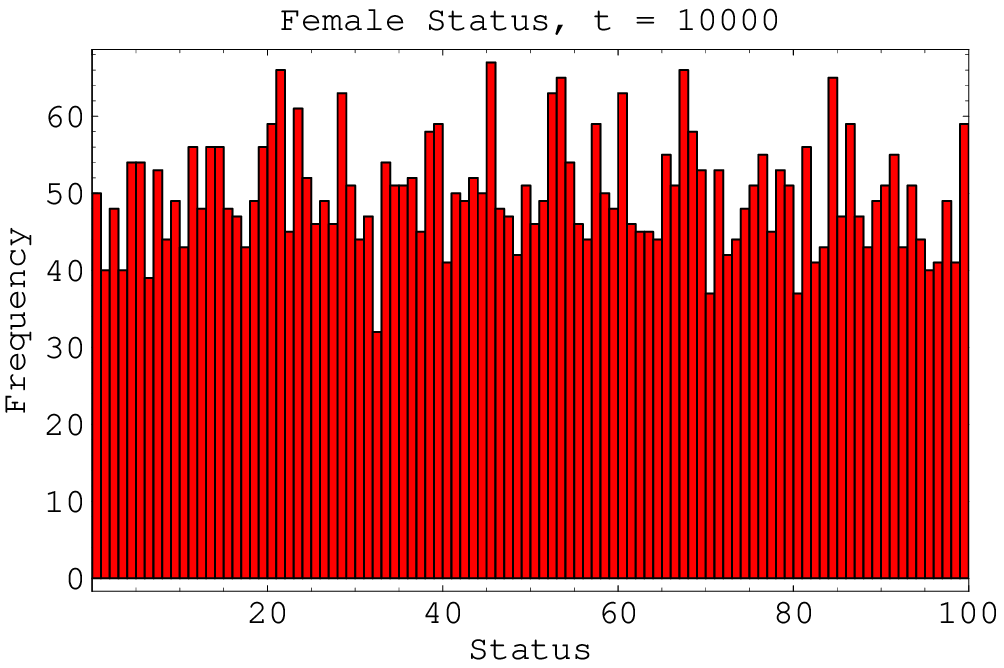}%
	\label{rational-cloning-noachieved.1.fHistogram.10000}%
}%
\\
\subfigure[Male status histo\-gram, for one run]{%
	\includegraphics[width=1.5in]{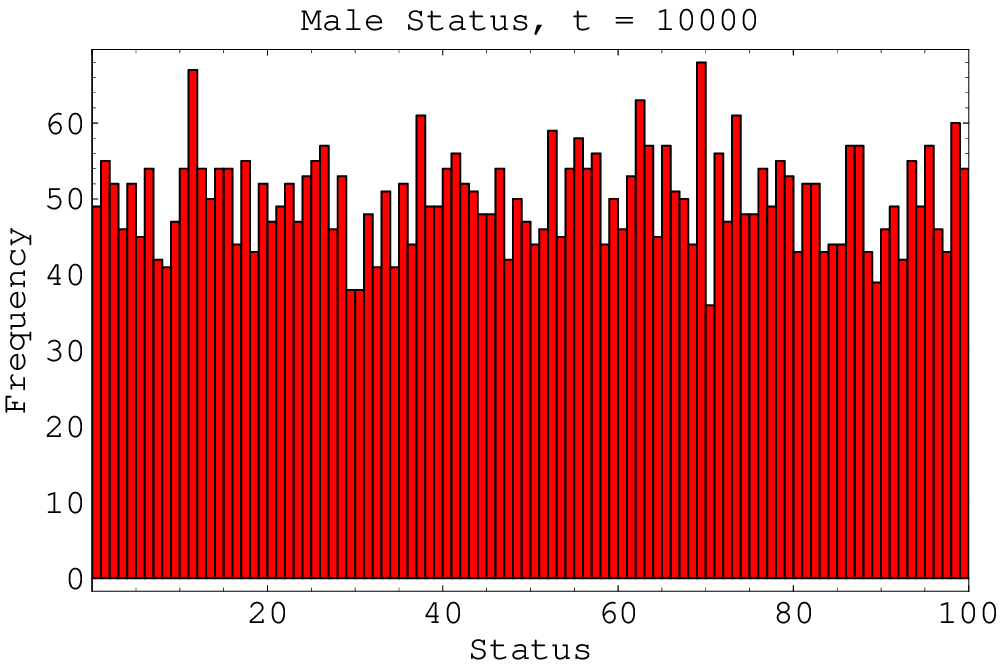}%
	\label{rational-cloning-noachieved.1.mHistogram.10000}%
}%
}%
\end{center}
\caption{Marriage frequency, hypergamy $h,$ and status histograms
at $t = 10,000$
marriages, for the model $\mathbf{M}_{1}$ from
Section~\ref{sec:rational}, using the rational strategy
$\mathbf{S}_{1}.$ Children were clones of their parents, and agents
did not achieve any status over their lifetimes. These plots changed
very little at later timesteps. The status histograms remained constant 
over the entire run.}
\label{fig:rational-cloning-noachieved.1.10000}%
\end{figure}

In addition, the marriage frequencies over the previous $10,000$
marriages were recorded for each combination of female status and male
status.  These statistics were calculated every $10,000$ marriages,
beginning at $t = 10,000$ marriages.  This is plotted for a single
typical run at $t = 10,000$ marriages in
Figure~\ref{fig:rational-cloning-noachieved.1.marriages.10000}; the
plots at other sample times are very similar.\footnote{The marriage
frequency plots in this version of the paper contain some artifacts
due to the default PDF distillation process used by
\texttt{arXiv.org}; because of this, they appear slightly fuzzier than
they should. For the original figures, see the version at the author's
homepage.}
In these plots, darker points represent more
marriages than lighter ones. (The frequency data were scaled
logarithmically before plotting, to increase the plots' legibility.)

To help analyze the marriage frequency data, a {\em hypergamy metric,}
$h,$ was also calculated every $10,000$ marriages. {\em Hypergamy} is a
situation in which an individual marries someone of higher status. The
opposite situation, where an individual marries someone of lower
status, is called {\em hypogamy}. The metric $h$ is defined as
follows:
\begin{equation}
h(s, t) = \frac{m_{+}(s, t) - m_{-}(s, t)}{m_{+}(s, t) + m_{-}(s, t) + m_{=}(s, t)} ,
\end{equation}
where
\begin{equation}
m_{+}(s, t) = \sum_{t' = t - \Delta t}^{t} \: \sum_{i = s + 1}^{s_{\mathrm{max}}} m_{s, i}(t')
\end{equation}
is the number of hypergamous marriages for agents with status $s$ over
the sampling period ending at time $t,$
\begin{equation}
m_{-}(s, t) = \sum_{t' = t - \Delta t}^{t} \: \sum_{i = s_{\mathrm{min}}}^{s - 1} m_{s, i}(t')
\end{equation}
is the number of hypogamous marriages for those agents over the same
period, and
\begin{equation}
m_{=}(s, t) = \sum_{t' = t - \Delta t}^{t} m_{s, s}(t') 
\end{equation}
is the number of marriages with agents of equal status. In these
equations, $s$ is an individual's status, $t$ is the number of
marriages that have elapsed, $\Delta t = 10,000$ is the sampling
interval, $s_{\mathrm{min}}$ and $s_{\mathrm{max}}$ are the minimum
and maximum possible statuses, respectively, and $m_{i, j}(t)$ is the
number of individuals of either sex having status $i$ and $j$ that
married at time $t$. The hypergamy metric $h$ ranges between $1$ and
$-1.$ If individuals with status $s$ have only hypergamous marriages,
i.e., they marry only individuals with higher status, then $h(s) = 1;$
conversely, if the individuals have only hypogamous marriages, then
$h(s) = -1.$ If the number of hypergamous marriages equals the number
of hypogamous marriages, or if the individuals only marry individuals
with exactly the same status $s,$ then $h(s) = 0.$ This value $h$ is
plotted for each status level in
Figure~\ref{fig:rational-cloning-noachieved.1.hypergamy.10000}, for a
single typical run, and in
Figure~\ref{fig:rational-cloning-noachieved.multiHypergamy.10000} for
all $50$ runs, to show the variation between runs. (If no marriages
occurred for a given status level, $h$ is undefined, and that point is
simply skipped.)

% rational-cloning-noachieved
%

This hypergamy metric $h$ can be used as a proxy for measuring the
degree of class endogamy: If the hypergamy changes quickly from a low
value to a high value as $s$ increases, then class endogamy is
present. The points at which these transitions occur represent class
boundaries. (The high values of $h$ near $s_{\mathrm{min}} = 0$ and
the low values near $s_{\mathrm{max}} = 99$ are artifacts of the
boundary conditions at these points.)  The metric does not work well
as a proxy for endogamy when the number of status levels in a class is
very close to $1,$ as will be evident later in
Section~\ref{sec:inherited}.

As can be seen in
Figure~\ref{fig:rational-cloning-noachieved.1.10000}, classes emerge
immediately when agents all use strategy $\mathbf{S}_{1},$ as
expected. This model produces $10$ classes when $\epsilon = 9$, while
archaic stratified societies probably only had two classes, in
general: nobles and commoners~\cite{flannery:1995}. This might seem to
invalidate the model; however it can be easily modified to produce
only two classes: Simply set $\epsilon = (s_{\mathrm{max}} -
s_{\mathrm{min}} + 1) / 2 - 1 = 49.$ Producing $10$ classes is, in
fact, a more rigorous test of the model than merely producing two
would be. In general, set $\epsilon = (s_{\mathrm{max}} -
s_{\mathrm{min}} + 1) / n - 1,$ where $n$ is the number of classes.

This simulation merely verifies what was already proven mathematically
in Section~\ref{sec:proof}. Furthermore, strategy $\mathbf{S}_{1}$ is
less than satisfying, since in order for an agent to calculate the
highest-status agent willing to marry it, it must first calculate what
class it lies in, using the procedure from the proof in
Section~\ref{sec:proof}. Hence, the agents are in essence forming
classes by calculating a priori what class they are in. (The same
applies to Burdett and Coles's
model~\cite{burdett:coles:1997,burdett:coles:2001}.) Is it possible
for the agents to {\em learn} what the highest-status agent willing to
marry them is, so that classes emerge without such calculations? If
so, this circularity could be broken.

\subsubsection{Strategy $\mathbf{S}_{2}$: Learning} \label{sec:learning}

The goal for this section is to show that there is a simple algorithm
$\mathbf{S}_{2}$ that agents can use to learn the status of the
highest-status agent willing to marry them.  Once such an algorithm
was found, no effort was made to improve on it or analyze it in
detail; this is merely an existence proof that such learning is
possible. A simpler or more efficient algorithm may well exist. The
algorithm is as follows: For each status level $s,$ keep a list of the
last $n_{\mathrm{e}}$ encounters that females with status $s$
experienced with males, along with their outcomes. Keep a similar set
of lists of males' encounters with females. Initially, no encounters
have occurred, and the lists are empty. Every time an encounter
occurs, update the appropriate male and female encounter lists. If
$n_{\mathrm{e}}$ encounters have already been recorded for a given
status and sex, then remove the oldest encounter before adding the
most recent one to the list. If the encounter list corresponding to an
agent's status $s$ is not full, the agent uses strategy
$\mathbf{S}_{3}$: Only accept suitors that have status at least equal
to $s - \epsilon,$ where $s$ is the agent's own status, and $\epsilon$
is some non-negative integer. This implicitly defines an interval with
radius $\epsilon$ around an agent's status $s,$ encompassing the
agent's potential mates. (Note that strategy $\mathbf{S}_{3}$ requires
that males and females both use the same status scale
$[s_{\mathrm{min}}, s_{\mathrm{max}}]$ to evaluate the opposite sex.)
Once an agent's encounter list is full, it switches to strategy
$\mathbf{S}_{1},$ described in Section~\ref{sec:rational}, where
$s'_{\mathrm{max}}(s)$ is defined to be the status of the
highest-status agent of the opposite sex that was willing to marry an
agent with status $s$ within the last $n_{\mathrm{e}}$ encounters. In
these simulations, $n_{\mathrm{e}} = 2 (s_{\mathrm{max}} -
s_{\mathrm{min}} + 1) = 200$ and $\epsilon = 9.$

\begin{figure}
\begin{center}
\subfigure[Marriage frequency, \newline for one run]{%
	\includegraphics[width=1.5in]{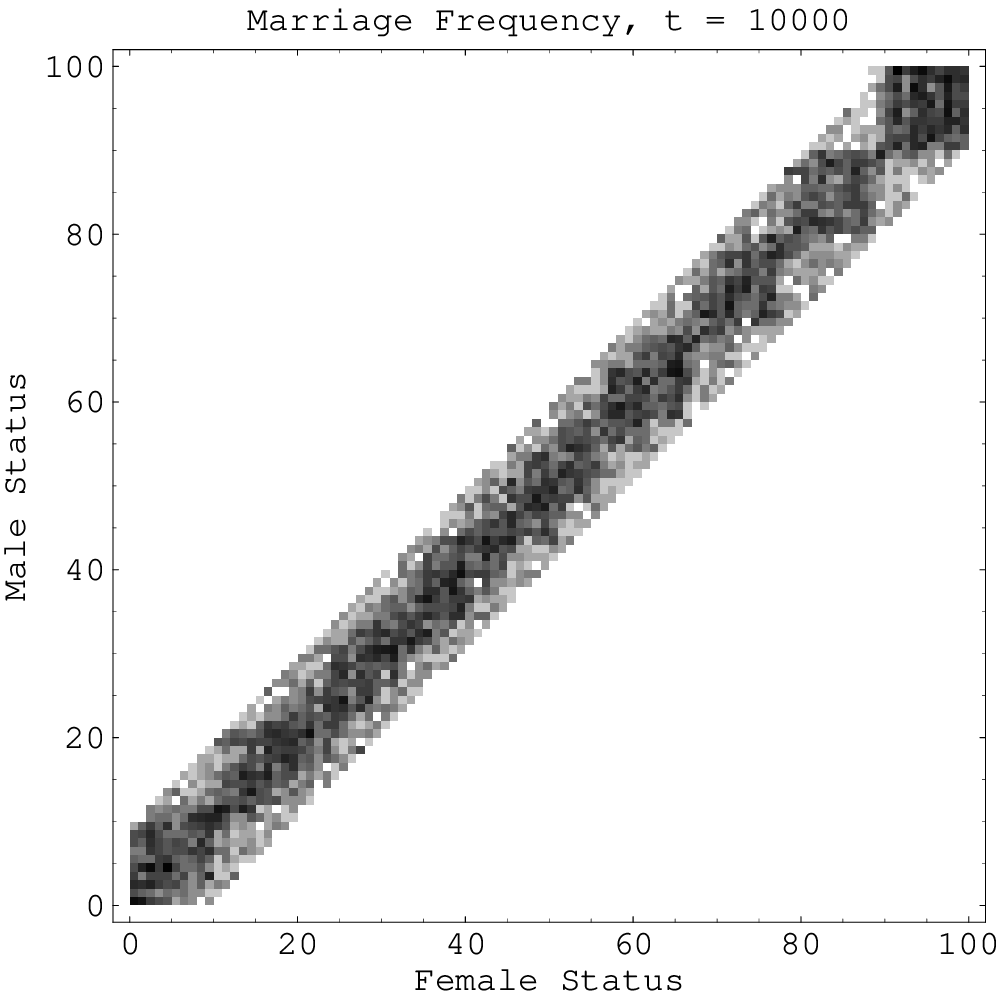}%
	\label{fig:learning-cloning-noachieved.1.marriages.10000}%
}%
\goodgap
\parbox{1.5in}{%
\subfigure[Hypergamy $h,$ for one run]{%
	\includegraphics[width=1.5in]{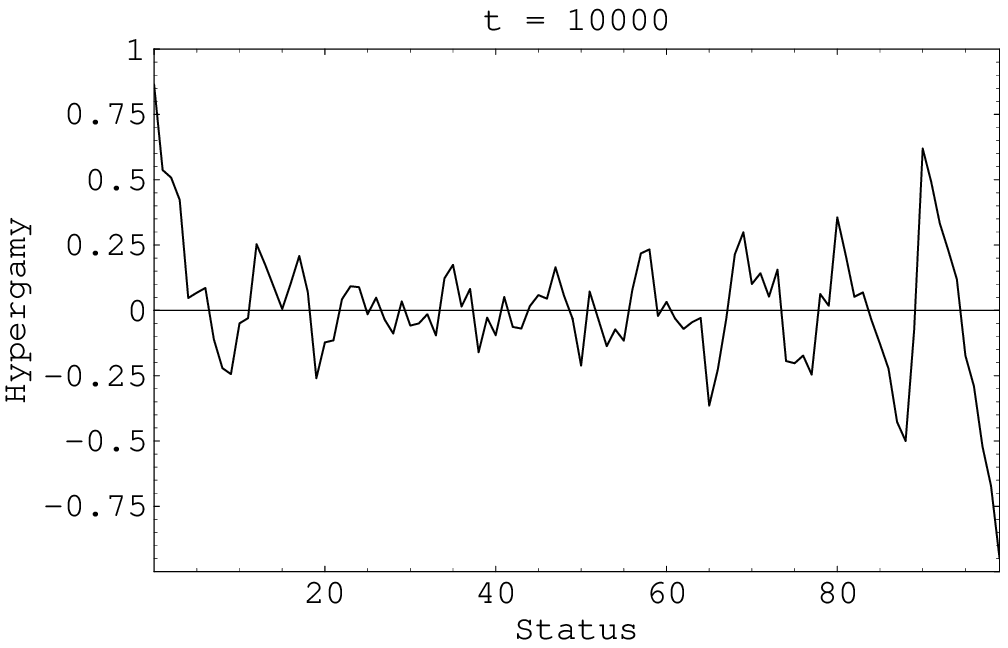}%
	\label{fig:learning-cloning-noachieved.1.hypergamy.10000}%
}%
\\
\subfigure[Hypergamy $h,$ for 50 runs]{%
	\includegraphics[width=1.5in]{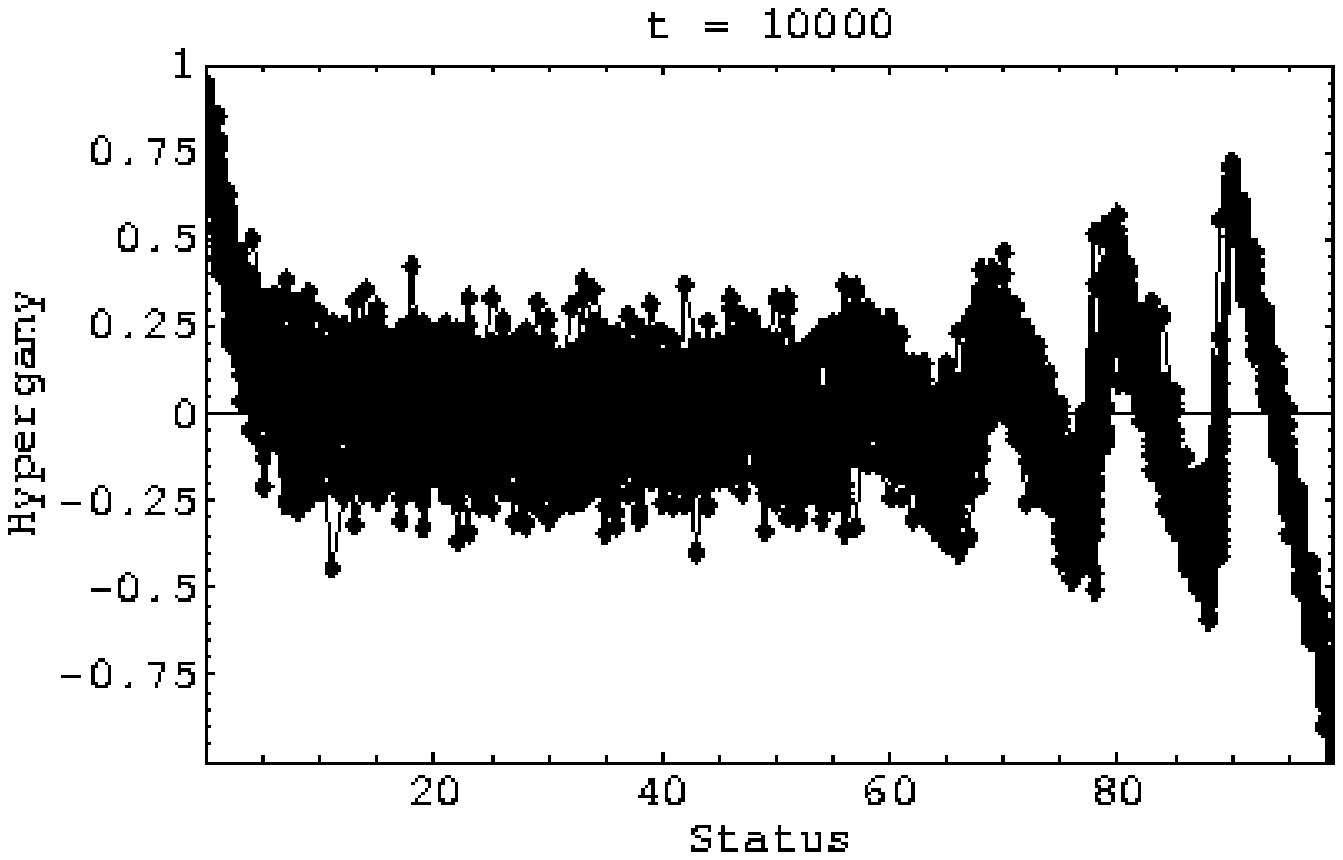}%
	\label{fig:learning-cloning-noachieved.multiHypergamy.10000}%
}%
}%
\goodgap
\parbox{1.5in}{%
\subfigure[Female status histo\-gram, for one run]{%
	\includegraphics[width=1.5in]{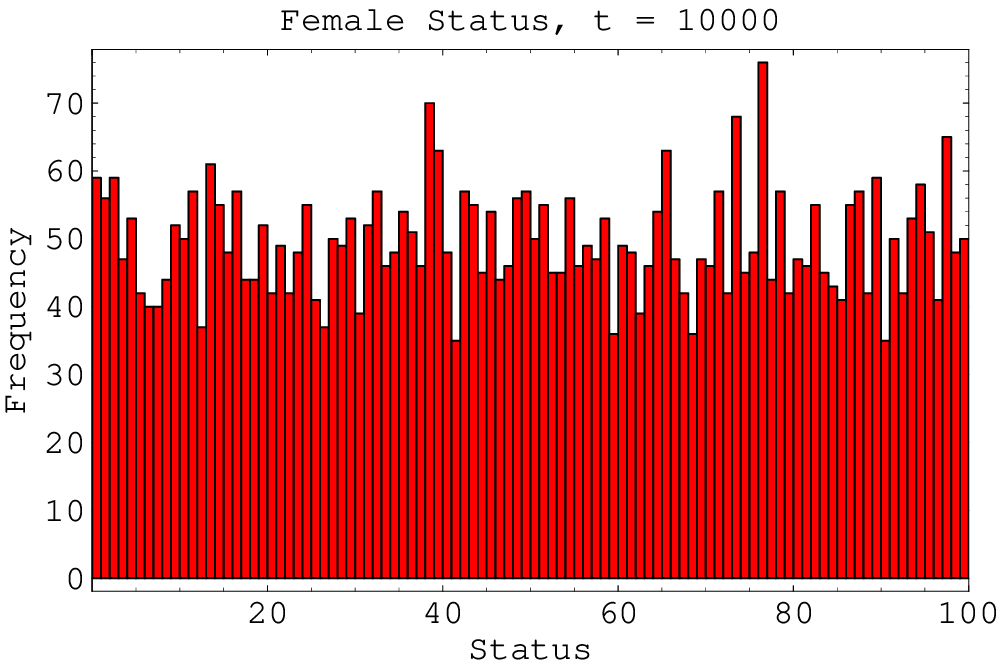}%
	\label{learning-cloning-noachieved.1.fHistogram.10000}%
}%
\\
\subfigure[Male status histo\-gram, for one run]{%
	\includegraphics[width=1.5in]{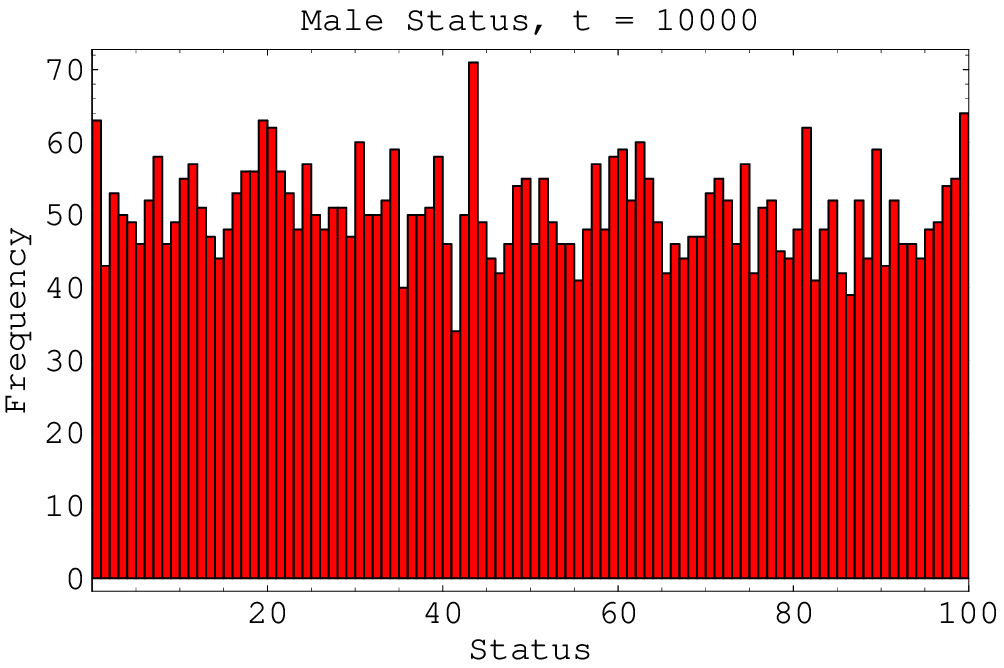}%
	\label{learning-cloning-noachieved.1.mHistogram.10000}%
}%
}%
\end{center}
\caption{Marriage frequency, hypergamy $h,$ and status histograms at
$t = 10,000$ marriages, for the model $\mathbf{M}_{2}$ from
Section~\ref{sec:learning}, using the learning algorithm
$\mathbf{S}_{2}.$ Children were clones of their parents, and agents
did not achieve any status over their lifetimes.}
\label{fig:learning-cloning-noachieved.1.10000}%
\end{figure}

\begin{figure}
\begin{center}
\subfigure[Marriage frequency, \newline for one run]{%
	\includegraphics[width=1.5in]{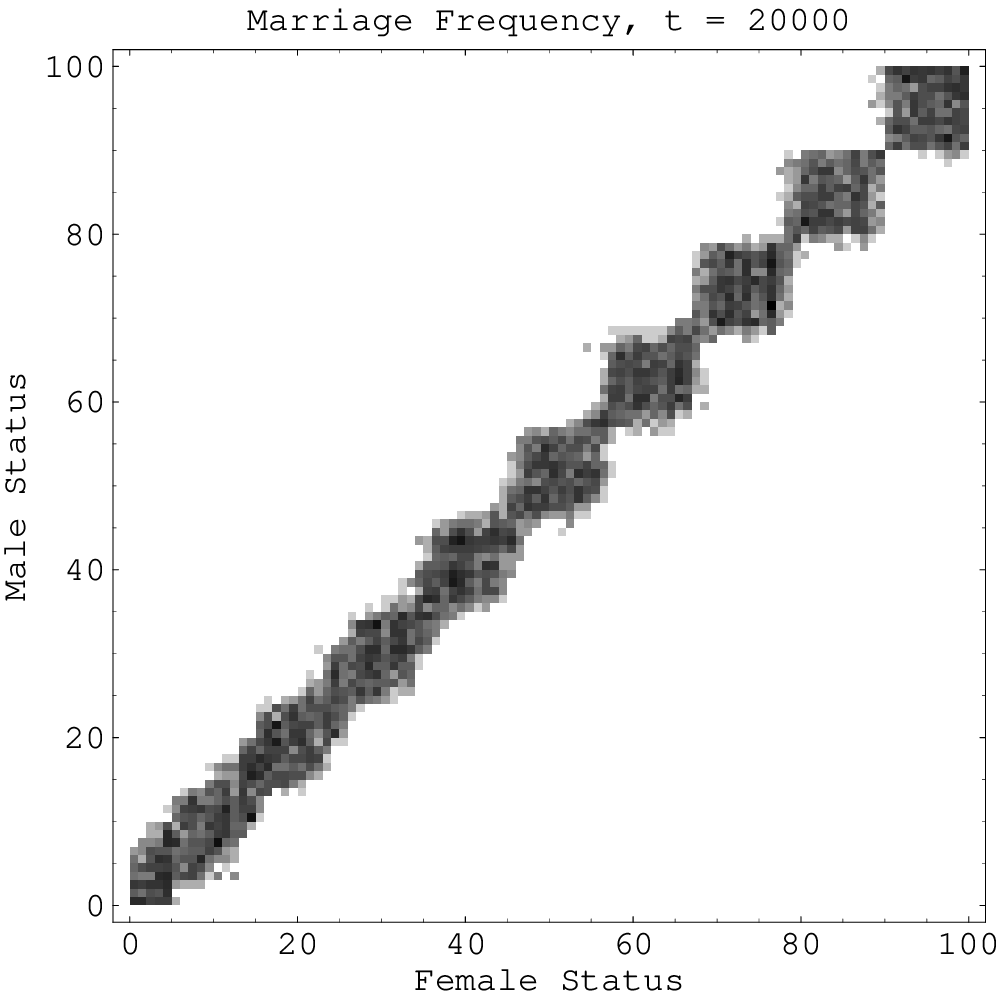}%
	\label{fig:learning-cloning-noachieved.1.marriages.20000}%
}%
\goodgap
\parbox{1.5in}{%
\subfigure[Hypergamy $h,$ for one run]{%
	\includegraphics[width=1.5in]{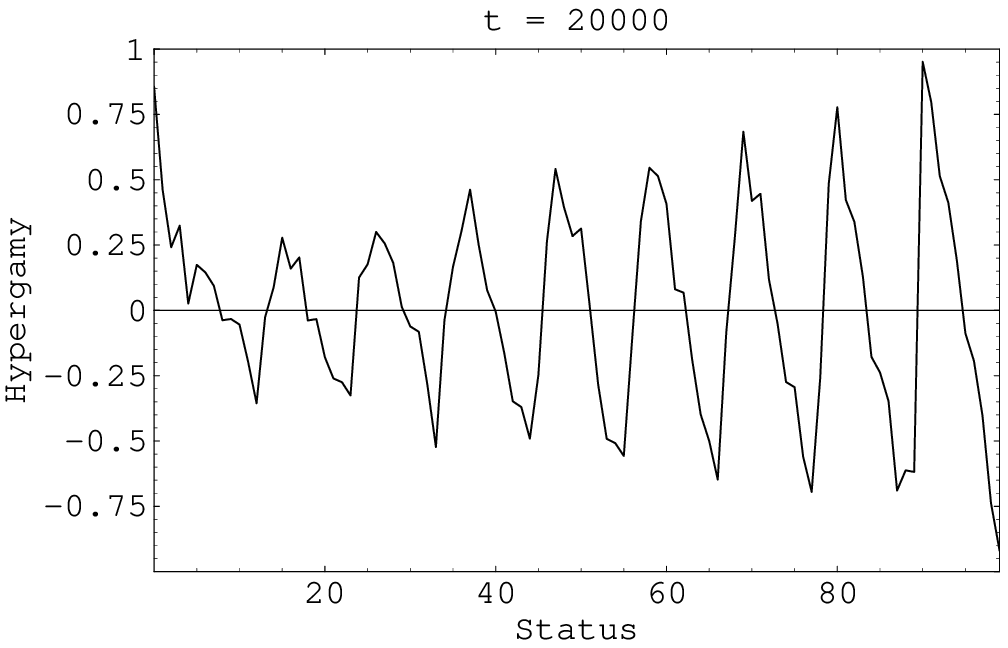}%
	\label{fig:learning-cloning-noachieved.1.hypergamy.20000}%
}%
\\
\subfigure[Hypergamy $h,$ for 50 runs]{%
	\includegraphics[width=1.5in]{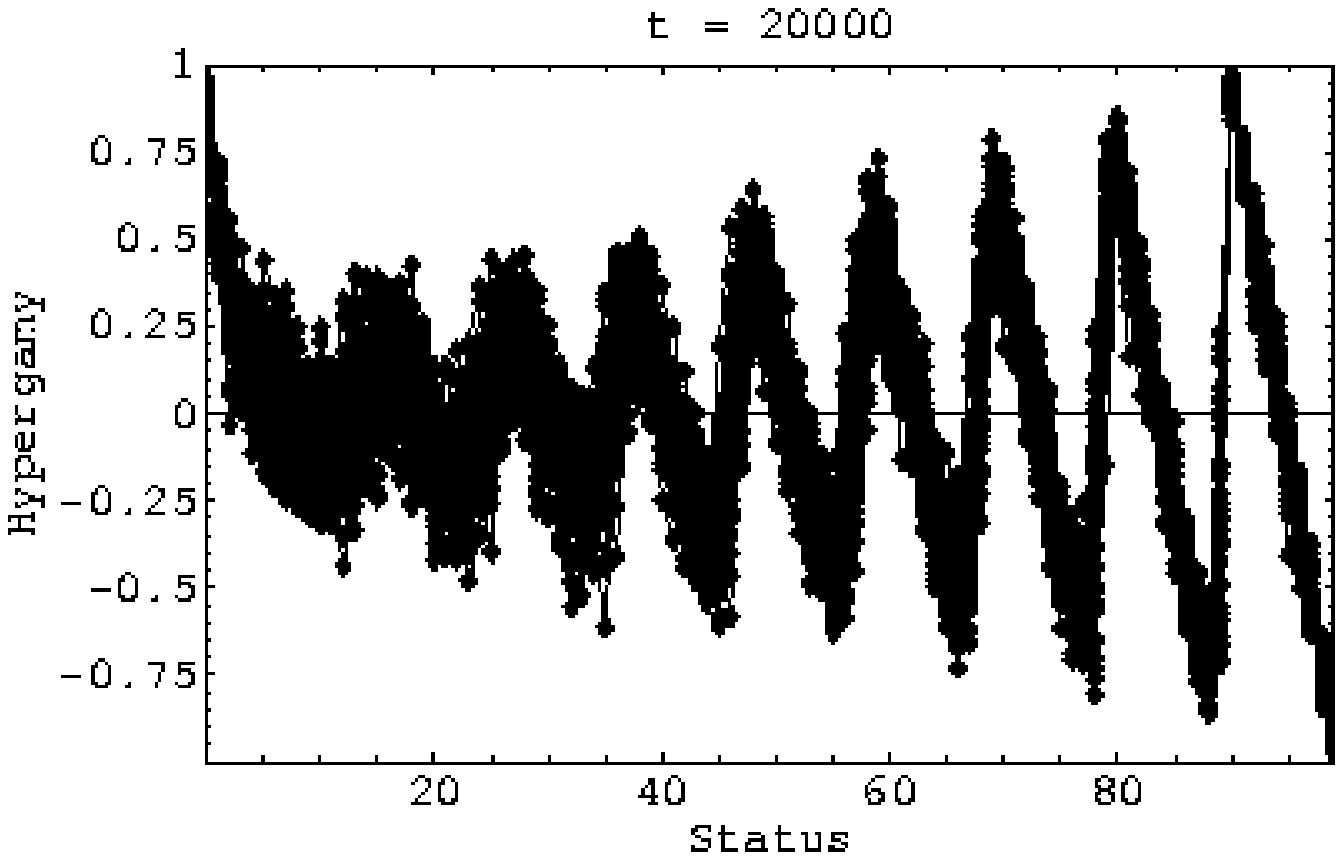}%
	\label{fig:learning-cloning-noachieved.multiHypergamy.20000}%
}%
}%
\goodgap
\parbox{1.5in}{%
\subfigure[Female status histo\-gram, for one run]{%
	\includegraphics[width=1.5in]{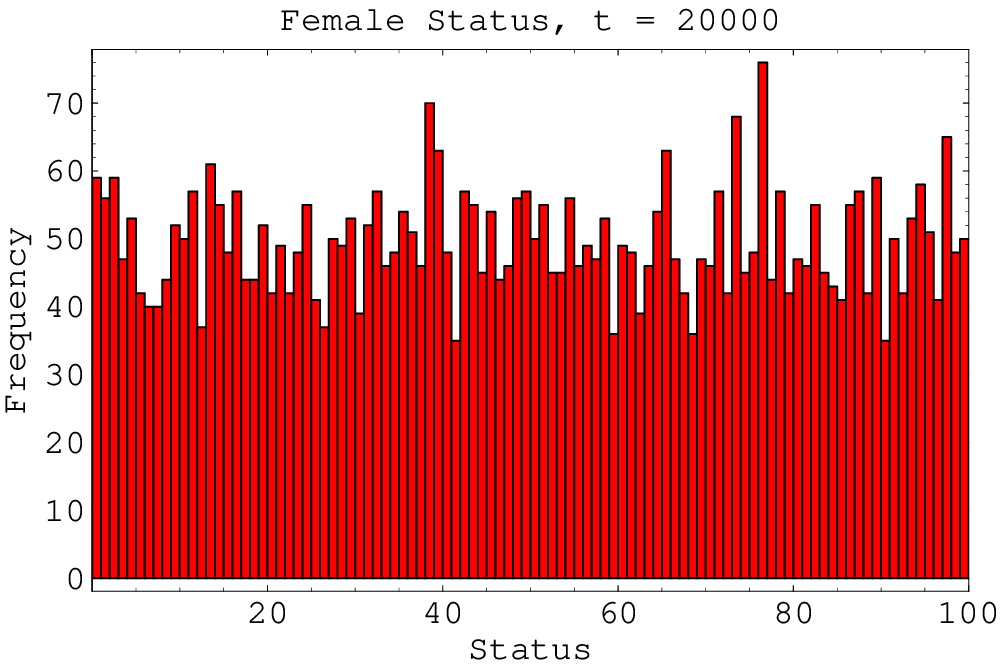}%
	\label{learning-cloning-noachieved.1.fHistogram.20000}%
}%
\\
\subfigure[Male status histo\-gram, for one run]{%
	\includegraphics[width=1.5in]{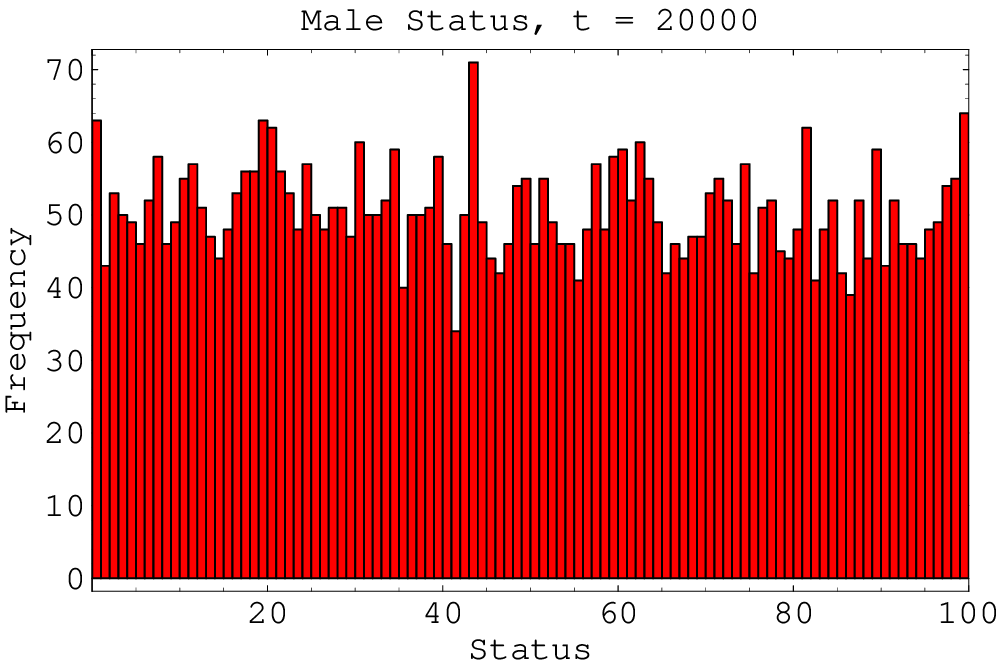}%
	\label{learning-cloning-noachieved.1.mHistogram.20000}%
}%
}%
\end{center}
\caption{Marriage frequency, hypergamy $h,$ and status histograms at
$t = 20,000$ marriages, for the model $\mathbf{M}_{2}$ from
Section~\ref{sec:learning}, using the learning algorithm
$\mathbf{S}_{2}.$ Children were clones of their parents, and agents
did not achieve any status over their lifetimes. These plots remained
very similar at later timesteps in the runs. The status histograms
remained constant.}
\label{fig:learning-cloning-noachieved.1.20000}%
\end{figure}

Model $\mathbf{M}_{1}$ from Section~\ref{sec:rational} was modified to
use the learning strategy $\mathbf{S}_{2},$ producing model
$\mathbf{M}_{2}.$ The same marriage frequency, hypergamy, and status
data were recorded as in Section~\ref{sec:rational}, and these are
plotted in
Figures~\ref{fig:learning-cloning-noachieved.1.10000}--\ref{fig:learning-cloning-noachieved.1.20000}
at $t = 10,000$ and $t = 20,000$ marriages. (These plots changed very
little after $20,000$ marriages.) The male and female status histograms
were constant during the entire run.

Class formation proceeds quickly in this model, with the
highest-status class evident after only $10,000$ marriages and others
forming within $20,000$ marriages; however it also stagnates quickly
after this point: Only about five clearly-defined classes formed, even
after $100,000$ marriages had occurred, compared with $10$ classes in
model $\mathbf{M}_{1}$ under the rational strategy $\mathbf{S}_{1}.$
Archaic societies probably only had two classes, nobles and
commoners~\cite{flannery:1995}; hence, this does not seem to be a
severe limitation of the model. (Note that class formation is
initiated among the highest-status agents and moves down along the
status scale, just as in the proof in Section~\ref{sec:proof}.)  Thus,
it is indeed easy for agents to learn $s'_{\mathrm{max}}(s).$ However,
the efficiency of this particular learning algorithm may depend on the
status distribution in the population, as well as the population
size. These factors are not investigated in this paper.

% learning-cloning-noachieved
%

In a real chiefdom, agents would not be starting with a blank slate at
$t = 0$ and would always have a history of past marriages to
consult. Thus, the initialization period required in this model, where
the agents use the interval around self strategy $\mathbf{S}_{3}$
before switching to the rational strategy $\mathbf{S}_{1},$ is an
artifact of the model itself.

\subsubsection{Strategy $\mathbf{S}_{3}$: Interval Around Self} \label{sec:self}

\begin{figure}
\begin{center}
\subfigure[Marriage frequency, \newline for one run]{%
	\includegraphics[width=1.5in]{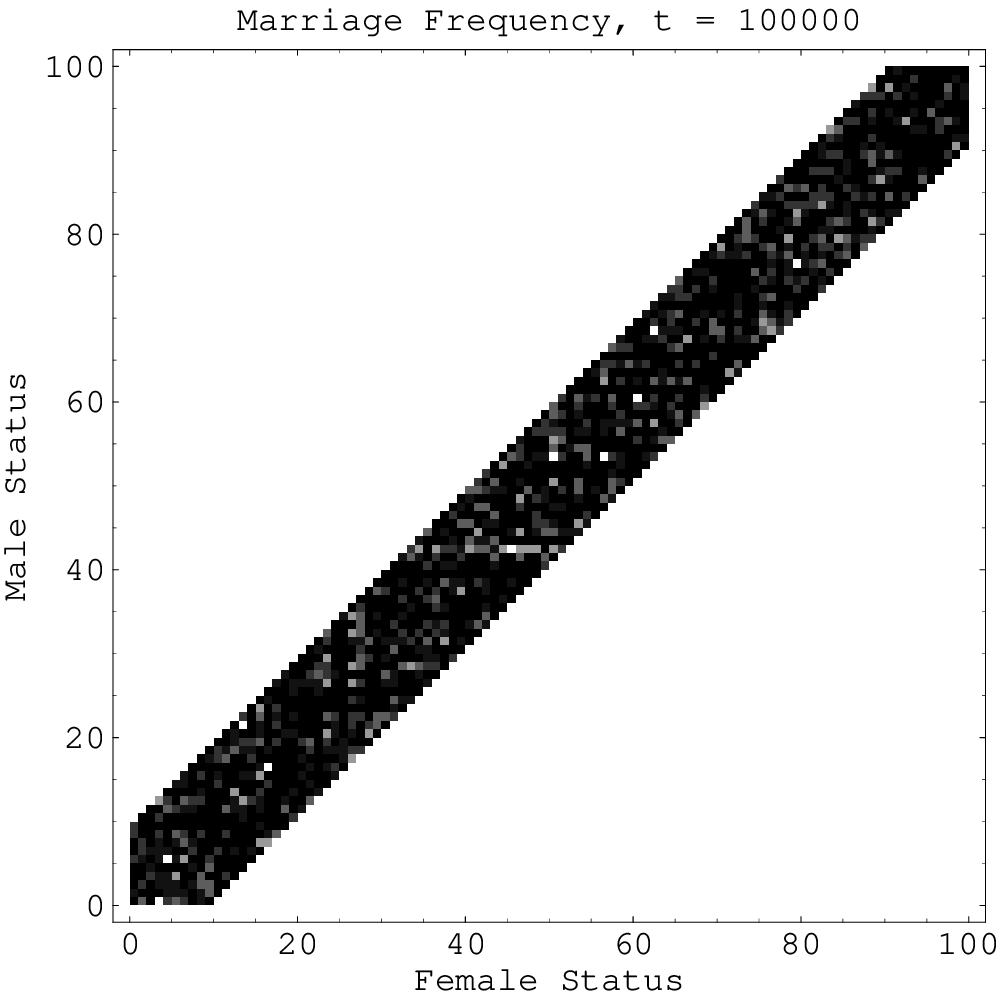}%
	\label{fig:self-cloning-noachieved.1.marriages.100000}%
}%
\goodgap
\parbox{1.5in}{%
\subfigure[Hypergamy $h,$ for one run]{%
	\includegraphics[width=1.5in]{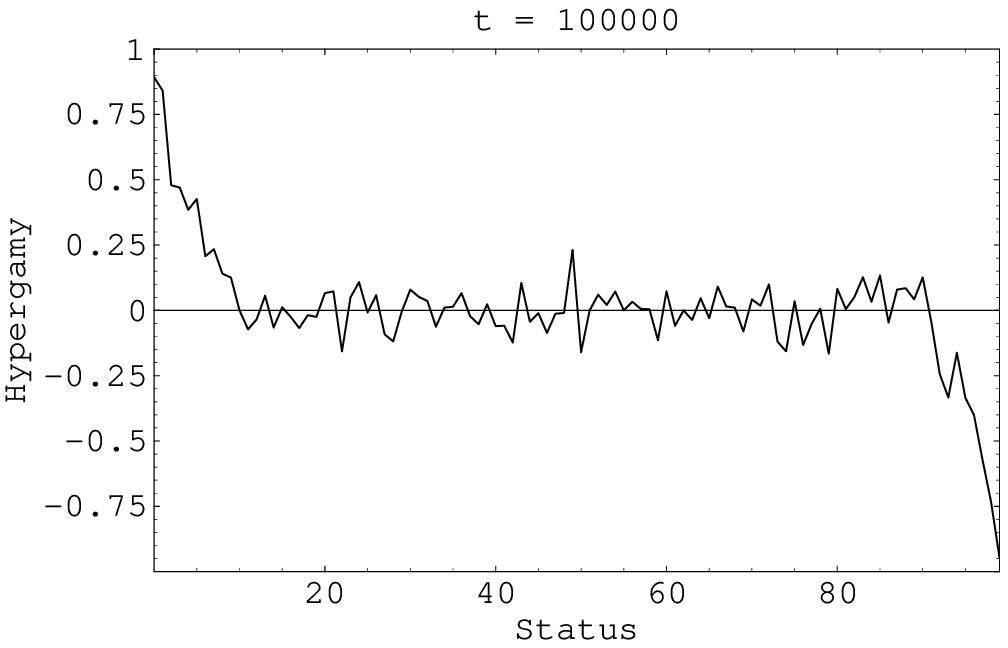}%
	\label{fig:self-cloning-noachieved.1.hypergamy.100000}%
}%
\\
\subfigure[Hypergamy $h,$ for 50 runs]{%
	\includegraphics[width=1.5in]{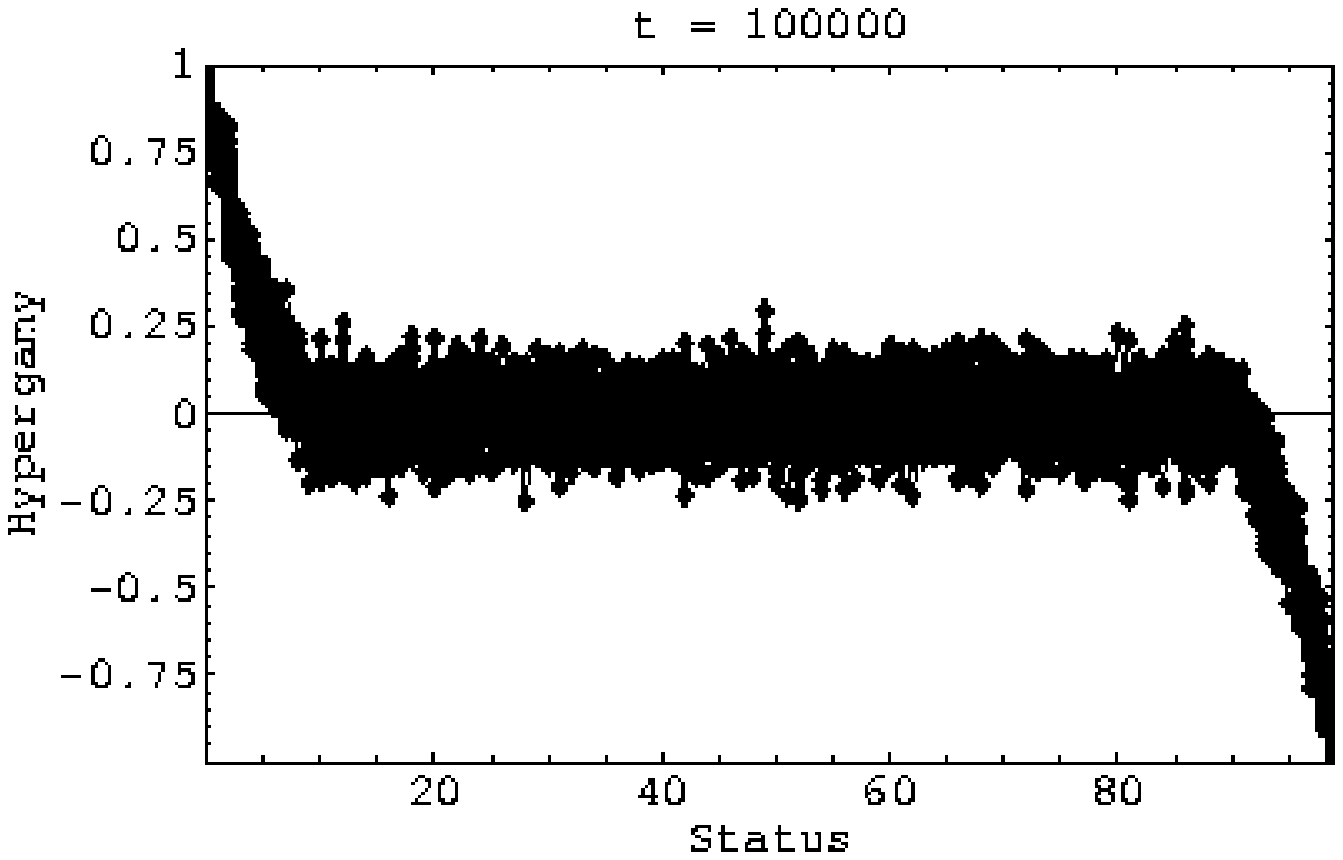}%
	\label{fig:self-cloning-noachieved.multiHypergamy.100000}%
}%
}%
\goodgap
\parbox{1.5in}{%
\subfigure[Female status histo\-gram, for one run]{%
	\includegraphics[width=1.5in]{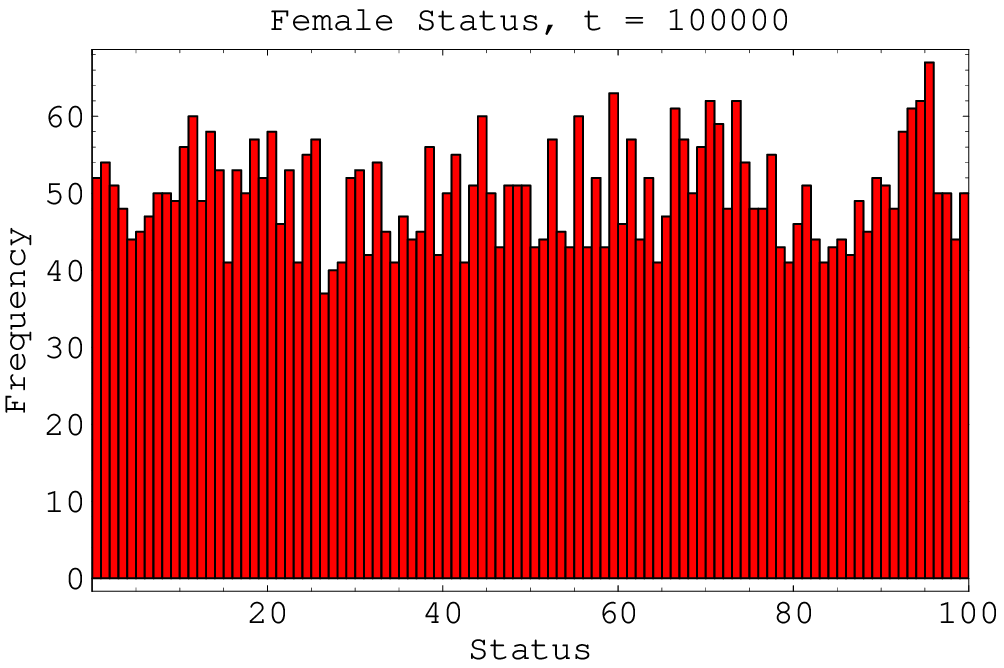}%
	\label{self-cloning-noachieved.1.fHistogram.100000}%
}%
\\
\subfigure[Male status histo\-gram, for one run]{%
	\includegraphics[width=1.5in]{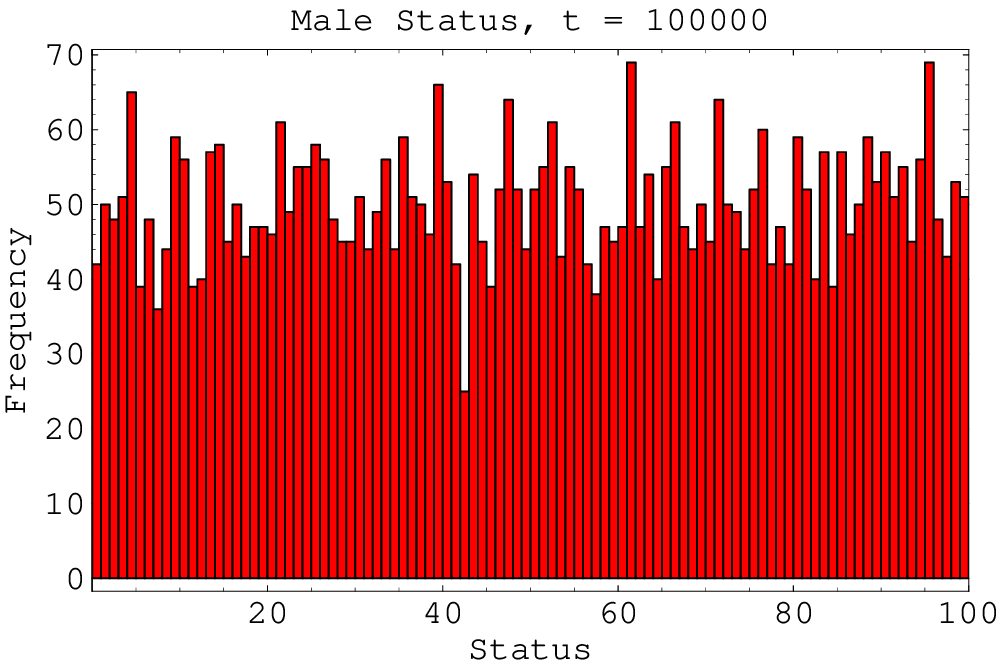}%
	\label{self-cloning-noachieved.1.mHistogram.100000}%
}%
}%
\end{center}
\caption{Marriage frequency, hypergamy $h,$ and status histograms at
$t = 100,000$ marriages, for the model $\mathbf{M}_{3}$ from
Section~\ref{sec:self}, using the interval around self strategy
$\mathbf{S}_{3}.$ Children were clones of their parents, and agents
did not achieve any status over their lifetimes. These plots were very
similar at all timesteps during the runs. The status histograms were
constant throughout the run.}
\label{fig:self-cloning-noachieved.1.100000}%
\end{figure}

Another plausible scenario would be for all of the agents to simply
use the strategy $\mathbf{S}_{3}$ described in
Section~\ref{sec:learning}, with $\epsilon = 9.$ This is a sort of
null hypothesis, $\mathbf{S}_{3}$ being the simplest possible strategy
that takes a suitor's status into consideration, relative to the
agent's own. Does this strategy also produce class endogamy? The model
$\mathbf{M}_{1}$ from Section~\ref{sec:rational} was modified to use
strategy $\mathbf{S}_{3},$ producing model $\mathbf{M}_{3},$ and the
results are plotted in
Figure~\ref{fig:self-cloning-noachieved.1.100000} at $t = 100,000$
marriages. As before, the male and female status histograms remain
constant over the entire run.  As can be seen, the married couples are
strongly correlated according to status, but no class endogamy occurs
within $100,000$ marriages under this strategy: Strategy
$\mathbf{S}_{3}$ by itself is not a mechanism for producing class
endogamy. Something further is needed.

% self-cloning-noachieved
%

\subsection{Inherited Status} \label{sec:inherited}

In the previous models, a child was simply the clone of one of its
parents. What happens if status is inherited from both parents? One
plausible type of inheritance is for both children simply to receive
the average of their parents' overall statuses. This has the virtue
that a high-status agent's offspring have a lower status if it marries
a low-status agent; conversely, a low-status agent who marries a
high-status agent has the status of its offspring raised. For example,
Kirch~\cite[pp. 31--36]{kirch:1984} writes that in Polynesian
chiefdoms, each person was ranked primarily according to his or her
genealogical distance from the founding ancestor, along the paternal
line. However, rank could also be transmitted along the maternal line,
and females outranked males in some cases. Thus, males could increase
the status of their offspring by marrying high-status females.

\begin{figure}
\begin{center}
\subfigure[Marriage frequency, \newline for one run]{%
	\includegraphics[width=1.5in]{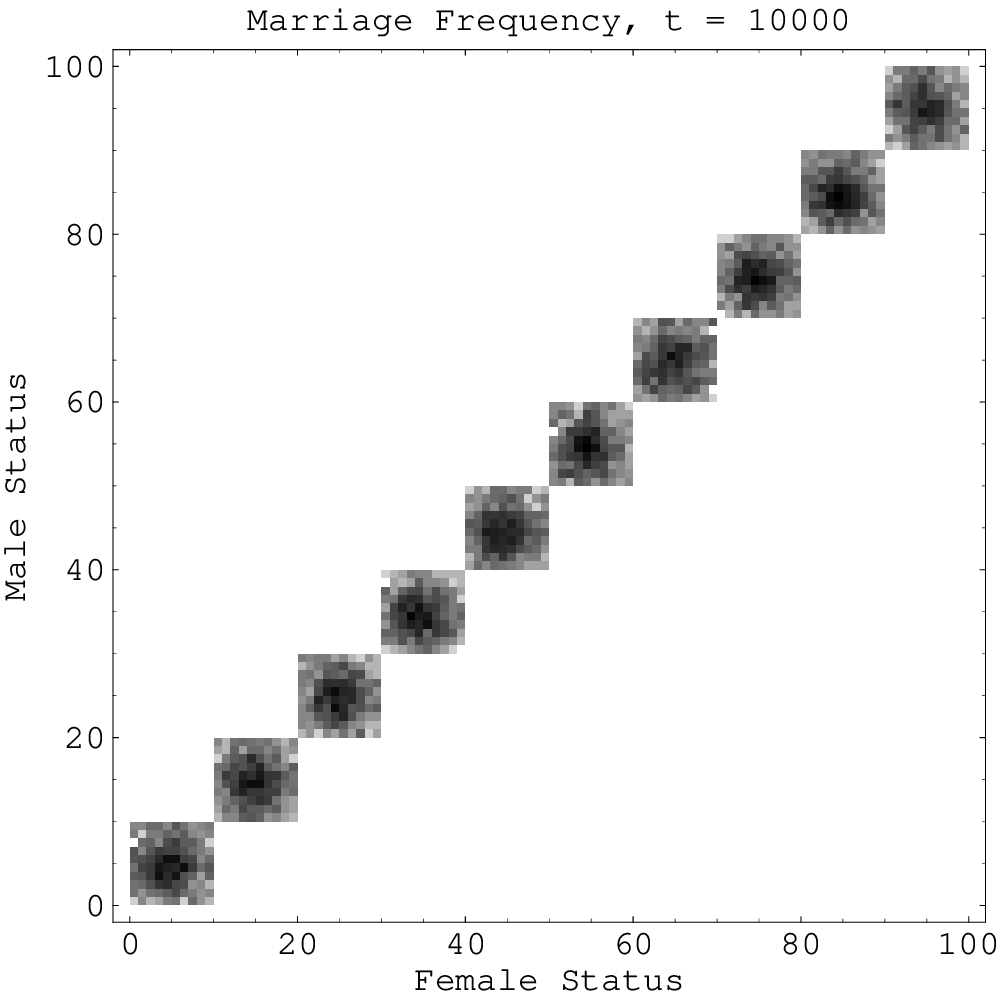}%
	\label{fig:rational-inherited-noachieved.1.marriages.10000}%
}%
\goodgap
\parbox{1.5in}{%
\subfigure[Hypergamy $h,$ for one run]{%
	\includegraphics[width=1.5in]{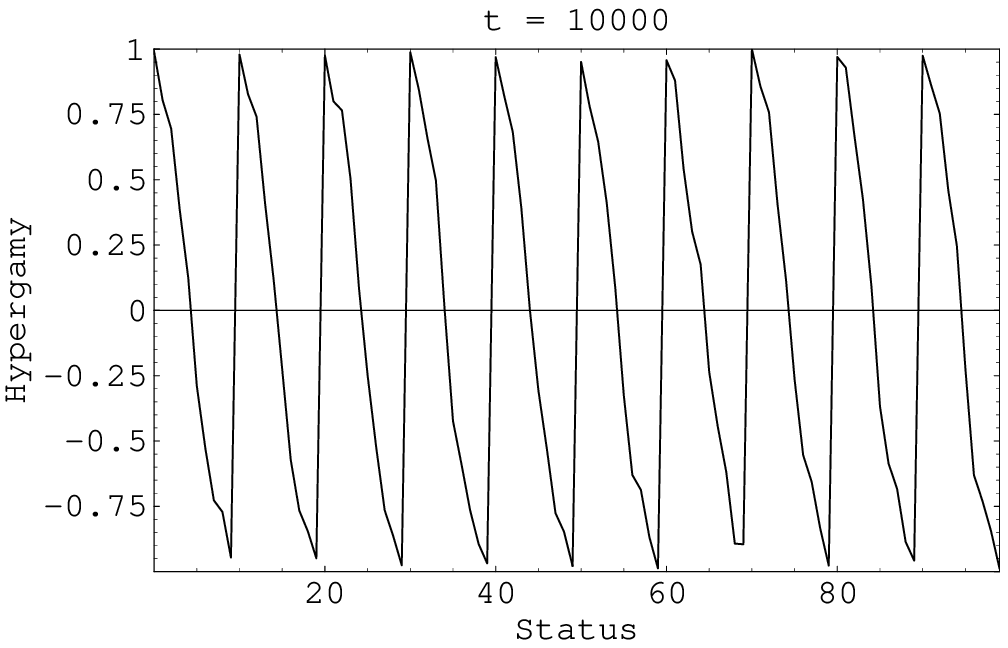}%
	\label{fig:rational-inherited-noachieved.1.hypergamy.10000}%
}%
\\
\subfigure[Hypergamy $h,$ for 50 runs]{%
	\includegraphics[width=1.5in]{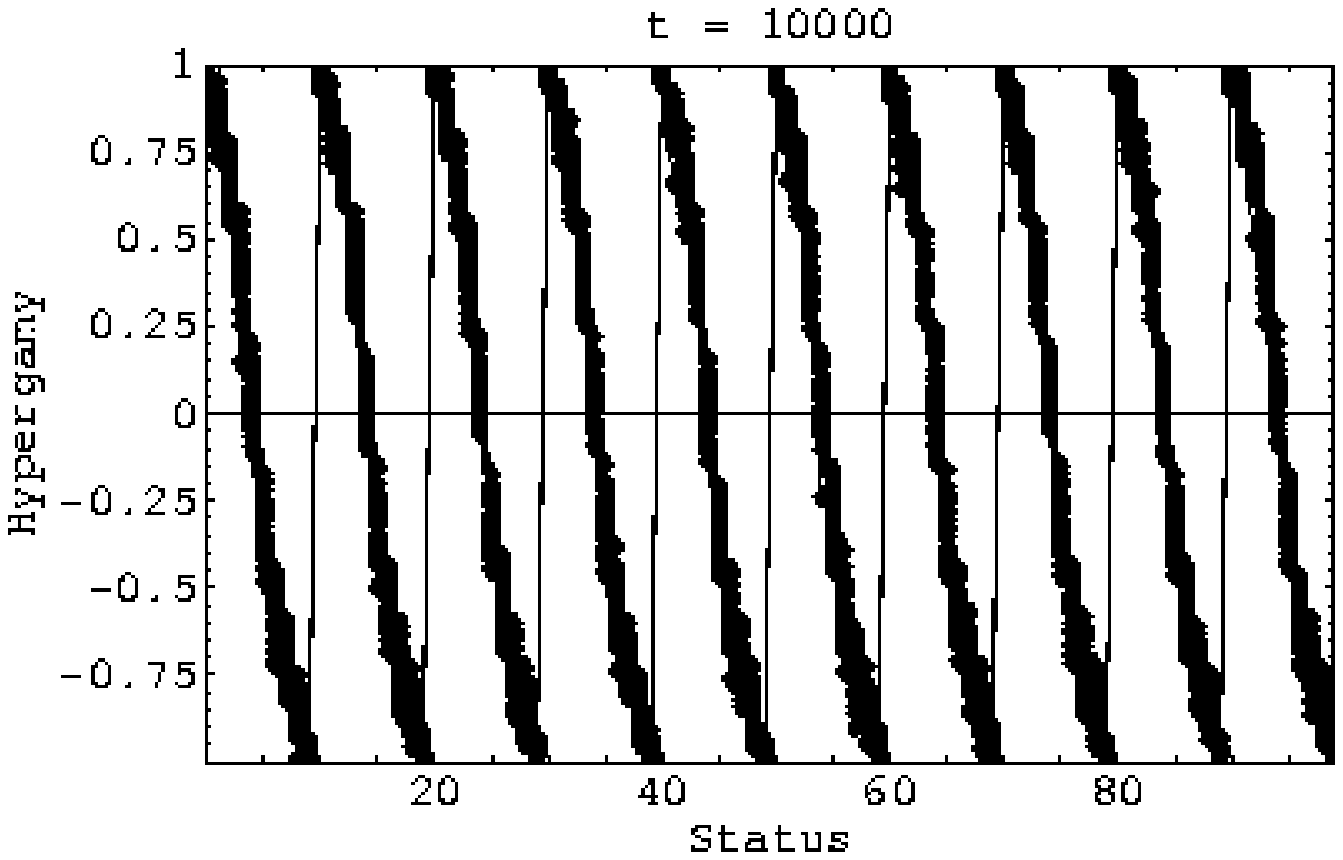}%
	\label{fig:rational-inherited-noachieved.multiHypergamy.10000}%
}%
}%
\goodgap
\parbox{1.5in}{%
\subfigure[Female status histo\-gram, for one run]{%
	\includegraphics[width=1.5in]{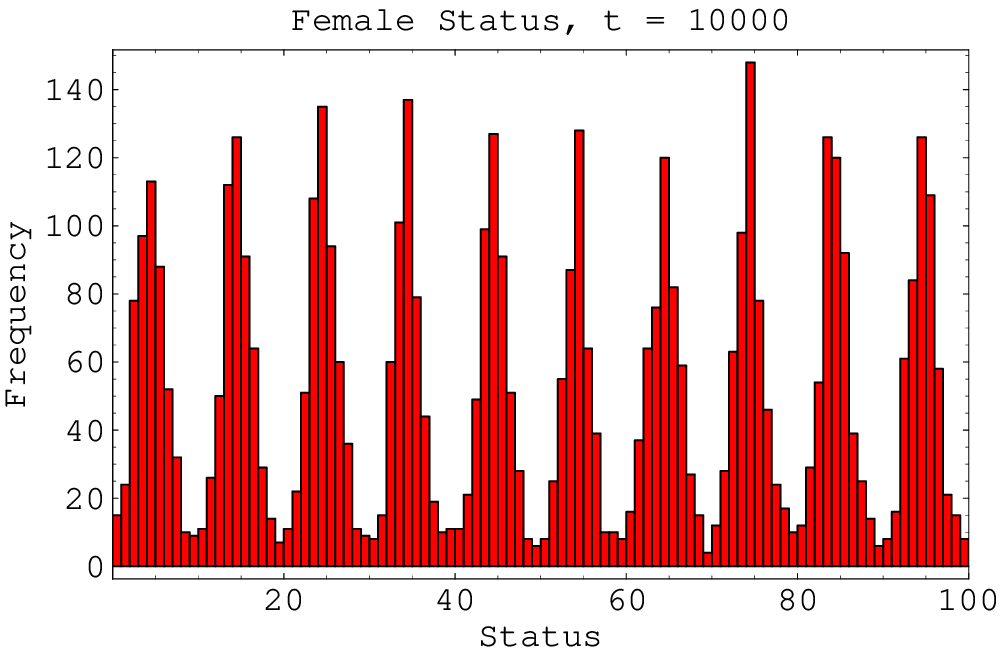}%
	\label{rational-inherited-noachieved.1.fHistogram.10000}%
}%
\\
\subfigure[Male status histo\-gram, for one run]{%
	\includegraphics[width=1.5in]{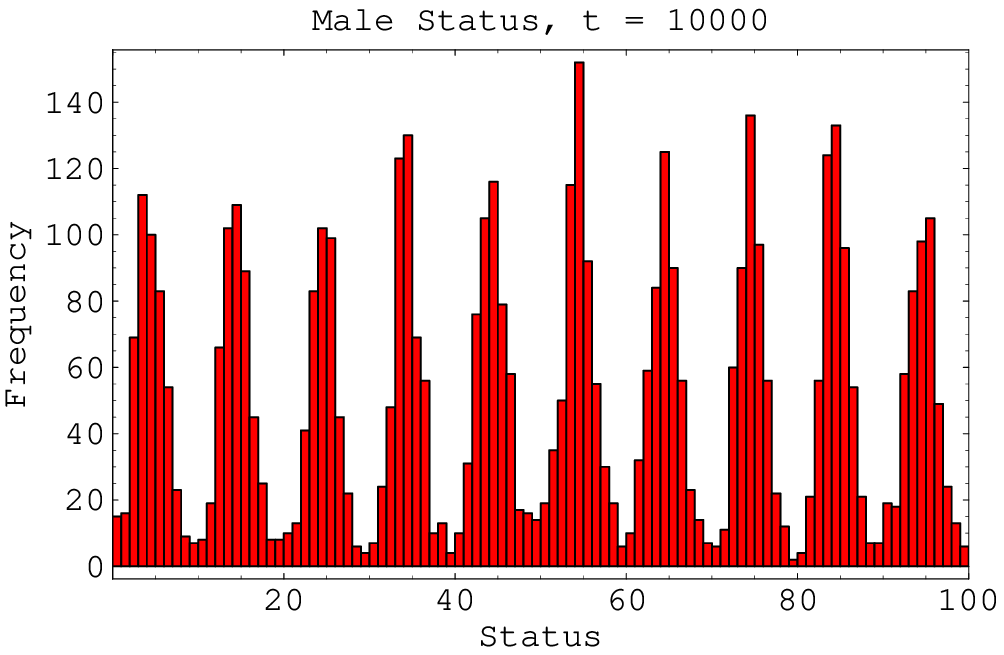}%
	\label{rational-inherited-noachieved.1.mHistogram.10000}%
}%
}%
\end{center}
\caption{Marriage frequency, hypergamy $h,$ and status histograms at
$t = 10,000$ marriages, for the model $\mathbf{M}_{4}$ from
Section~\ref{sec:inherited}, using the rational strategy
$\mathbf{S}_{1}.$ Children inherited the average of their parents'
statuses, and agents did not achieve any status over their lifetimes.}
\label{fig:rational-inherited-noachieved.1.10000}%
\end{figure}

\begin{figure}
\begin{center}
\subfigure[Marriage frequency, \newline for one run]{%
	\includegraphics[width=1.5in]{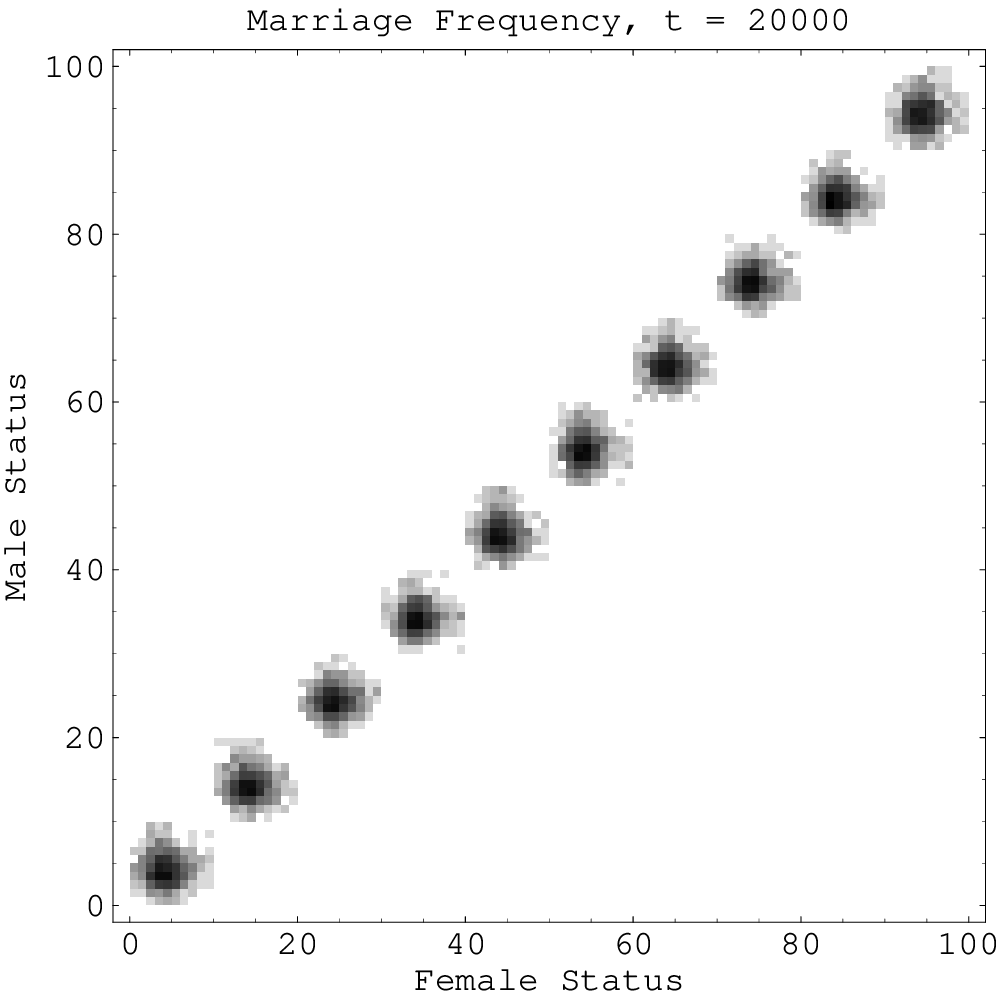}%
	\label{fig:rational-inherited-noachieved.1.marriages.20000}%
}%
\goodgap
\parbox{1.5in}{%
\subfigure[Hypergamy $h,$ for one run]{%
	\includegraphics[width=1.5in]{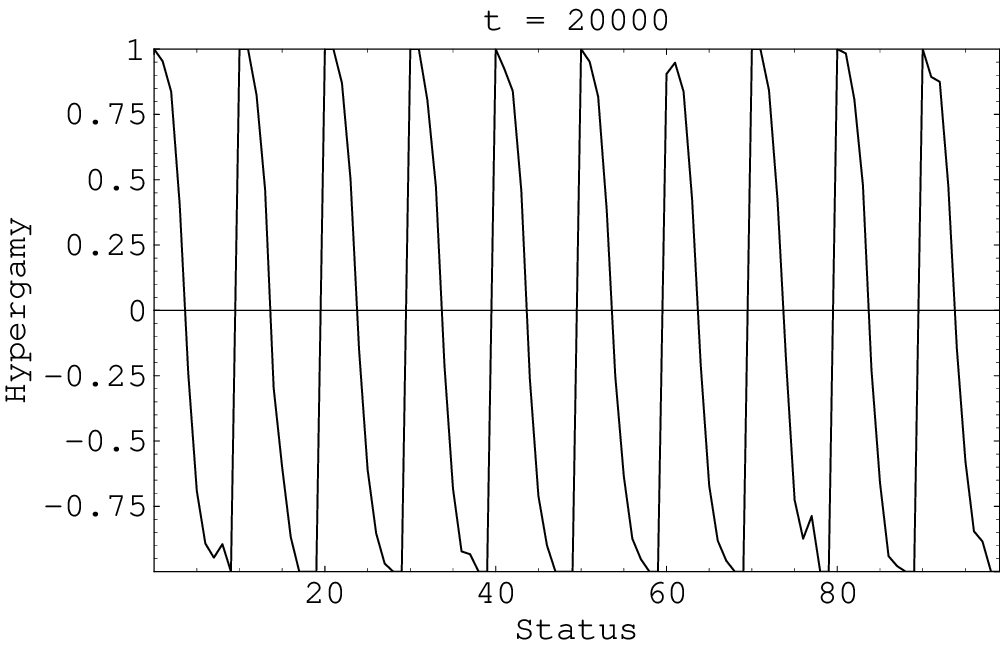}%
	\label{fig:rational-inherited-noachieved.1.hypergamy.20000}%
}%
\\
\subfigure[Hypergamy $h,$ for 50 runs]{%
	\includegraphics[width=1.5in]{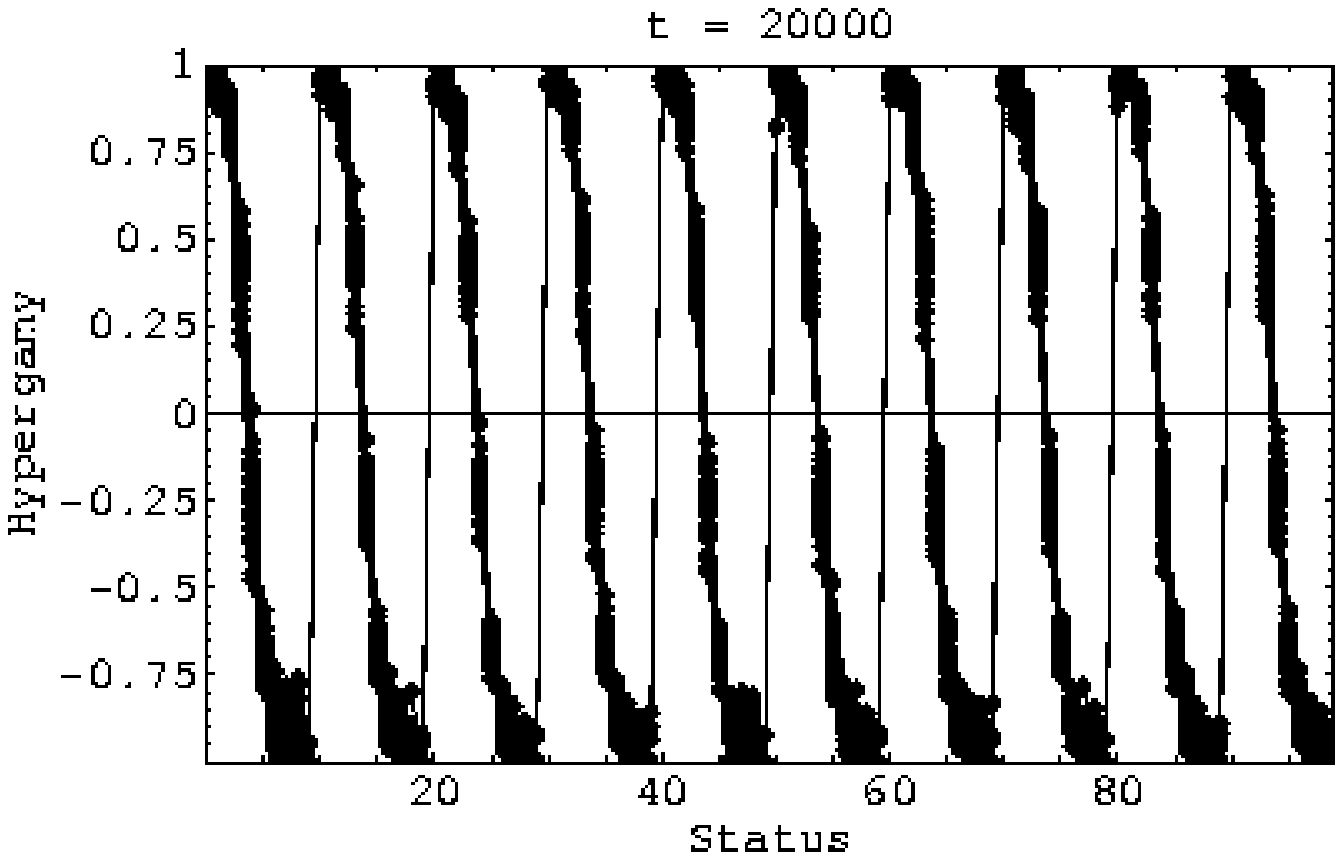}%
	\label{fig:rational-inherited-noachieved.multiHypergamy.20000}%
}%
}%
\goodgap
\parbox{1.5in}{%
\subfigure[Female status histo\-gram, for one run]{%
	\includegraphics[width=1.5in]{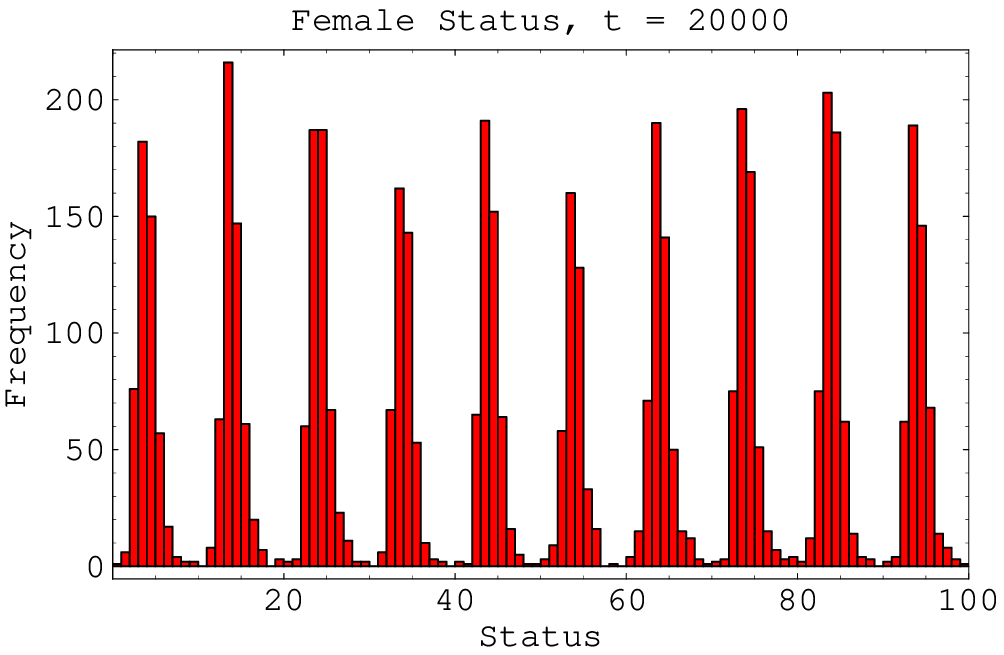}%
	\label{rational-inherited-noachieved.1.fHistogram.20000}%
}%
\\
\subfigure[Male status histo\-gram, for one run]{%
	\includegraphics[width=1.5in]{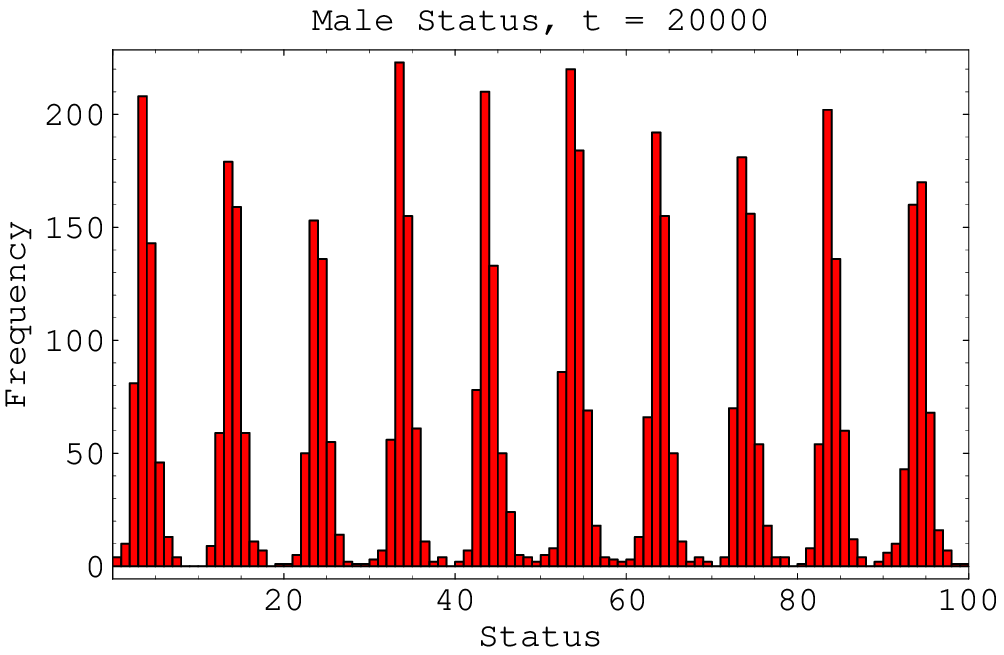}%
	\label{rational-inherited-noachieved.1.mHistogram.20000}%
}%
}%
\end{center}
\caption{Marriage frequency, hypergamy $h,$ and status histograms at
$t = 20,000$ marriages, for the model $\mathbf{M}_{4}$ from
Section~\ref{sec:inherited}, using the rational strategy
$\mathbf{S}_{1}.$ Children inherited the average of their parents'
statuses, and agents did not achieve any status over their lifetimes.}
\label{fig:rational-inherited-noachieved.1.20000}%
\end{figure}

\begin{figure}
\begin{center}
\subfigure[Marriage frequency, \newline for one run]{%
	\includegraphics[width=1.5in]{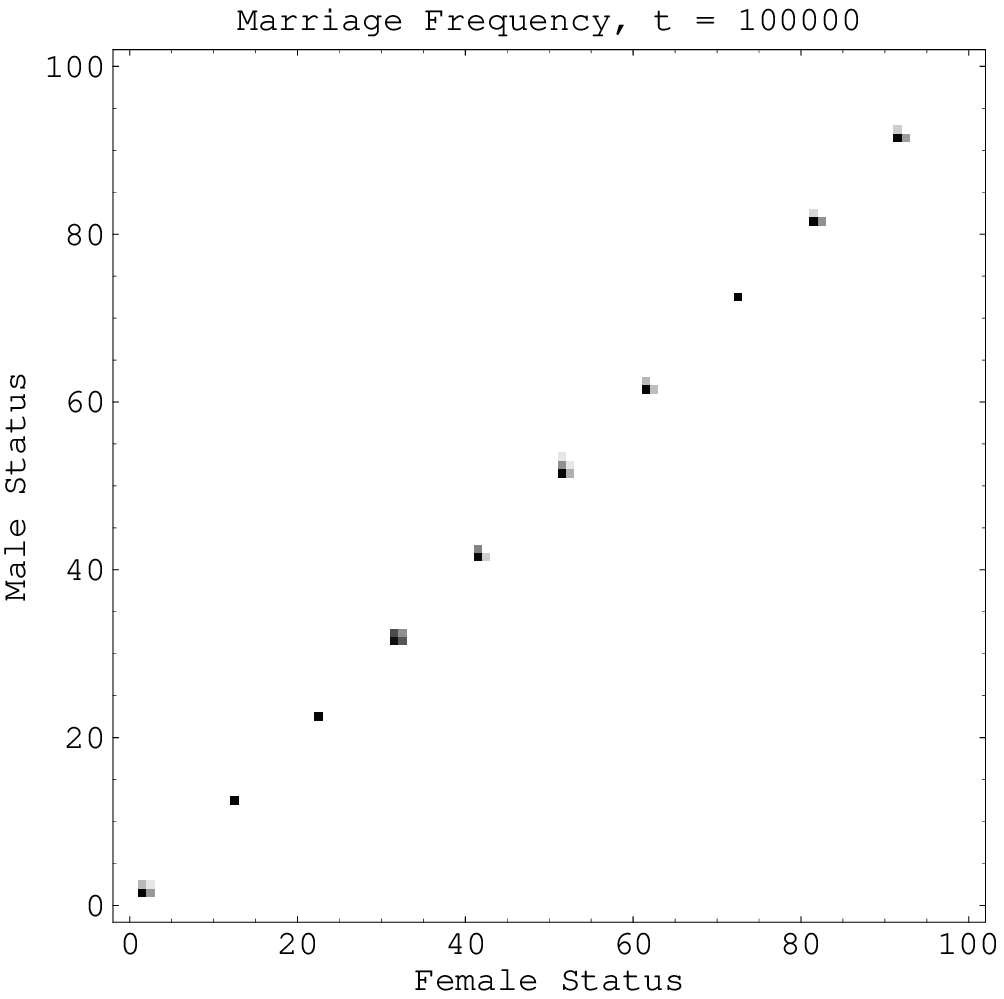}%
	\label{fig:rational-inherited-noachieved.1.marriages.100000}%
}%
\goodgap
\parbox{1.5in}{%
\subfigure[Hypergamy $h,$ for one run]{%
	\includegraphics[width=1.5in]{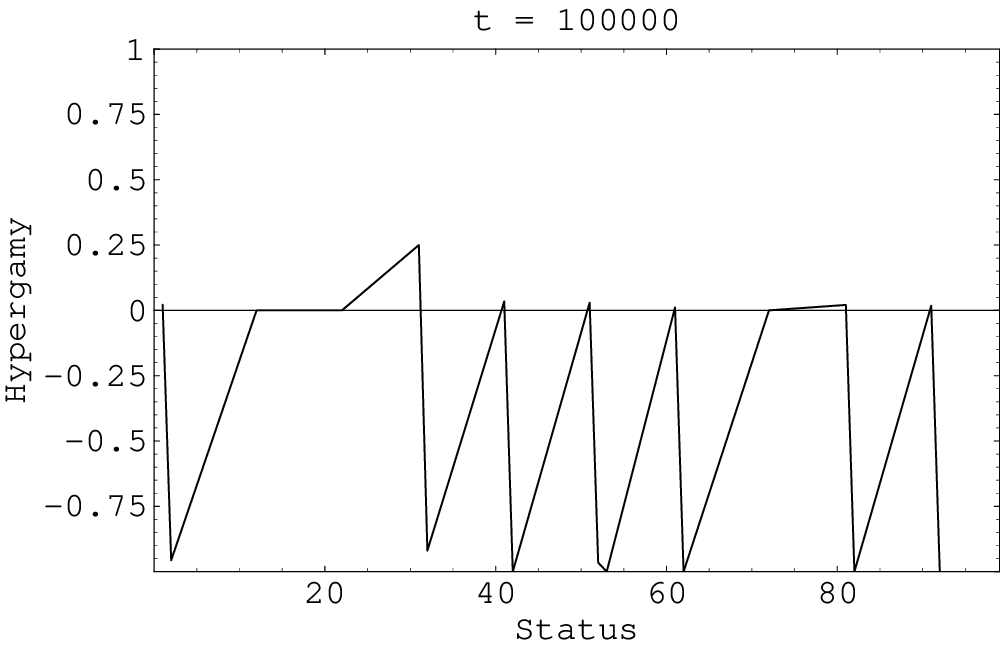}%
	\label{fig:rational-inherited-noachieved.1.hypergamy.100000}%
}%
\\
\subfigure[Hypergamy $h,$ for 50 runs]{%
	\includegraphics[width=1.5in]{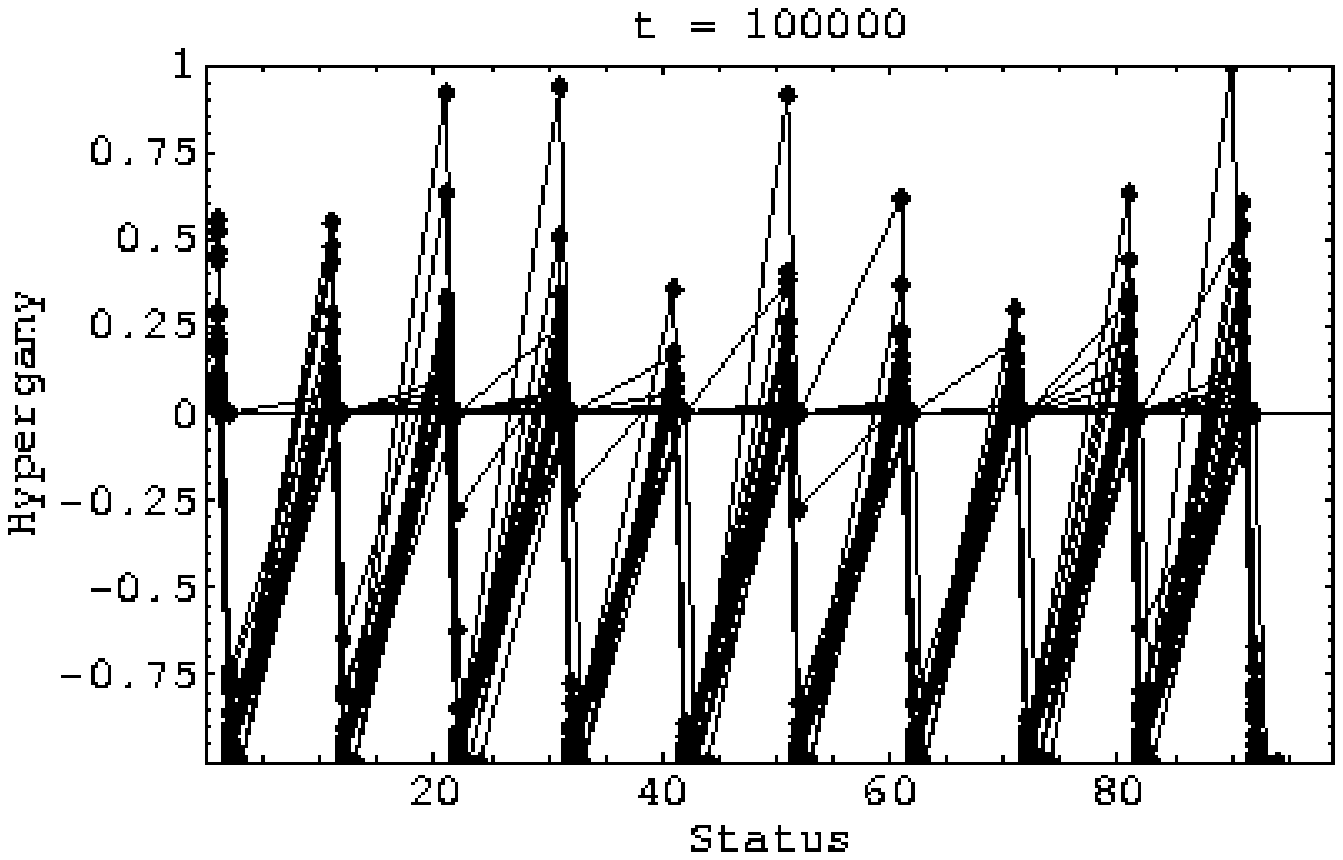}%
	\label{fig:rational-inherited-noachieved.multiHypergamy.100000}%
}%
}%
\goodgap
\parbox{1.5in}{%
\subfigure[Female status histo\-gram, for one run]{%
	\includegraphics[width=1.5in]{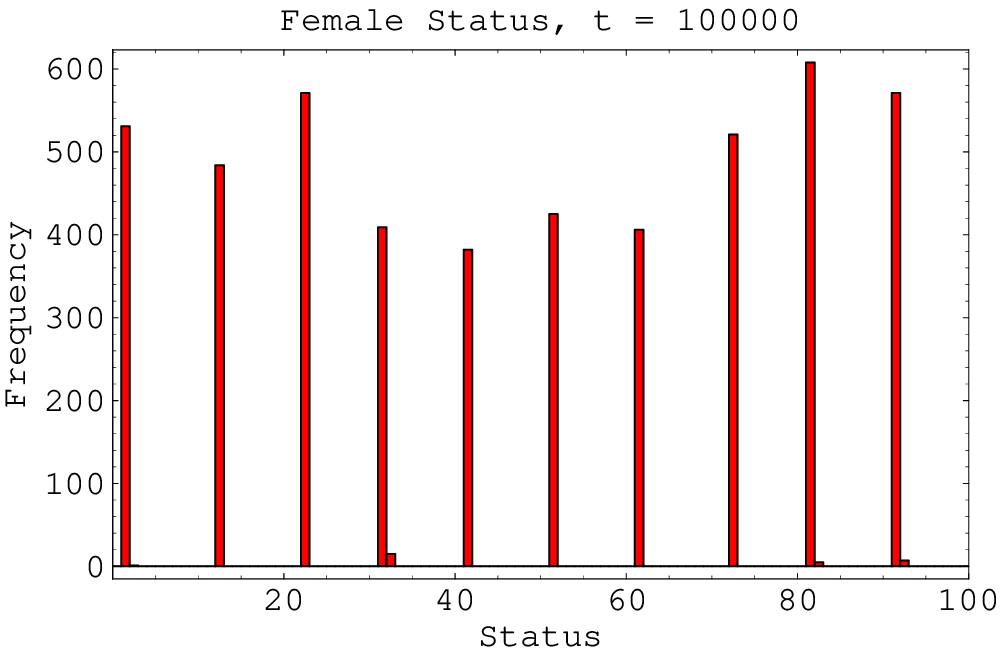}%
	\label{rational-inherited-noachieved.1.fHistogram.100000}%
}%
\\
\subfigure[Male status histo\-gram, for one run]{%
	\includegraphics[width=1.5in]{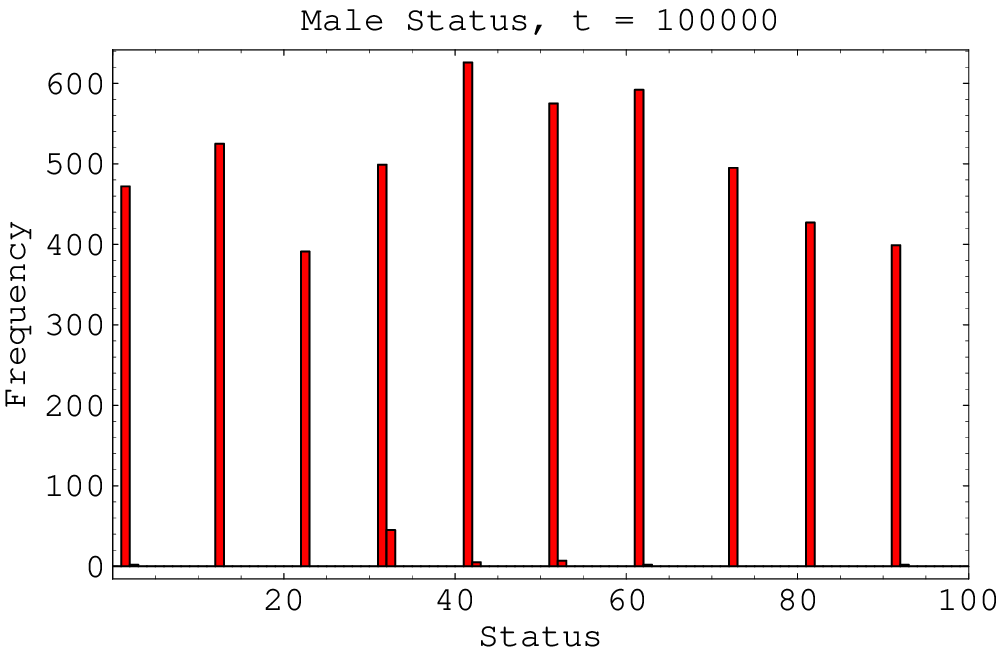}%
	\label{rational-inherited-noachieved.1.mHistogram.100000}%
}%
}%
\end{center}
\caption{Marriage frequency, hypergamy $h,$ and status histograms at
$t = 100,000$ marriages, for the model $\mathbf{M}_{4}$ from
Section~\ref{sec:inherited}, using the rational strategy
$\mathbf{S}_{1}.$ Children inherited the average of their parents'
statuses, and agents did not achieve any status over their lifetimes.}
\label{fig:rational-inherited-noachieved.1.100000}%
\end{figure}

\begin{figure}
\begin{center}
\subfigure[Marriage frequency, \newline for one run]{%
	\includegraphics[width=1.5in]{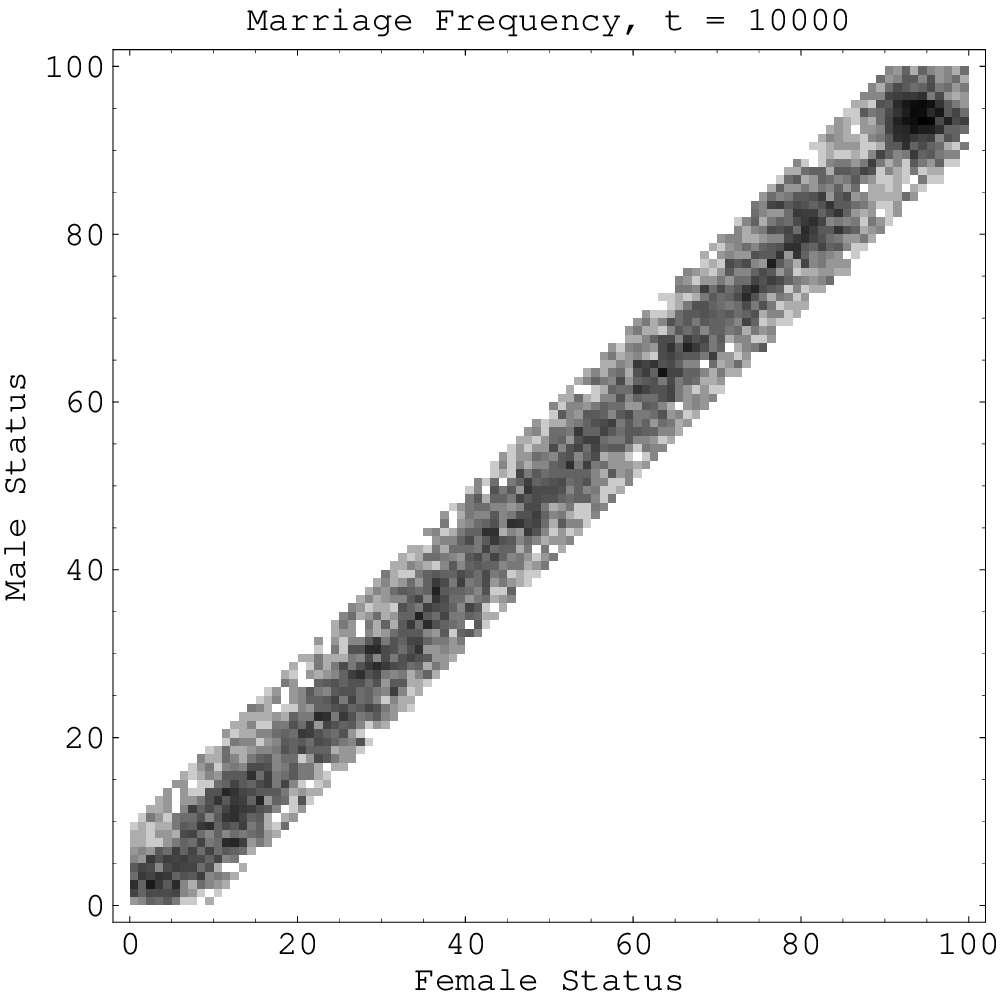}%
	\label{fig:learning-inherited-noachieved.1.marriages.10000}%
}%
\goodgap
\parbox{1.5in}{%
\subfigure[Hypergamy $h,$ for one run]{%
	\includegraphics[width=1.5in]{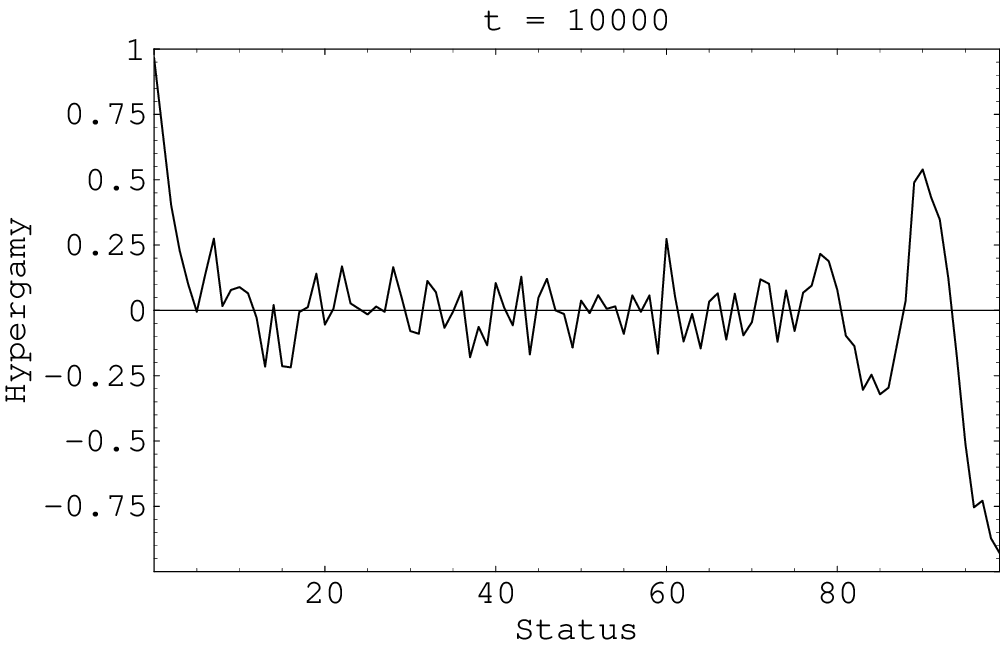}%
	\label{fig:learning-inherited-noachieved.1.hypergamy.10000}%
}%
\\
\subfigure[Hypergamy $h,$ for 50 runs]{%
	\includegraphics[width=1.5in]{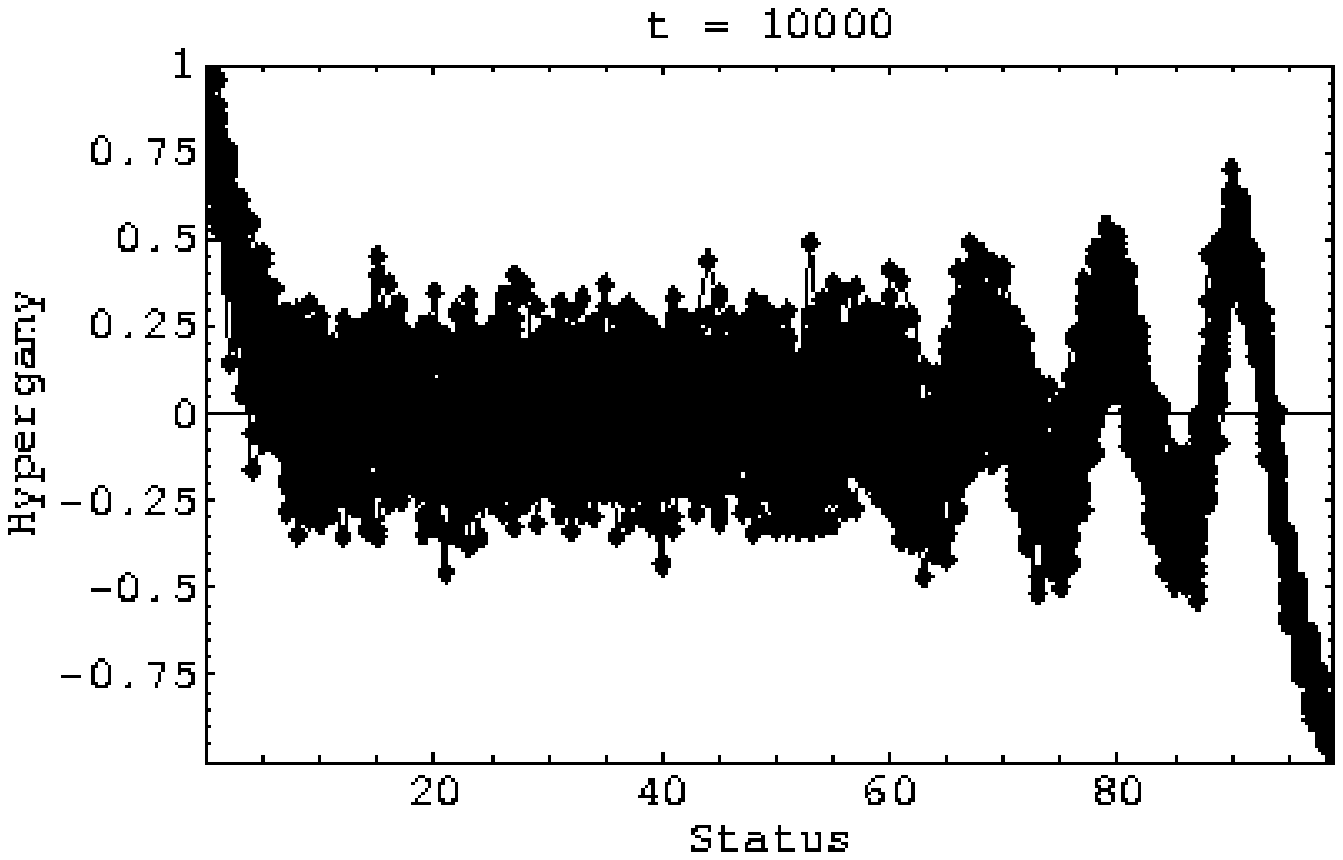}%
	\label{fig:learning-inherited-noachieved.multiHypergamy.10000}%
}%
}%
\goodgap
\parbox{1.5in}{%
\subfigure[Female status histo\-gram, for one run]{%
	\includegraphics[width=1.5in]{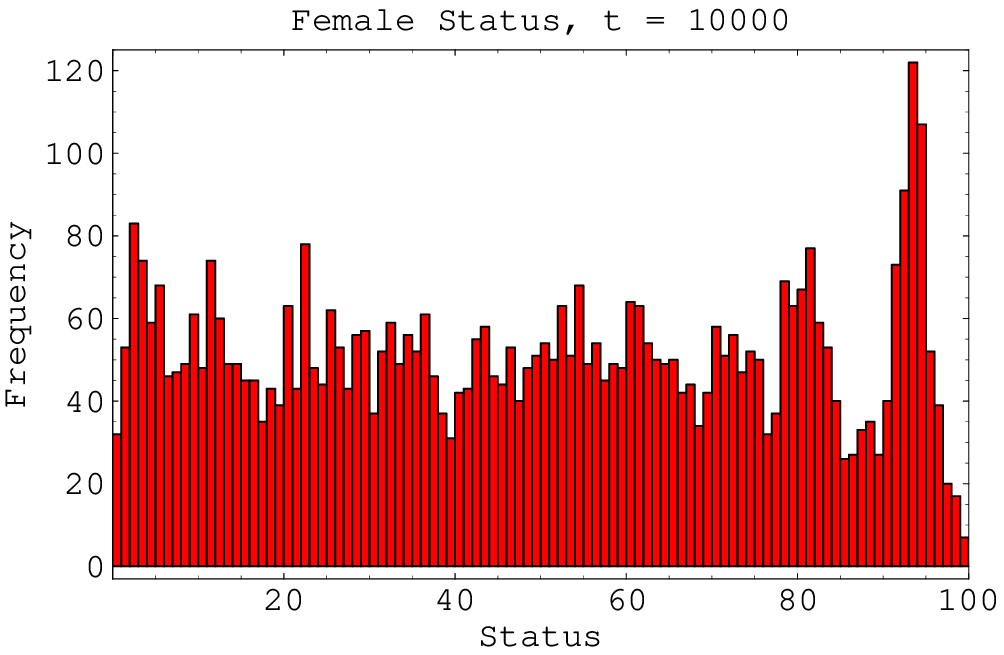}%
	\label{learning-inherited-noachieved.1.fHistogram.10000}%
}%
\\
\subfigure[Male status histo\-gram, for one run]{%
	\includegraphics[width=1.5in]{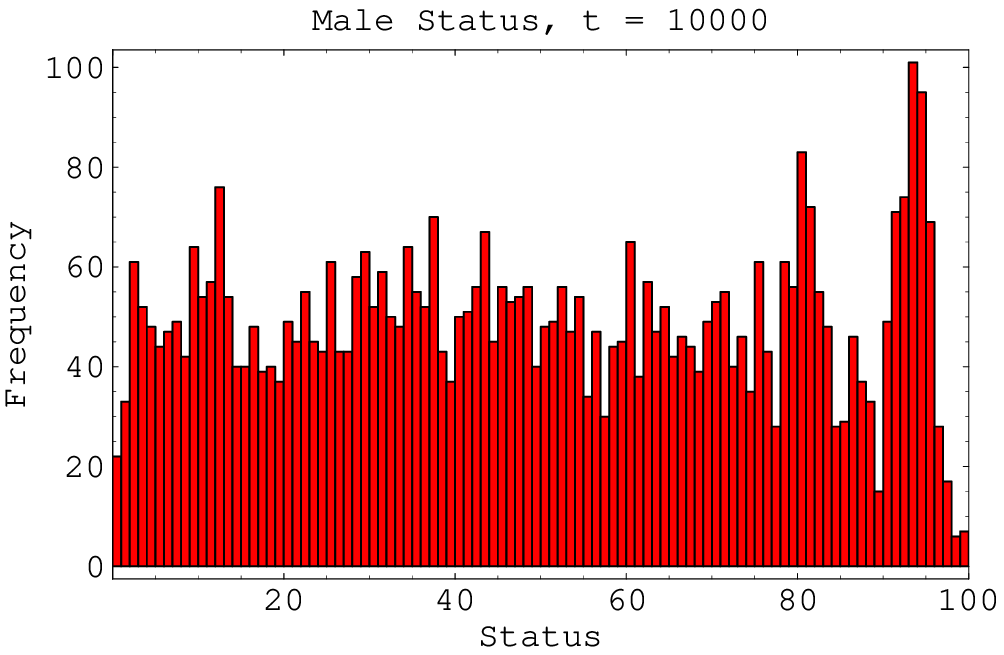}%
	\label{learning-inherited-noachieved.1.mHistogram.10000}%
}%
}%
\end{center}
\caption{Marriage frequency, hypergamy $h,$ and status histograms at
$t = 10,000$ marriages, for the model $\mathbf{M}_{5}$ from
Section~\ref{sec:inherited}, using the learning algorithm
$\mathbf{S}_{2}.$ Children inherited the average of their parents'
statuses, and agents did not achieve any status over their lifetimes.}
\label{fig:learning-inherited-noachieved.1.10000}%
\end{figure}

\begin{figure}
\begin{center}
\subfigure[Marriage frequency, \newline for one run]{%
	\includegraphics[width=1.5in]{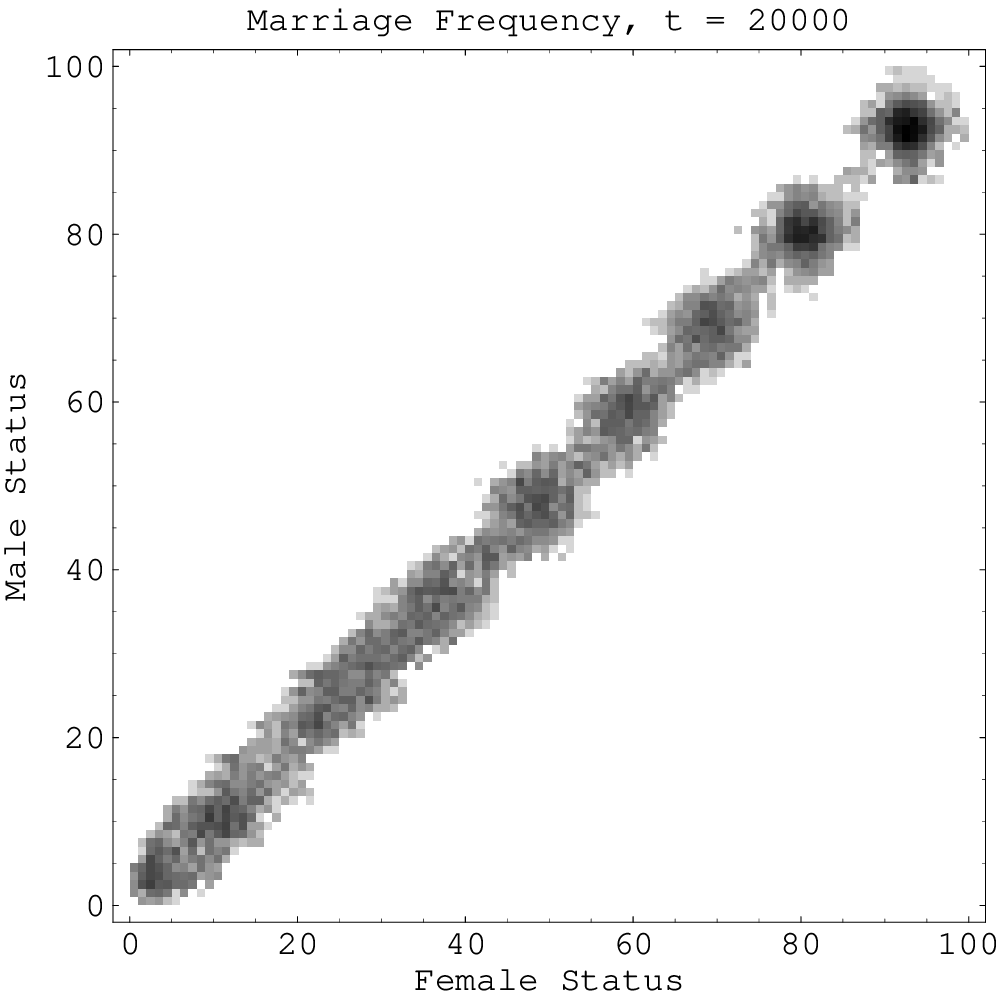}%
	\label{fig:learning-inherited-noachieved.1.marriages.20000}%
}%
\goodgap
\parbox{1.5in}{%
\subfigure[Hypergamy $h,$ for one run]{%
	\includegraphics[width=1.5in]{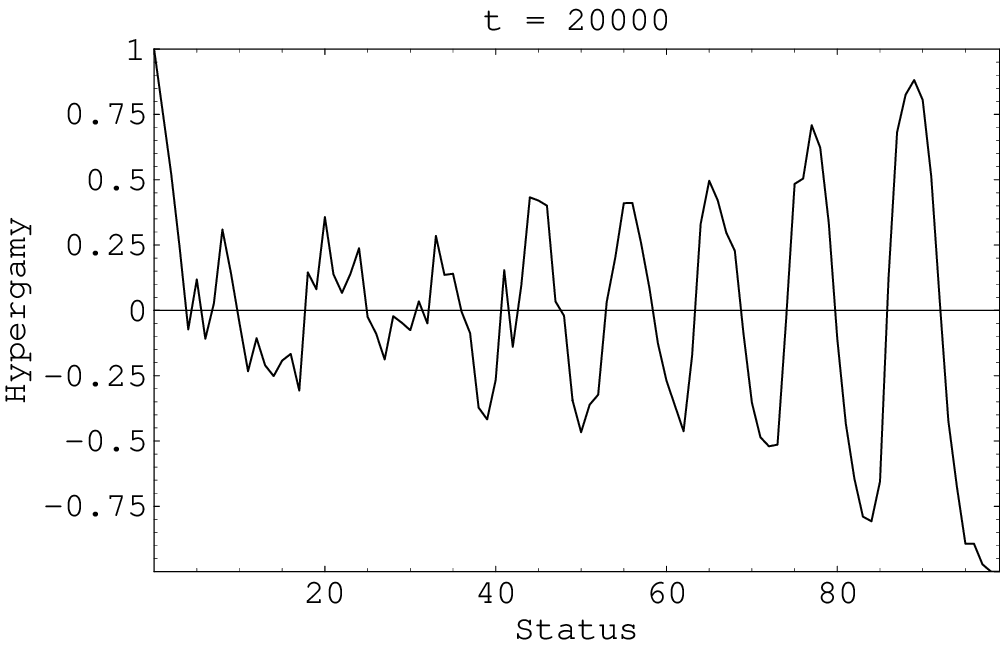}%
	\label{fig:learning-inherited-noachieved.1.hypergamy.20000}%
}%
\\
\subfigure[Hypergamy $h,$ for 50 runs]{%
	\includegraphics[width=1.5in]{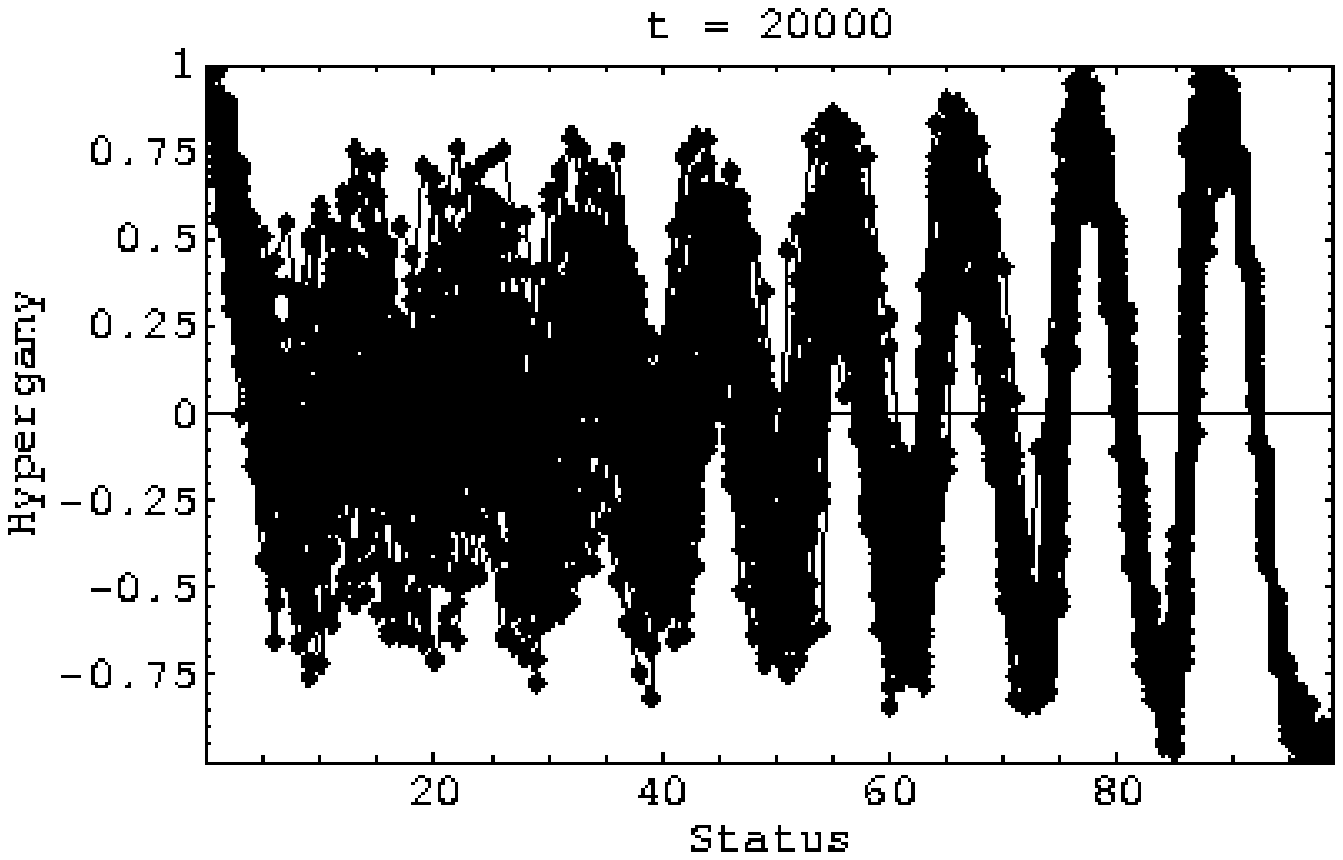}%
	\label{fig:learning-inherited-noachieved.multiHypergamy.20000}%
}%
}%
\goodgap
\parbox{1.5in}{%
\subfigure[Female status histo\-gram, for one run]{%
	\includegraphics[width=1.5in]{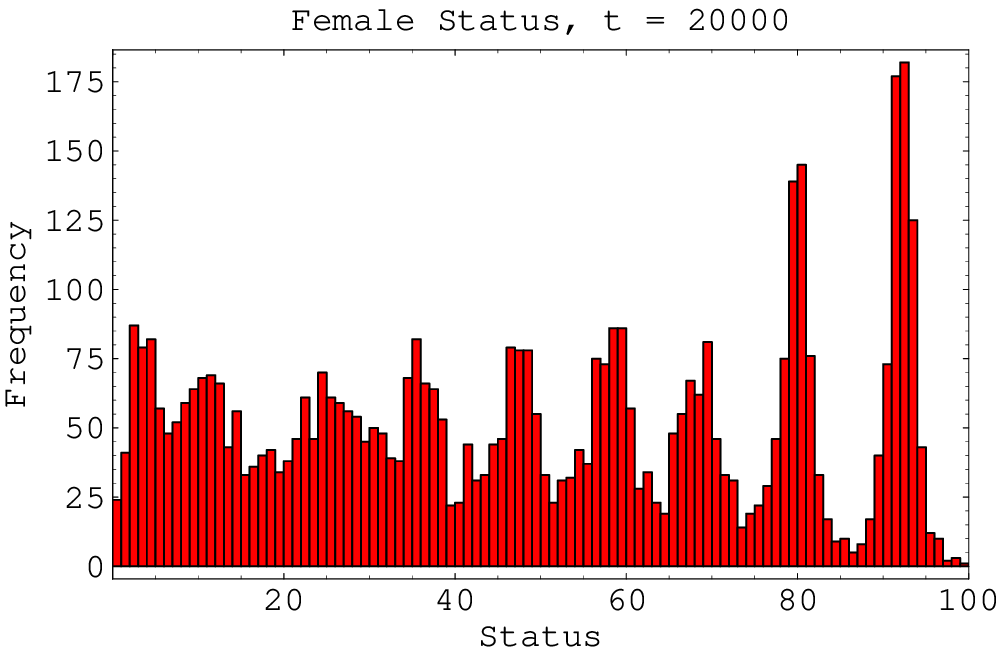}%
	\label{learning-inherited-noachieved.1.fHistogram.20000}%
}%
\\
\subfigure[Male status histo\-gram, for one run]{%
	\includegraphics[width=1.5in]{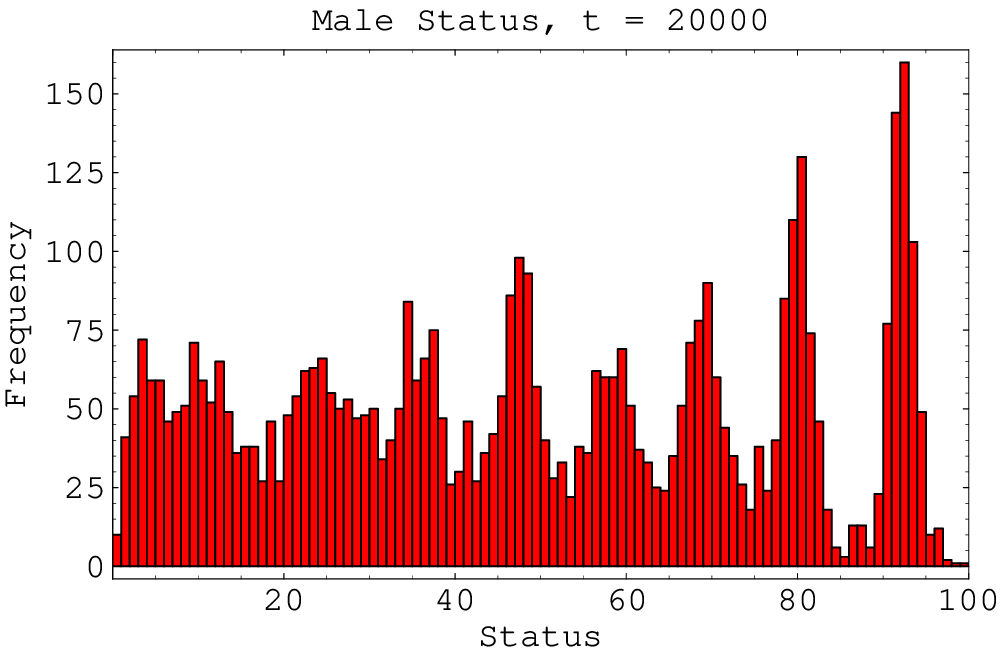}%
	\label{learning-inherited-noachieved.1.mHistogram.20000}%
}%
}%
\end{center}
\caption{Marriage frequency, hypergamy $h,$ and status histograms at
$t = 20,000$ marriages, for the model $\mathbf{M}_{5}$ from
Section~\ref{sec:inherited}, using the learning algorithm
$\mathbf{S}_{2}.$ Children inherited the average of their parents'
statuses, and agents did not achieve any status over their lifetimes.}
\label{fig:learning-inherited-noachieved.1.20000}%
\end{figure}

\begin{figure}
\begin{center}
\subfigure[Marriage frequency, \newline for one run]{%
	\includegraphics[width=1.5in]{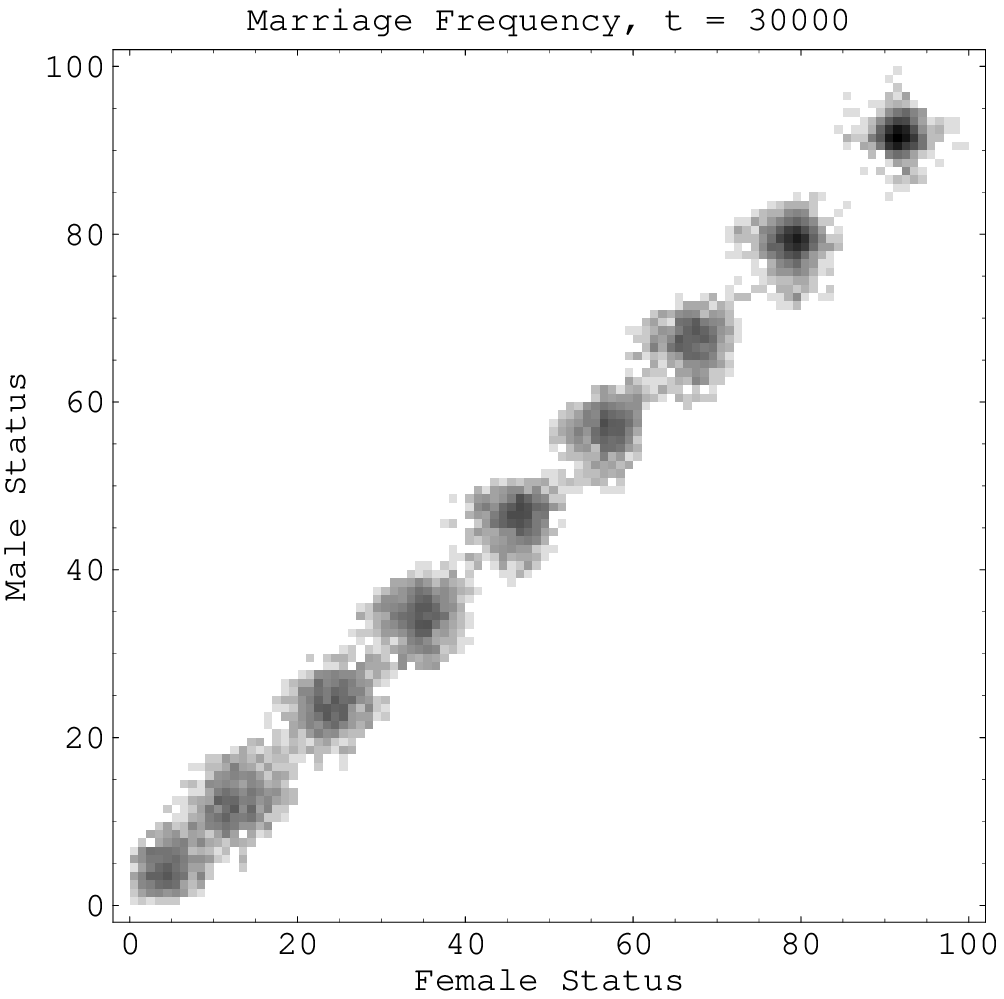}%
	\label{fig:learning-inherited-noachieved.1.marriages.30000}%
}%
\goodgap
\parbox{1.5in}{%
\subfigure[Hypergamy $h,$ for one run]{%
	\includegraphics[width=1.5in]{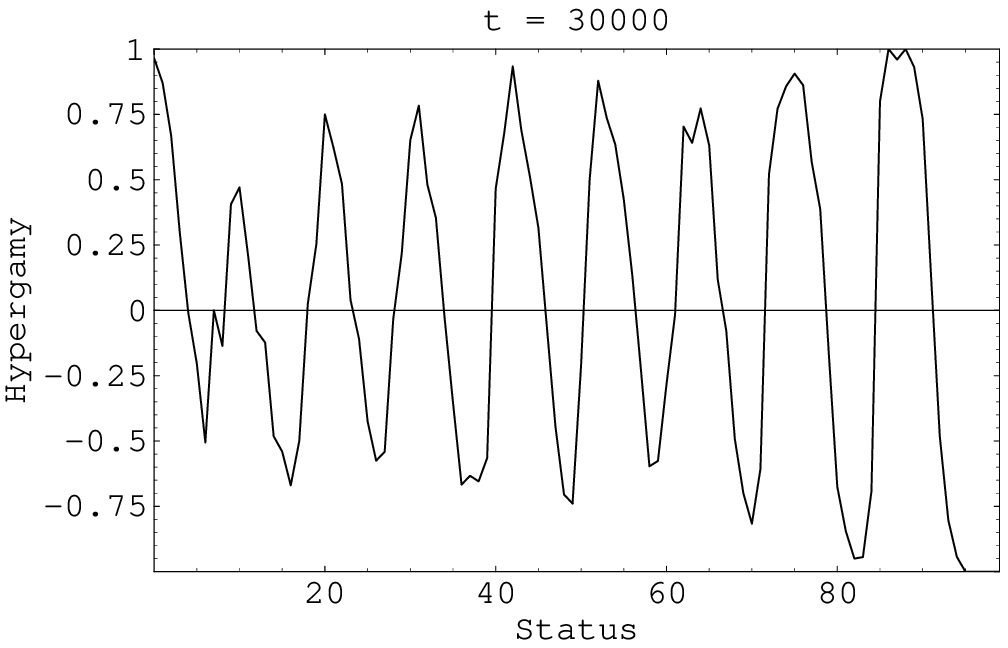}%
	\label{fig:learning-inherited-noachieved.1.hypergamy.30000}%
}%
\\
\subfigure[Hypergamy $h,$ for 50 runs]{%
	\includegraphics[width=1.5in]{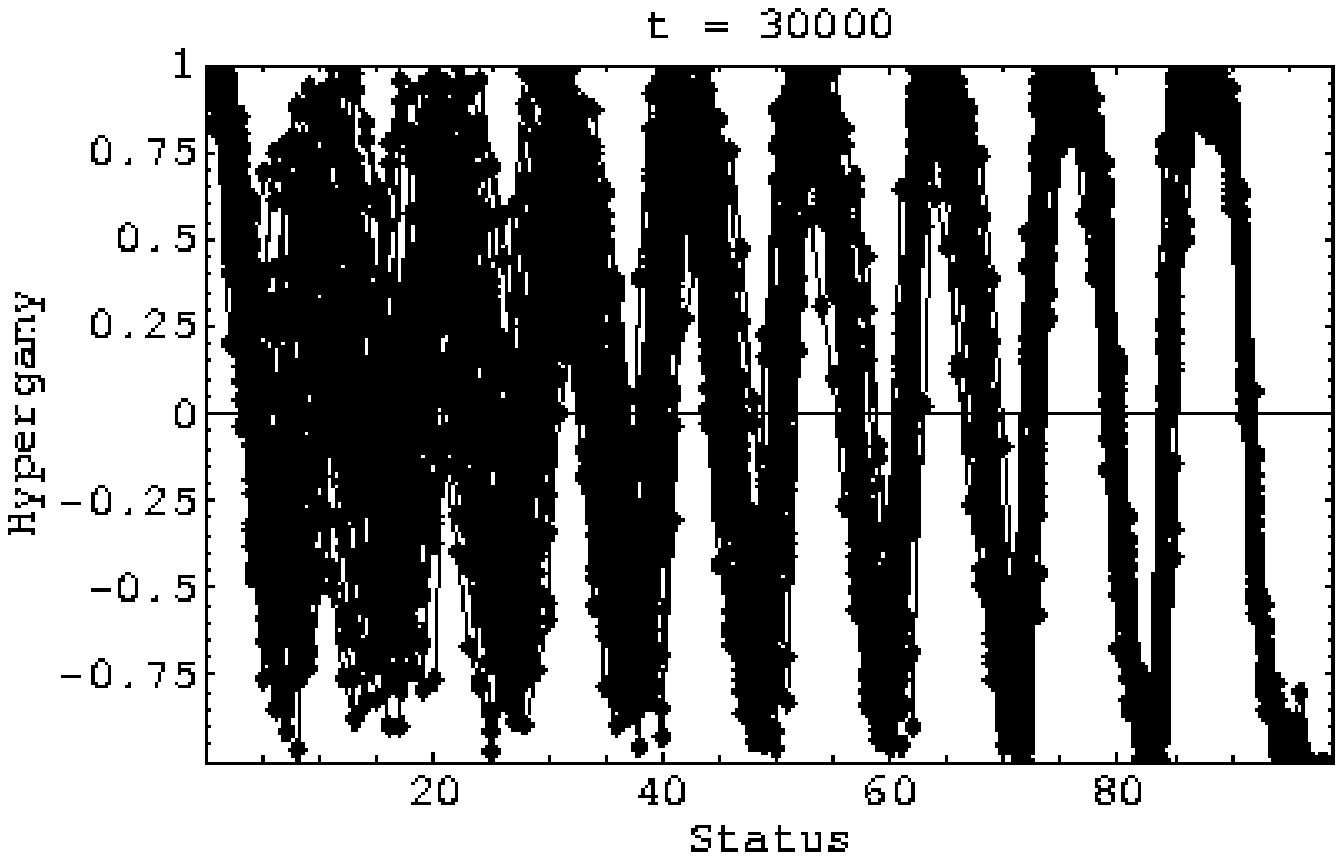}%
	\label{fig:learning-inherited-noachieved.multiHypergamy.30000}%
}%
}%
\goodgap
\parbox{1.5in}{%
\subfigure[Female status histo\-gram, for one run]{%
	\includegraphics[width=1.5in]{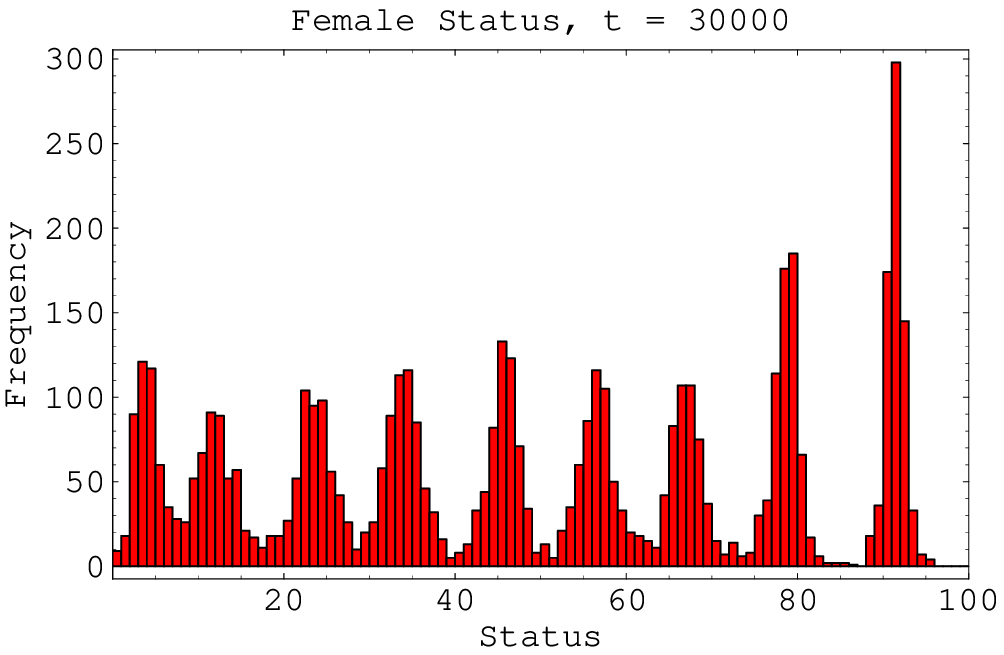}%
	\label{learning-inherited-noachieved.1.fHistogram.30000}%
}%
\\
\subfigure[Male status histo\-gram, for one run]{%
	\includegraphics[width=1.5in]{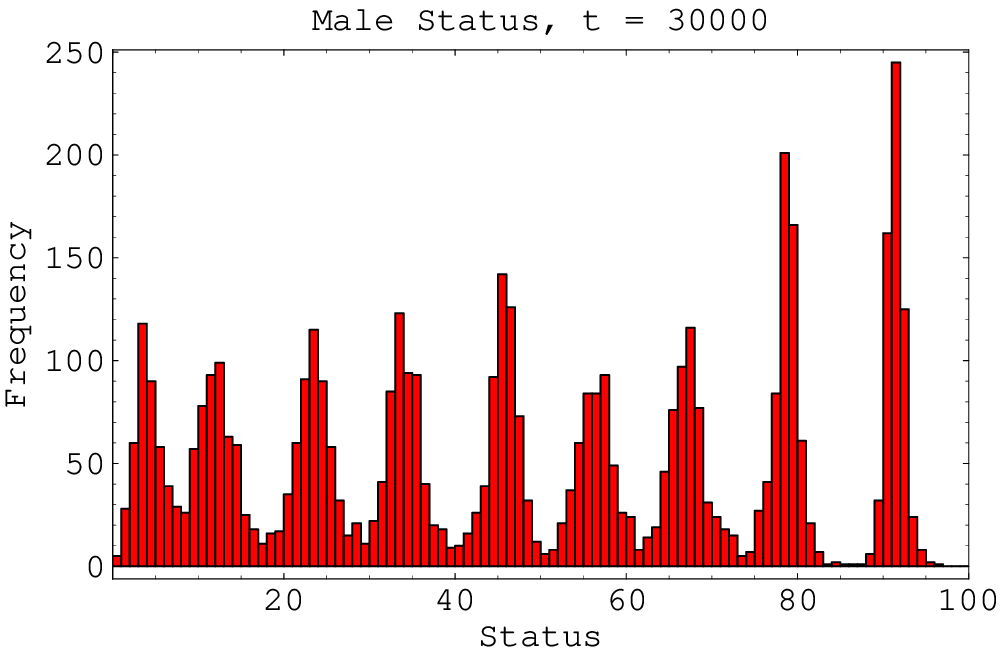}%
	\label{learning-inherited-noachieved.1.mHistogram.30000}%
}%
}%
\end{center}
\caption{Marriage frequency, hypergamy $h,$ and status histograms at
$t = 30,000$ marriages, for the model $\mathbf{M}_{5}$ from
Section~\ref{sec:inherited}, using the learning algorithm
$\mathbf{S}_{2}.$ Children inherited the average of their parents'
statuses, and agents did not achieve any status over their lifetimes.}
\label{fig:learning-inherited-noachieved.1.30000}%
\end{figure}

\begin{figure}
\begin{center}
\subfigure[Marriage frequency, \newline for one run]{%
	\includegraphics[width=1.5in]{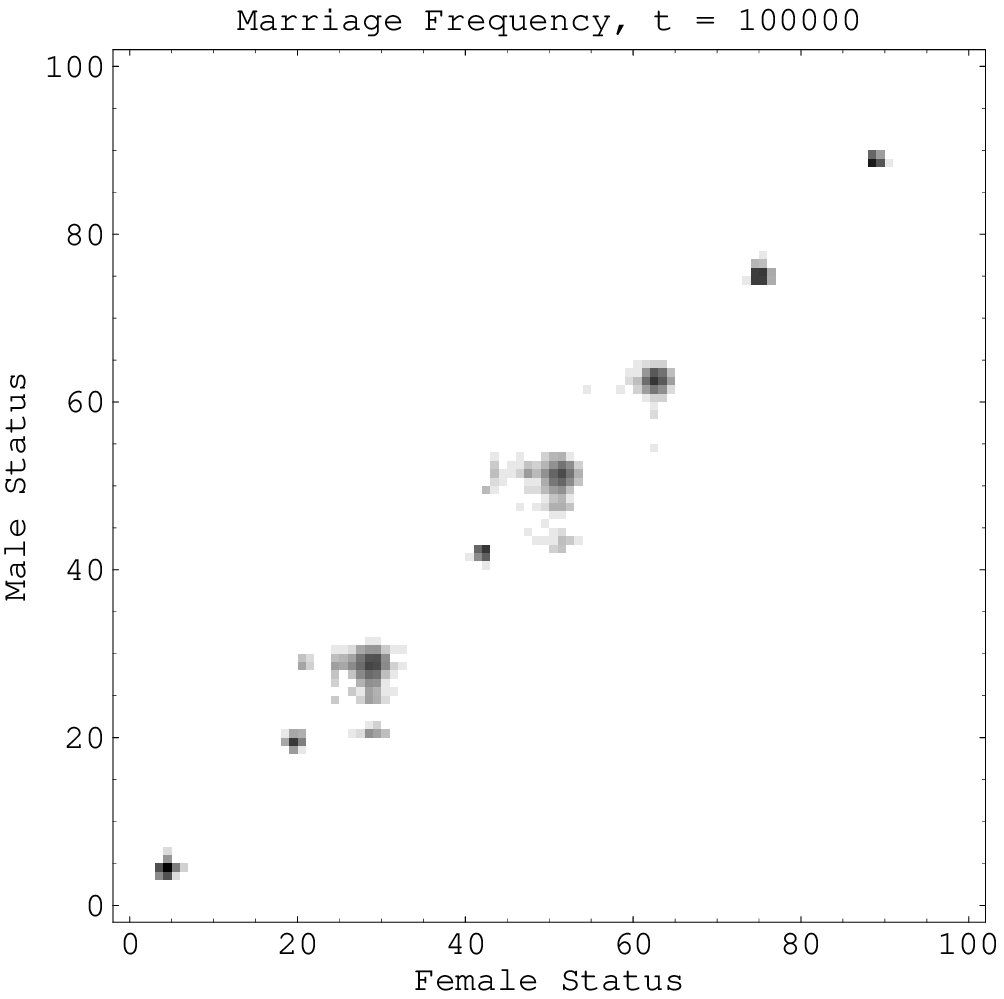}%
	\label{fig:learning-inherited-noachieved.1.marriages.100000}%
}%
\goodgap
\parbox{1.5in}{%
\subfigure[Hypergamy $h,$ for one run]{%
	\includegraphics[width=1.5in]{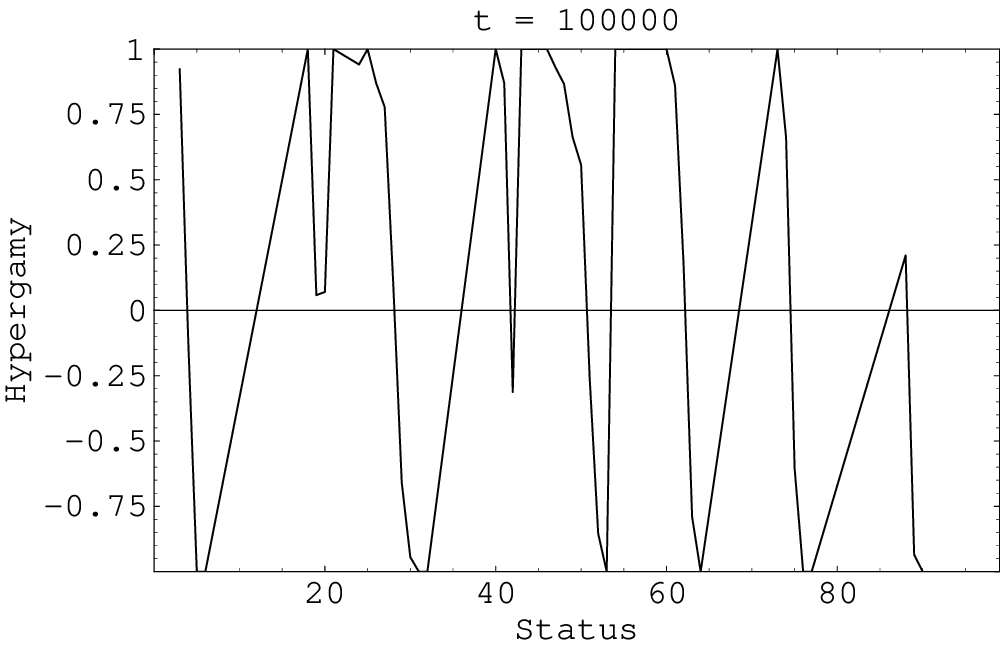}%
	\label{fig:learning-inherited-noachieved.1.hypergamy.100000}%
}%
\\
\subfigure[Hypergamy $h,$ for 50 runs]{%
	\includegraphics[width=1.5in]{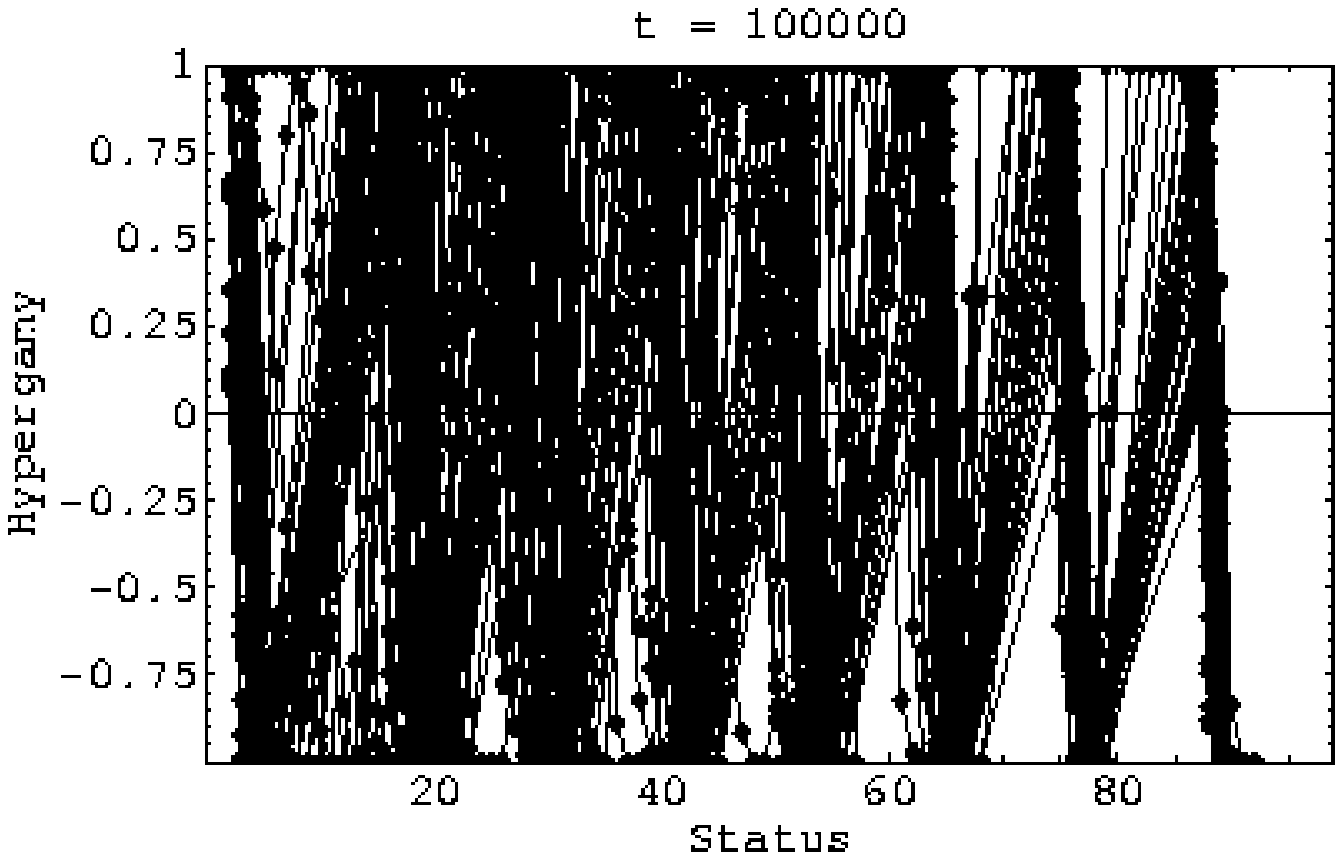}%
	\label{fig:learning-inherited-noachieved.multiHypergamy.100000}%
}%
}%
\goodgap
\parbox{1.5in}{%
\subfigure[Female status histo\-gram, for one run]{%
	\includegraphics[width=1.5in]{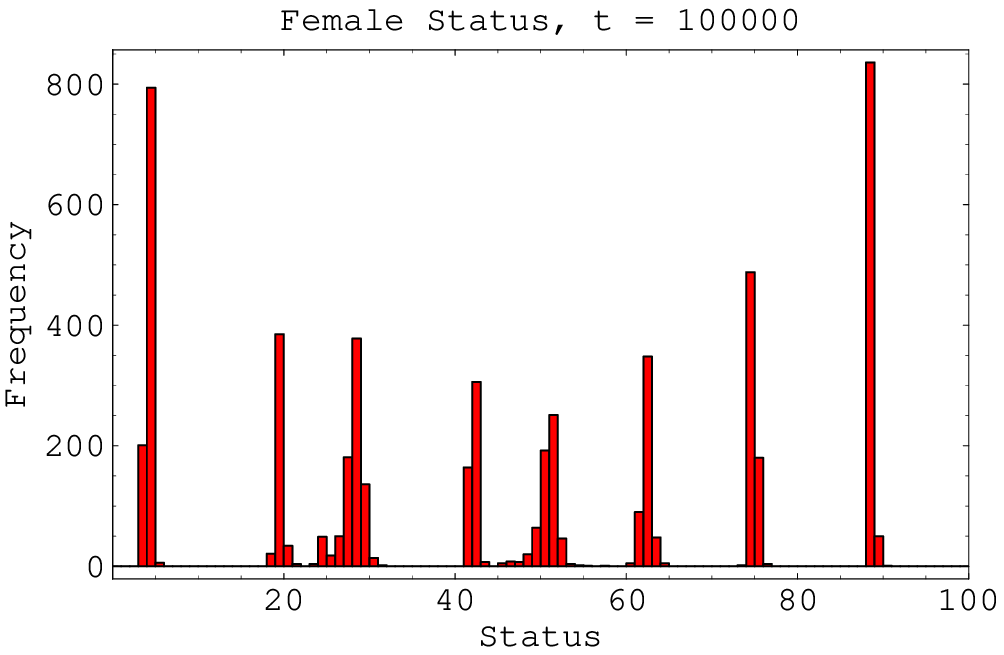}%
	\label{learning-inherited-noachieved.1.fHistogram.100000}%
}%
\\
\subfigure[Male status histo\-gram, for one run]{%
	\includegraphics[width=1.5in]{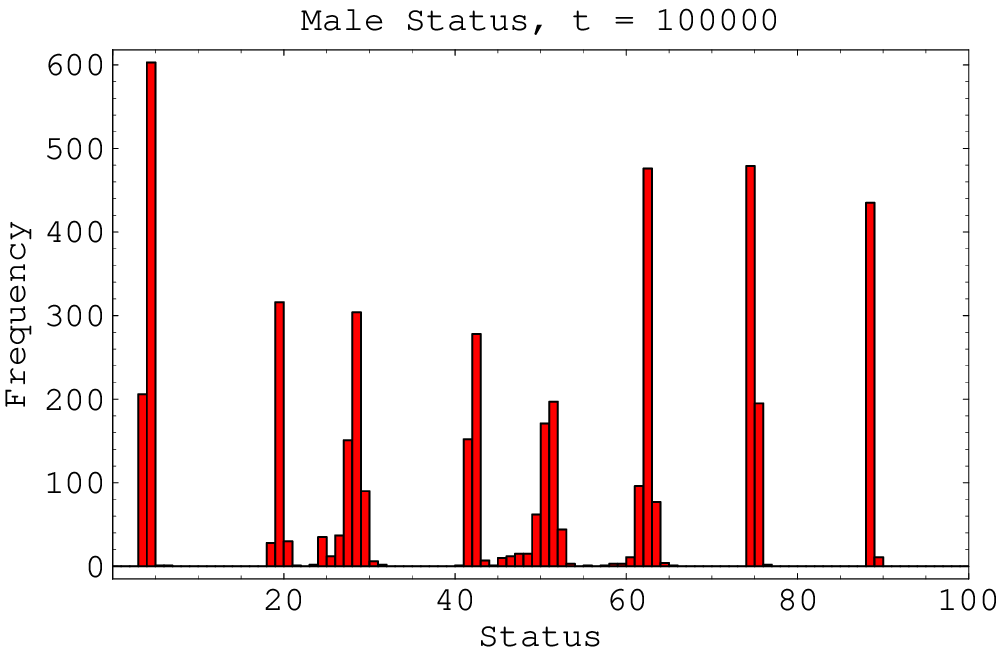}%
	\label{learning-inherited-noachieved.1.mHistogram.100000}%
}%
}%
\end{center}
\caption{Marriage frequency, hypergamy $h,$ and status histograms at
$t = 100,000$ marriages, for the model $\mathbf{M}_{5}$ from
Section~\ref{sec:inherited}, using the learning algorithm
$\mathbf{S}_{2}.$ Children inherited the average of their parents'
statuses, and agents did not achieve any status over their lifetimes.}
\label{fig:learning-inherited-noachieved.1.100000}%
\end{figure}

\begin{figure}
\begin{center}
\subfigure[Marriage frequency, \newline for one run]{%
	\includegraphics[width=1.5in]{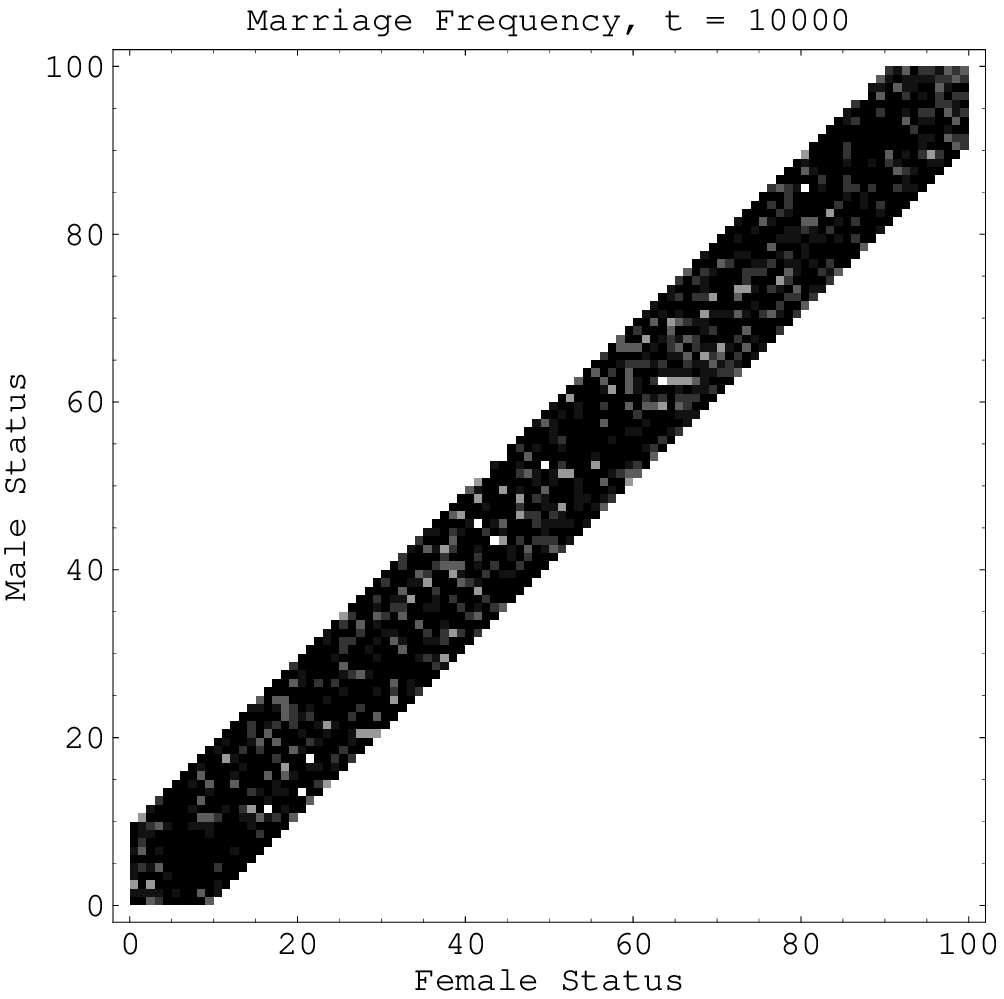}%
	\label{fig:self-inherited-noachieved.1.marriages.10000}%
}%
\goodgap
\parbox{1.5in}{%
\subfigure[Hypergamy $h,$ for one run]{%
	\includegraphics[width=1.5in]{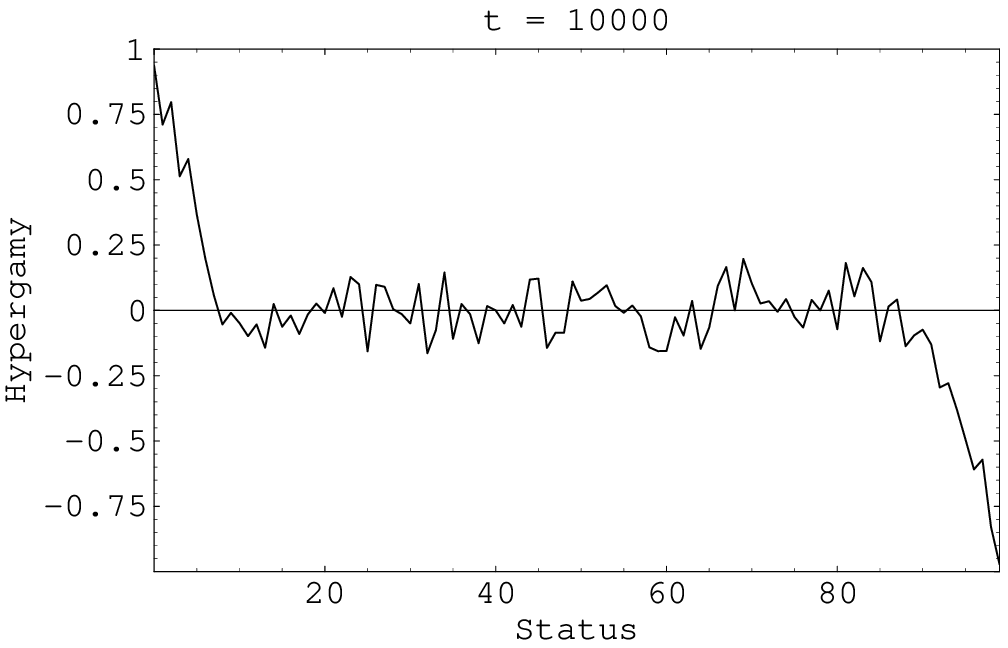}%
	\label{fig:self-inherited-noachieved.1.hypergamy.10000}%
}%
\\
\subfigure[Hypergamy $h,$ for 50 runs]{%
	\includegraphics[width=1.5in]{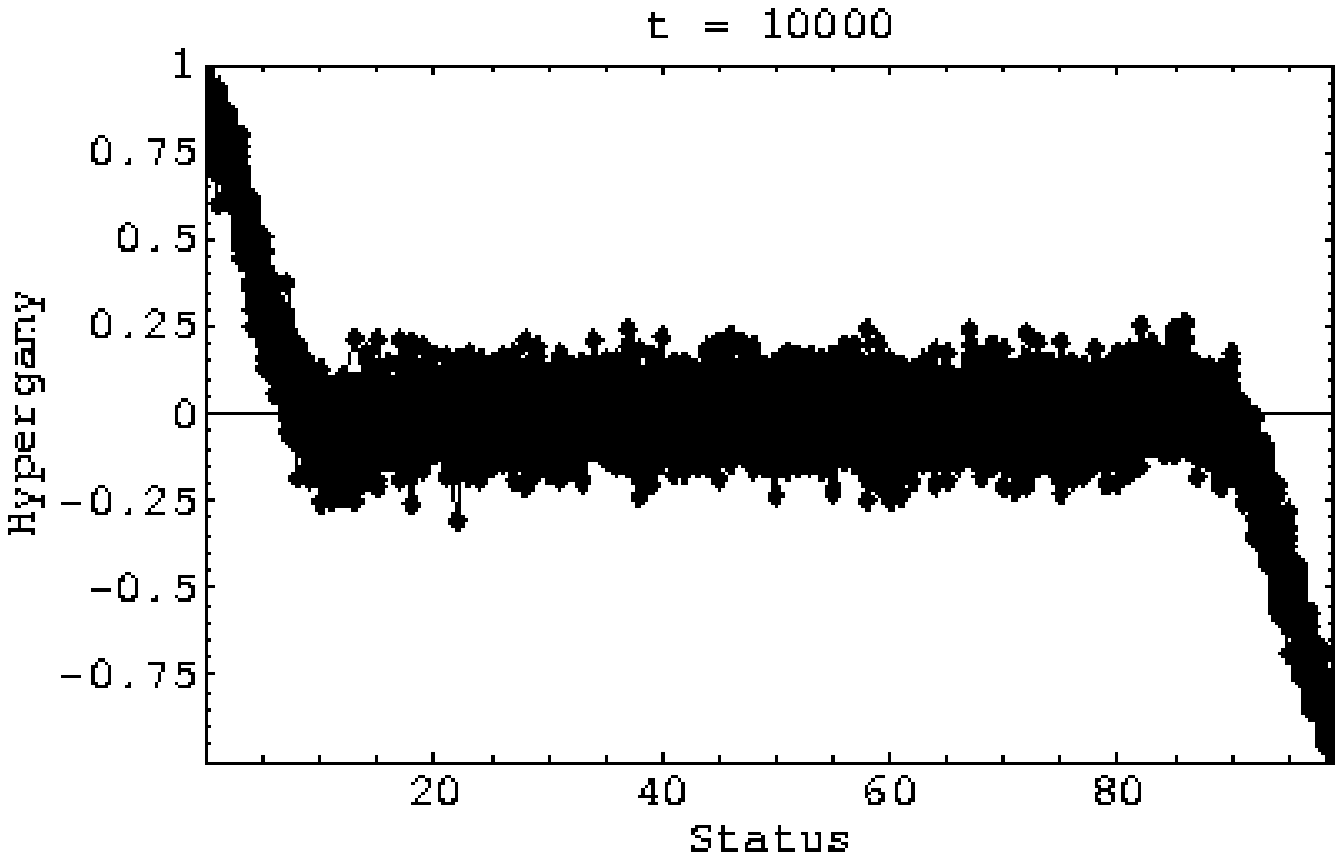}%
	\label{fig:self-inherited-noachieved.multiHypergamy.10000}%
}%
}%
\goodgap
\parbox{1.5in}{%
\subfigure[Female status histo\-gram, for one run]{%
	\includegraphics[width=1.5in]{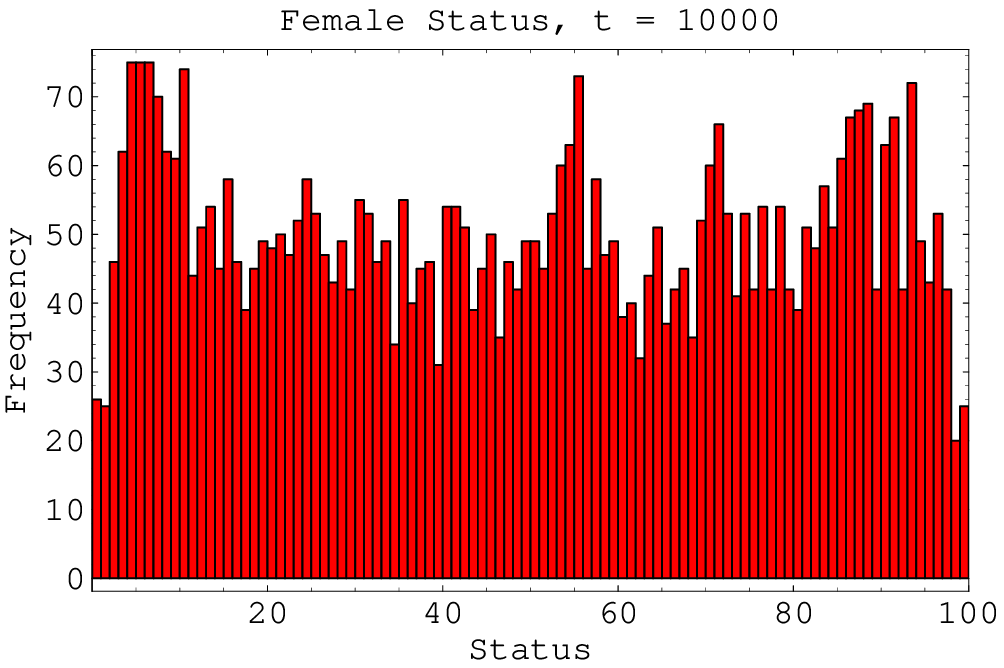}%
	\label{self-inherited-noachieved.1.fHistogram.10000}%
}%
\\
\subfigure[Male status histo\-gram, for one run]{%
	\includegraphics[width=1.5in]{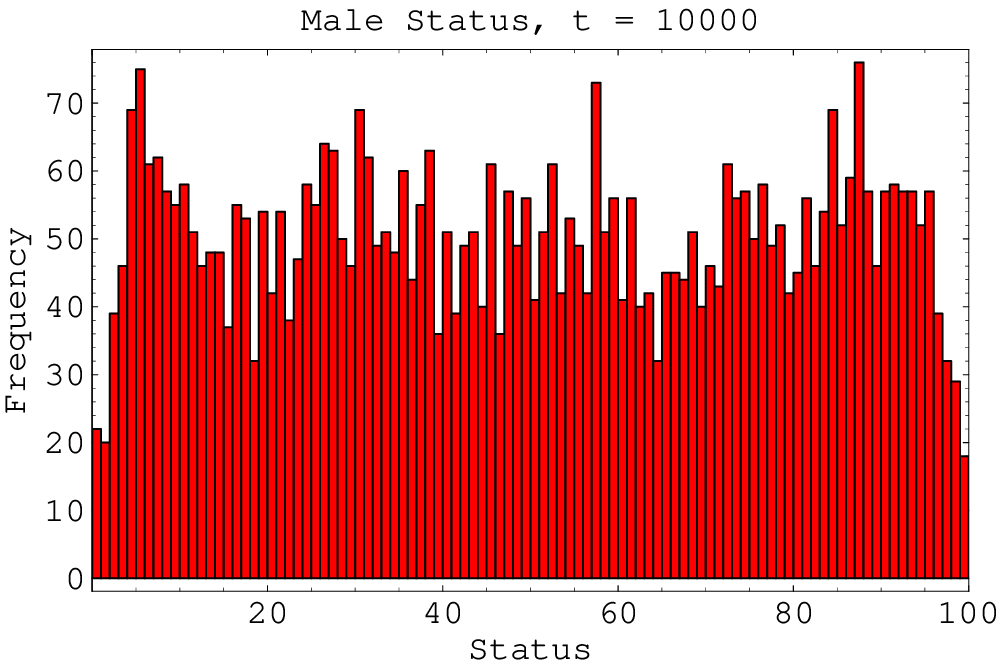}%
	\label{self-inherited-noachieved.1.mHistogram.10000}%
}%
}%
\end{center}
\caption{Marriage frequency, hypergamy $h,$ and status histograms at
$t = 10,000$ marriages, for the model $\mathbf{M}_{6}$ from
Section~\ref{sec:inherited}, using the interval around self strategy
$\mathbf{S}_{3}.$ Children inherited the average of their parents'
statuses, and agents did not achieve any status over their lifetimes.}
\label{fig:self-inherited-noachieved.1.10000}%
\end{figure}

\begin{figure}
\begin{center}
\subfigure[Marriage frequency, \newline for one run]{%
	\includegraphics[width=1.5in]{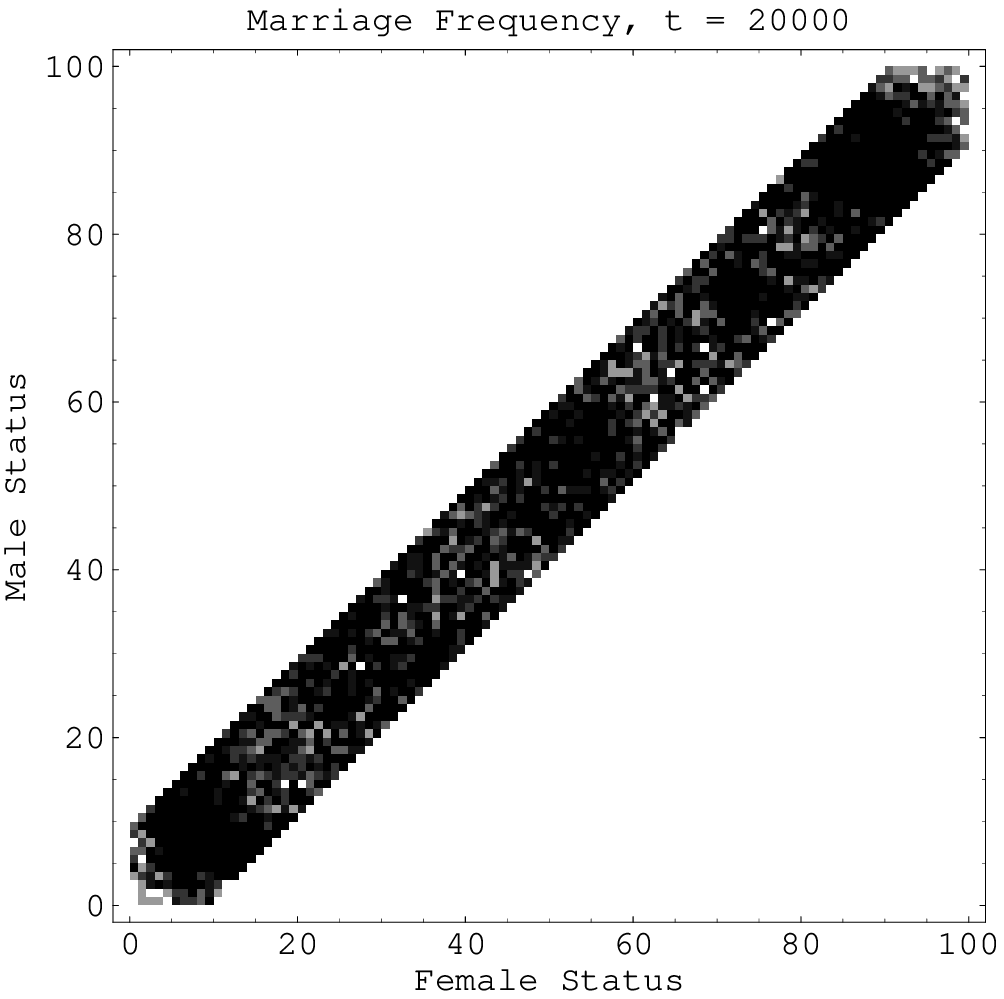}%
	\label{fig:self-inherited-noachieved.1.marriages.20000}%
}%
\goodgap
\parbox{1.5in}{%
\subfigure[Hypergamy $h,$ for one run]{%
	\includegraphics[width=1.5in]{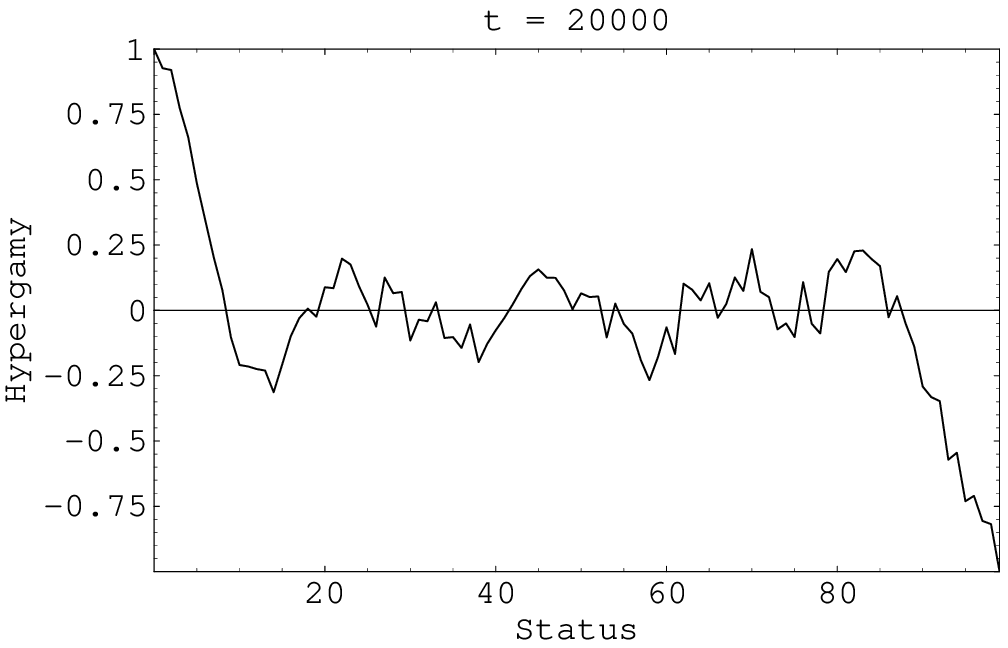}%
	\label{fig:self-inherited-noachieved.1.hypergamy.20000}%
}%
\\
\subfigure[Hypergamy $h,$ for 50 runs]{%
	\includegraphics[width=1.5in]{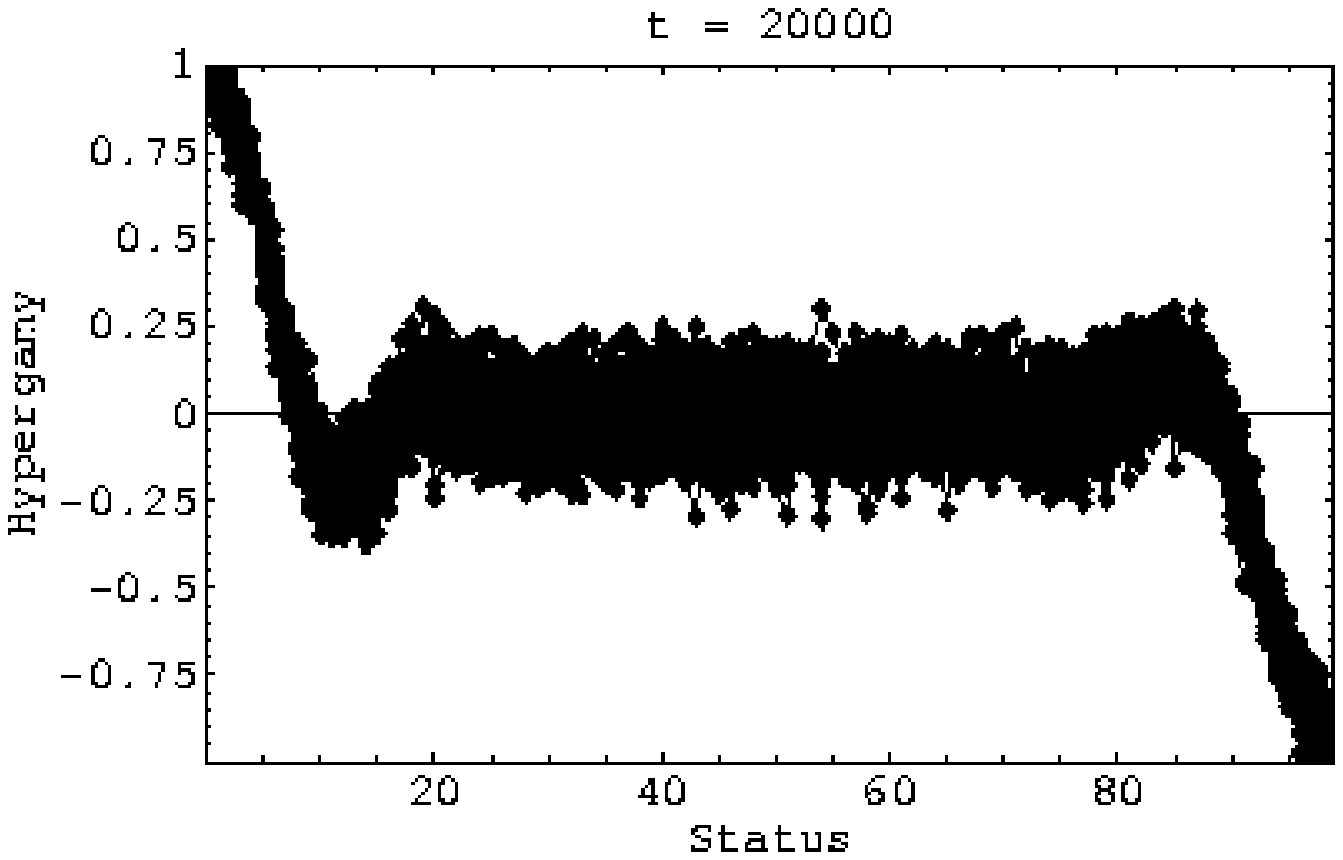}%
	\label{fig:self-inherited-noachieved.multiHypergamy.20000}%
}%
}%
\goodgap
\parbox{1.5in}{%
\subfigure[Female status histo\-gram, for one run]{%
	\includegraphics[width=1.5in]{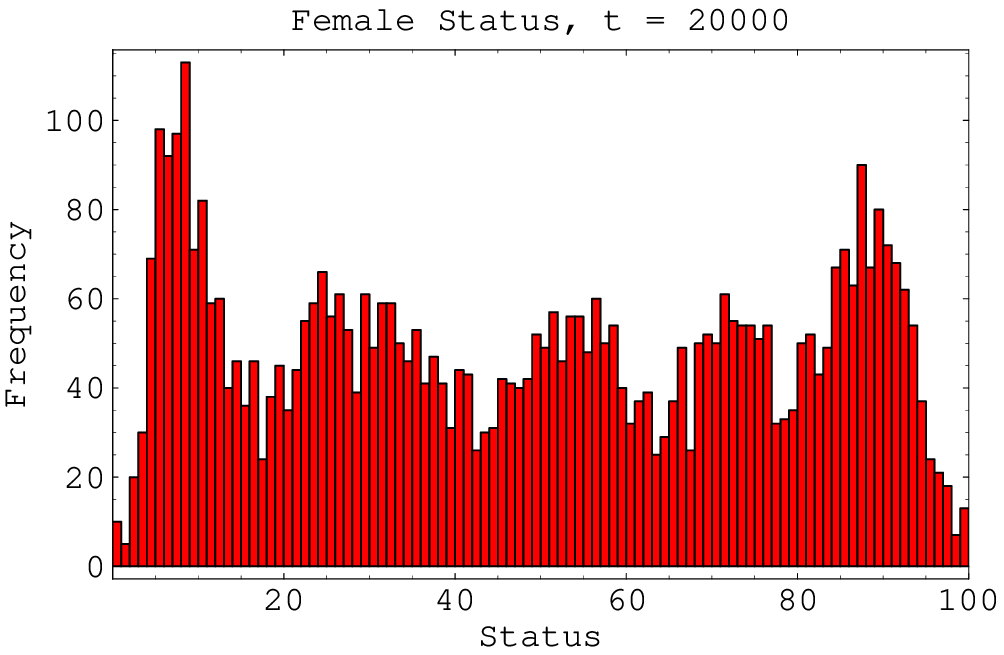}%
	\label{self-inherited-noachieved.1.fHistogram.20000}%
}%
\\
\subfigure[Male status histo\-gram, for one run]{%
	\includegraphics[width=1.5in]{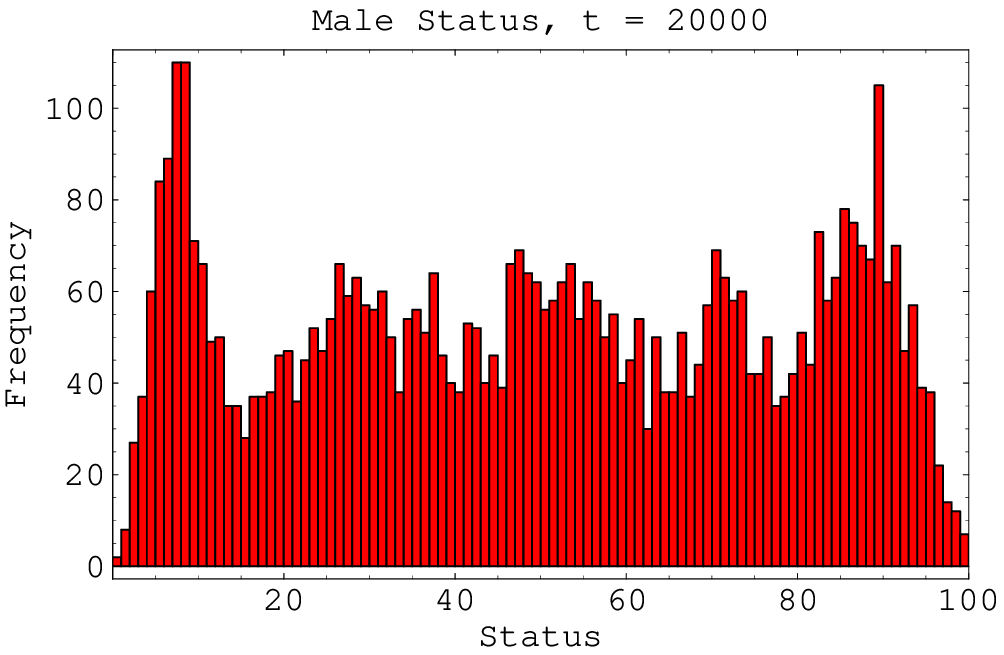}%
	\label{self-inherited-noachieved.1.mHistogram.20000}%
}%
}%
\end{center}
\caption{Marriage frequency, hypergamy $h,$ and status histograms at
$t = 20,000$ marriages, for the model $\mathbf{M}_{6}$ from
Section~\ref{sec:inherited}, using the interval around self strategy
$\mathbf{S}_{3}.$ Children inherited the average of their parents'
statuses, and agents did not achieve any status over their lifetimes.}
\label{fig:self-inherited-noachieved.1.20000}%
\end{figure}

\begin{figure}
\begin{center}
\subfigure[Marriage frequency, \newline for one run]{%
	\includegraphics[width=1.5in]{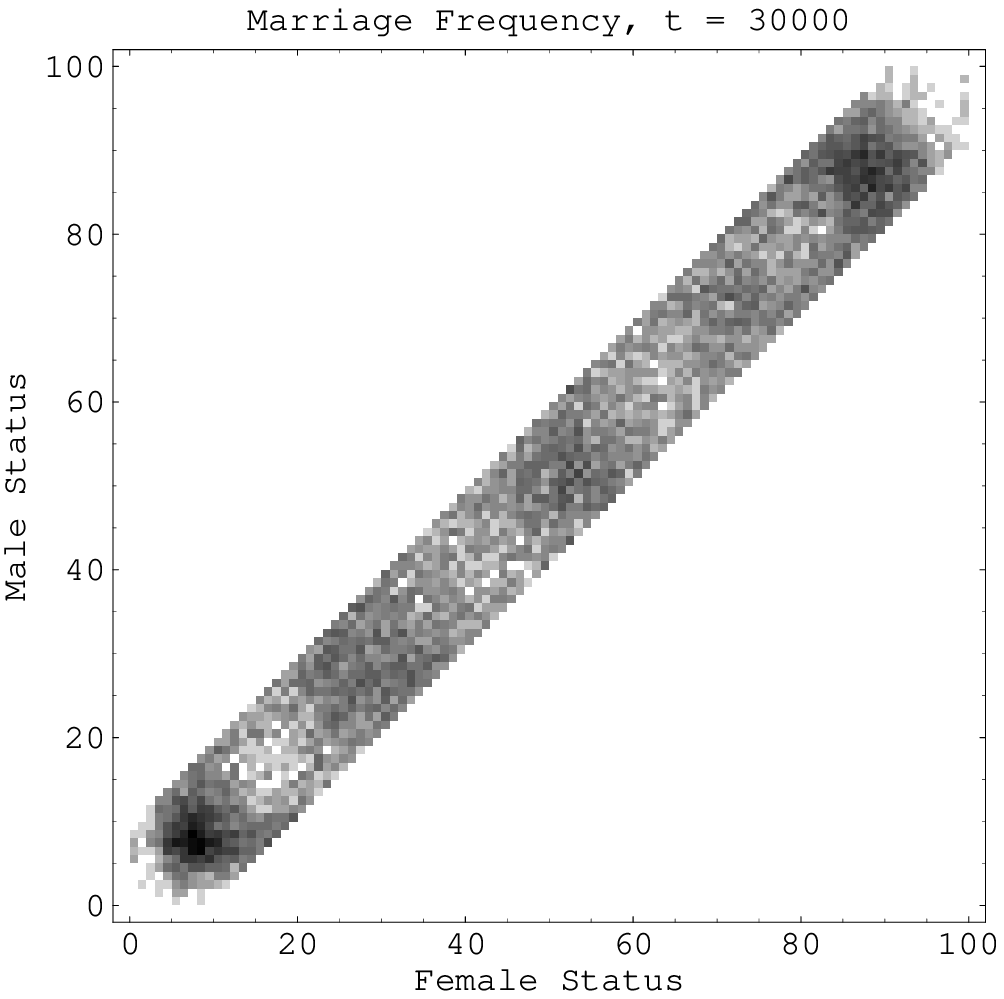}%
	\label{fig:self-inherited-noachieved.1.marriages.30000}%
}%
\goodgap
\parbox{1.5in}{%
\subfigure[Hypergamy $h,$ for one run]{%
	\includegraphics[width=1.5in]{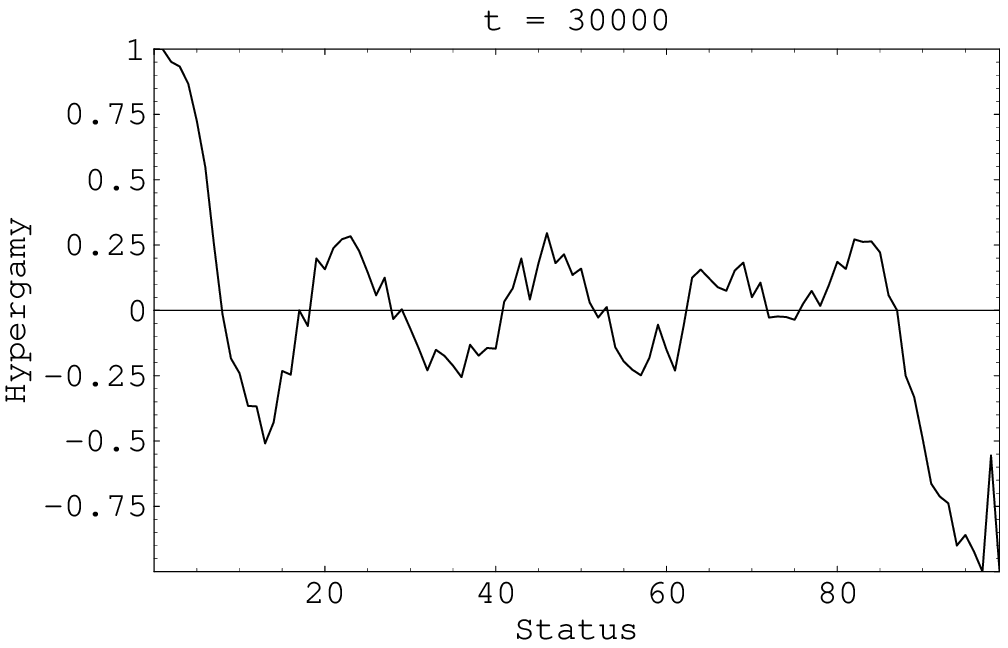}%
	\label{fig:self-inherited-noachieved.1.hypergamy.30000}%
}%
\\
\subfigure[Hypergamy $h,$ for 50 runs]{%
	\includegraphics[width=1.5in]{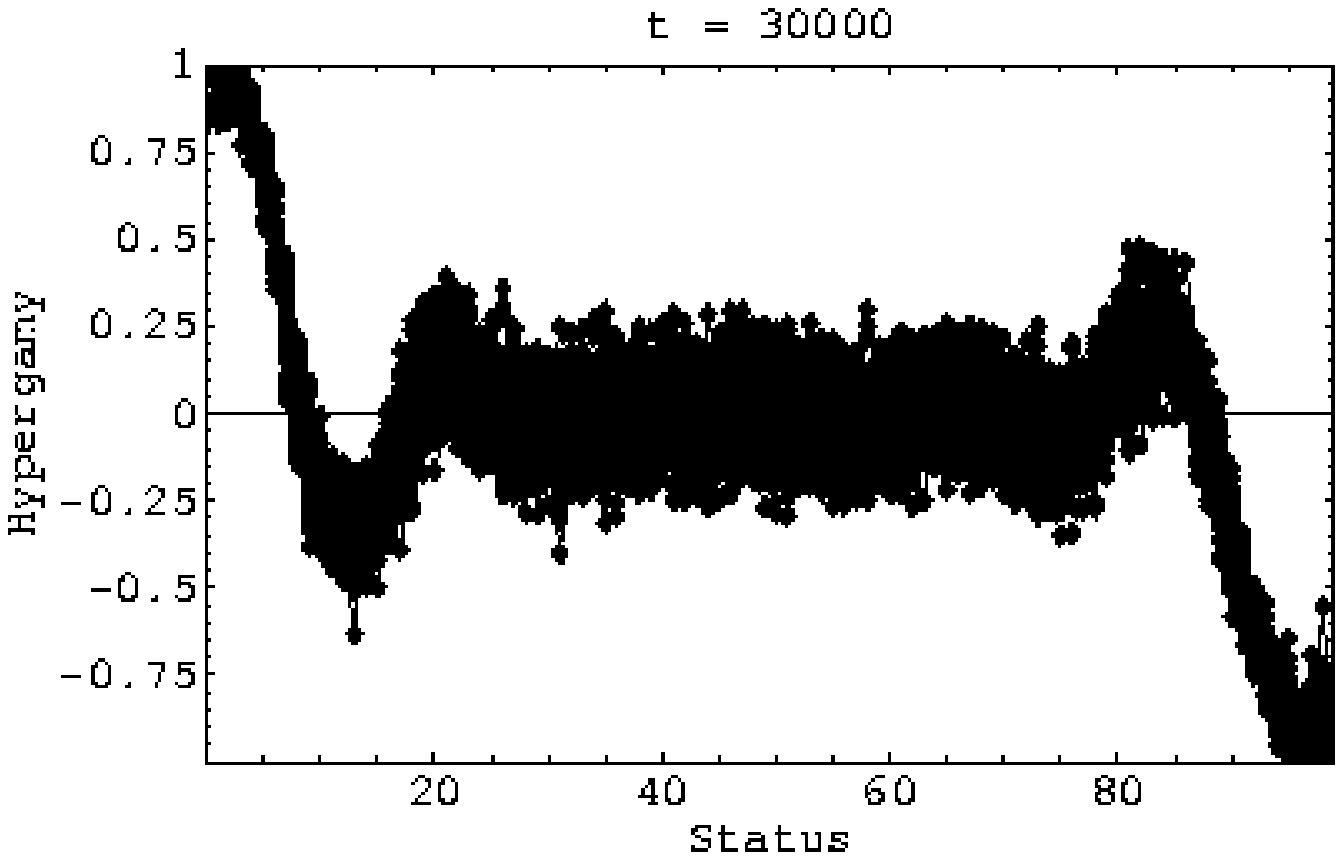}%
	\label{fig:self-inherited-noachieved.multiHypergamy.30000}%
}%
}%
\goodgap
\parbox{1.5in}{%
\subfigure[Female status histo\-gram, for one run]{%
	\includegraphics[width=1.5in]{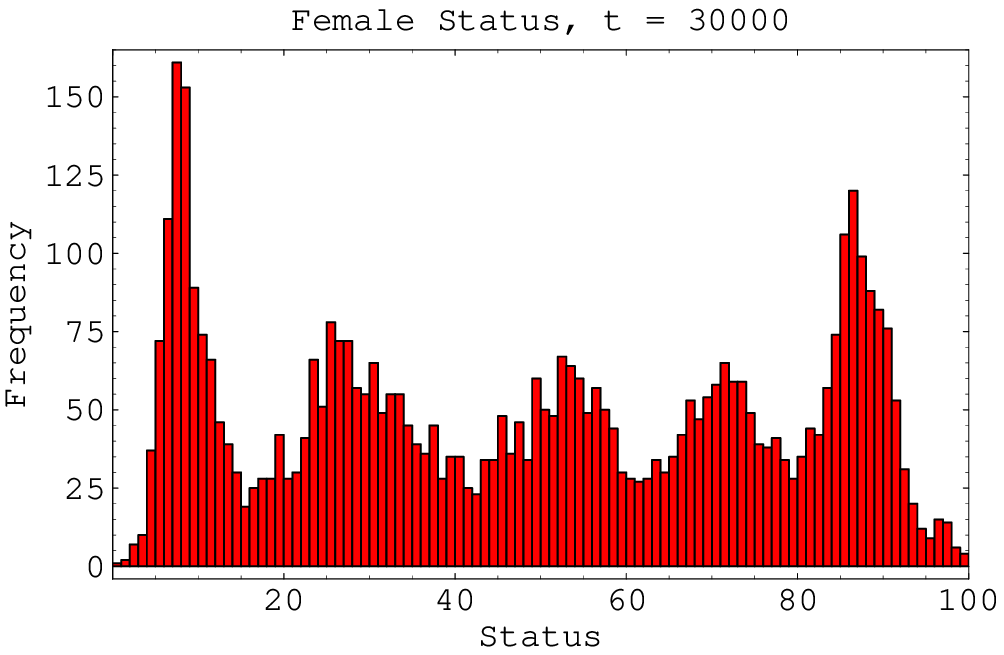}%
	\label{self-inherited-noachieved.1.fHistogram.30000}%
}%
\\
\subfigure[Male status histo\-gram, for one run]{%
	\includegraphics[width=1.5in]{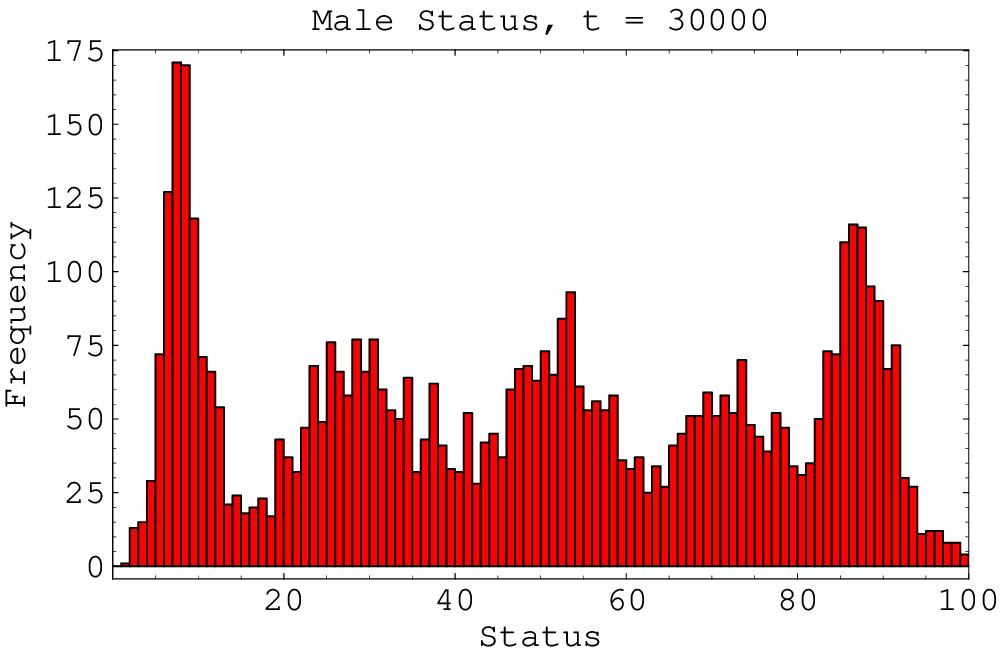}%
	\label{self-inherited-noachieved.1.mHistogram.30000}%
}%
}%
\end{center}
\caption{Marriage frequency, hypergamy $h,$ and status histograms at
$t = 30,000$ marriages, for the model $\mathbf{M}_{6}$ from
Section~\ref{sec:inherited}, using the interval around self strategy
$\mathbf{S}_{3}.$ Children inherited the average of their parents'
statuses, and agents did not achieve any status over their lifetimes.}
\label{fig:self-inherited-noachieved.1.30000}%
\end{figure}

\begin{figure}
\begin{center}
\subfigure[Marriage frequency, \newline for one run]{%
	\includegraphics[width=1.5in]{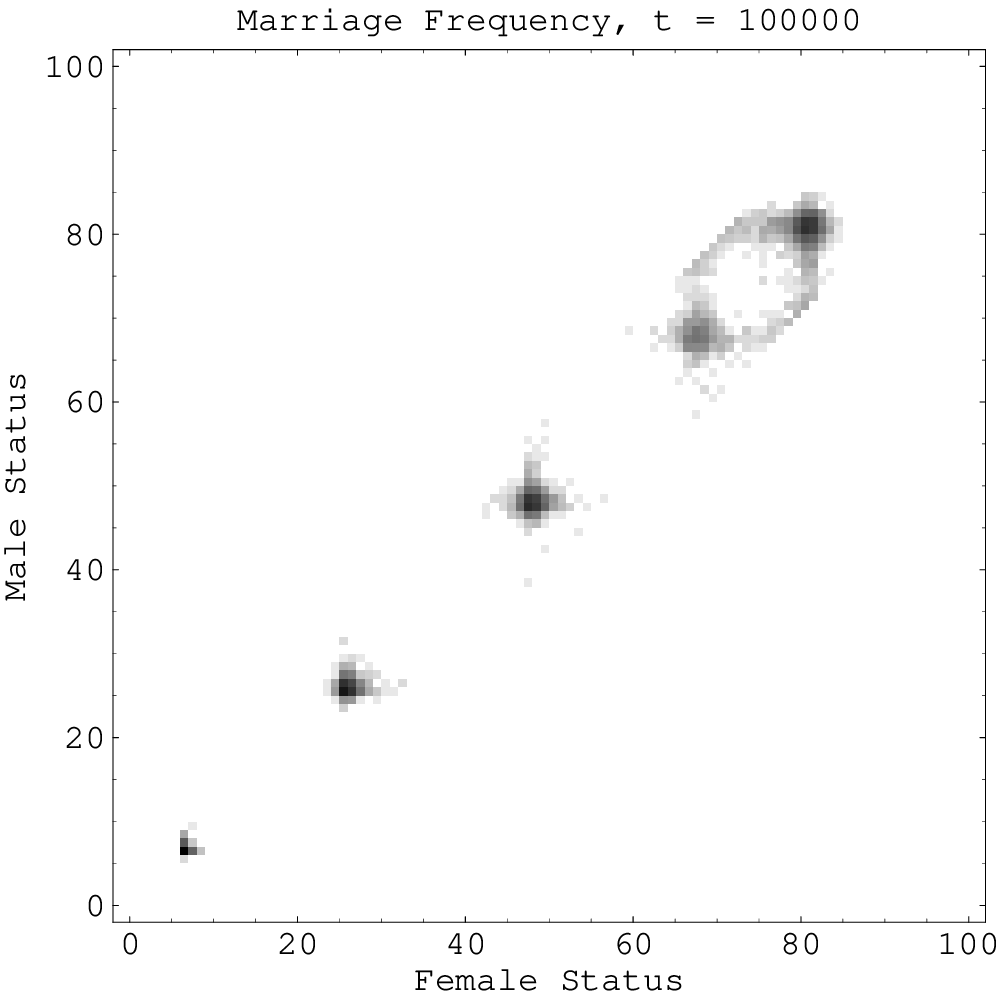}%
	\label{fig:self-inherited-noachieved.1.marriages.100000}%
}%
\goodgap
\parbox{1.5in}{%
\subfigure[Hypergamy $h,$ for one run]{%
	\includegraphics[width=1.5in]{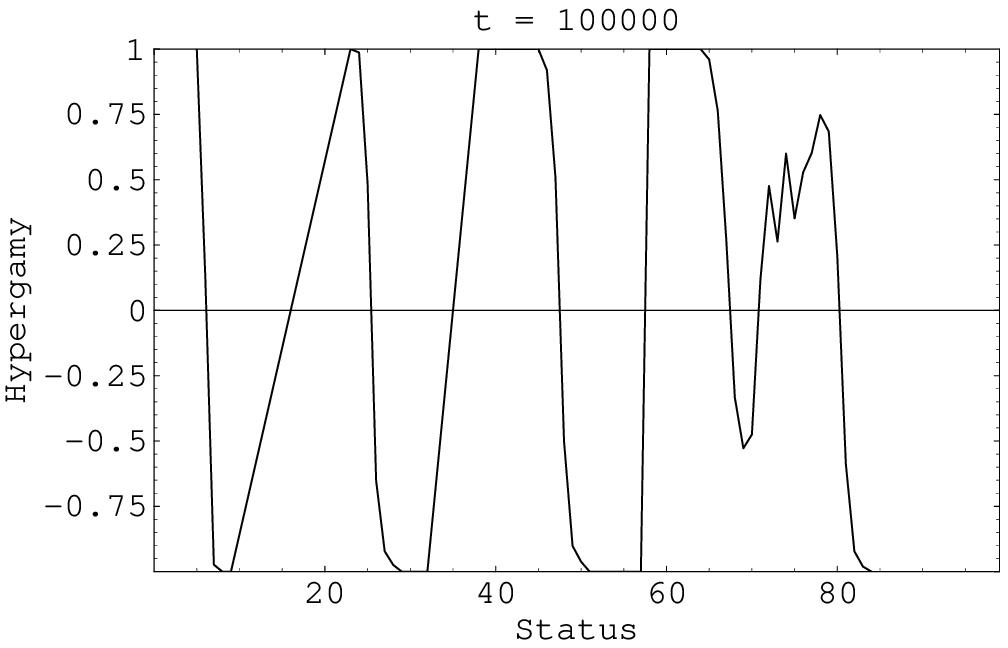}%
	\label{fig:self-inherited-noachieved.1.hypergamy.100000}%
}%
\\
\subfigure[Hypergamy $h,$ for 50 runs]{%
	\includegraphics[width=1.5in]{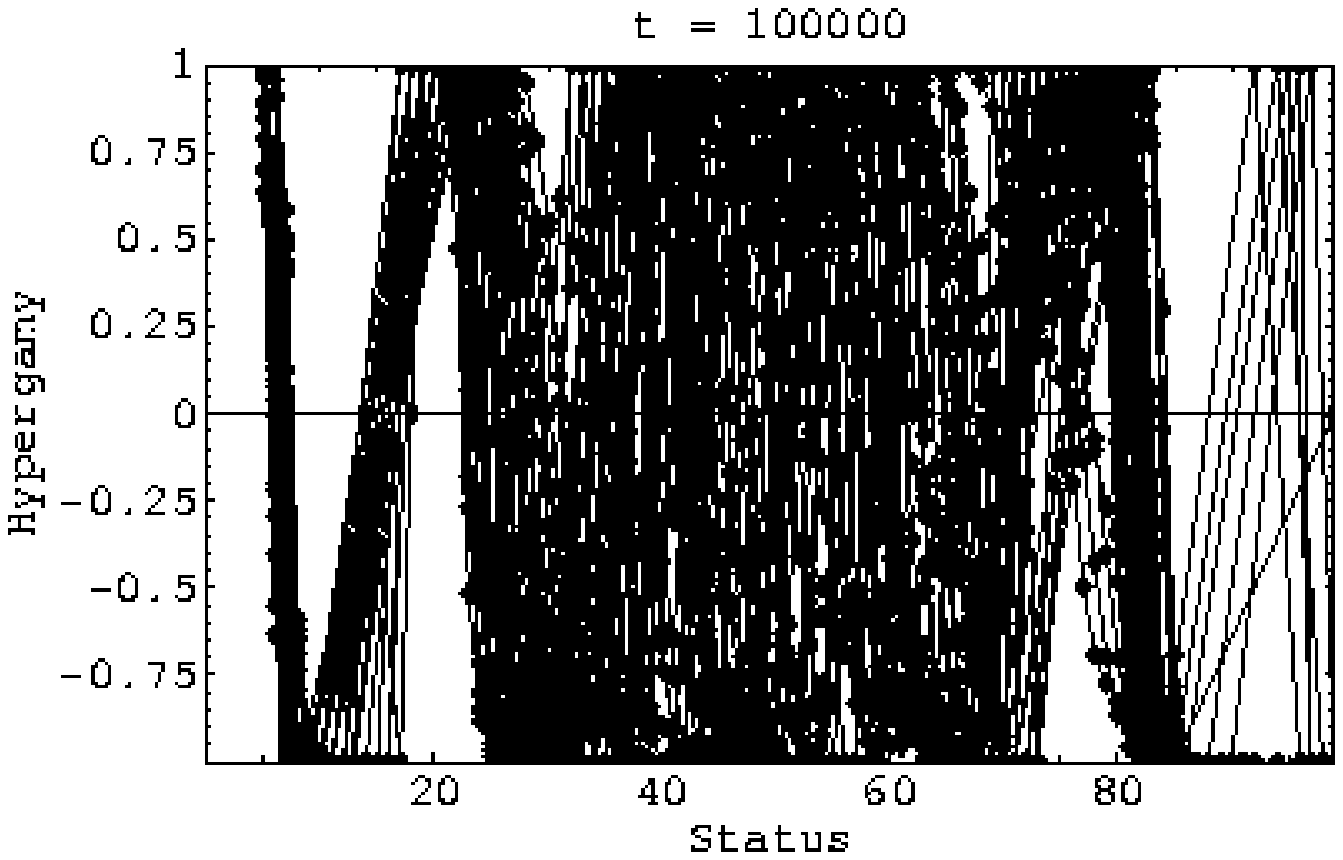}%
	\label{fig:self-inherited-noachieved.multiHypergamy.100000}%
}%
}%
\goodgap
\parbox{1.5in}{%
\subfigure[Female status histo\-gram, for one run]{%
	\includegraphics[width=1.5in]{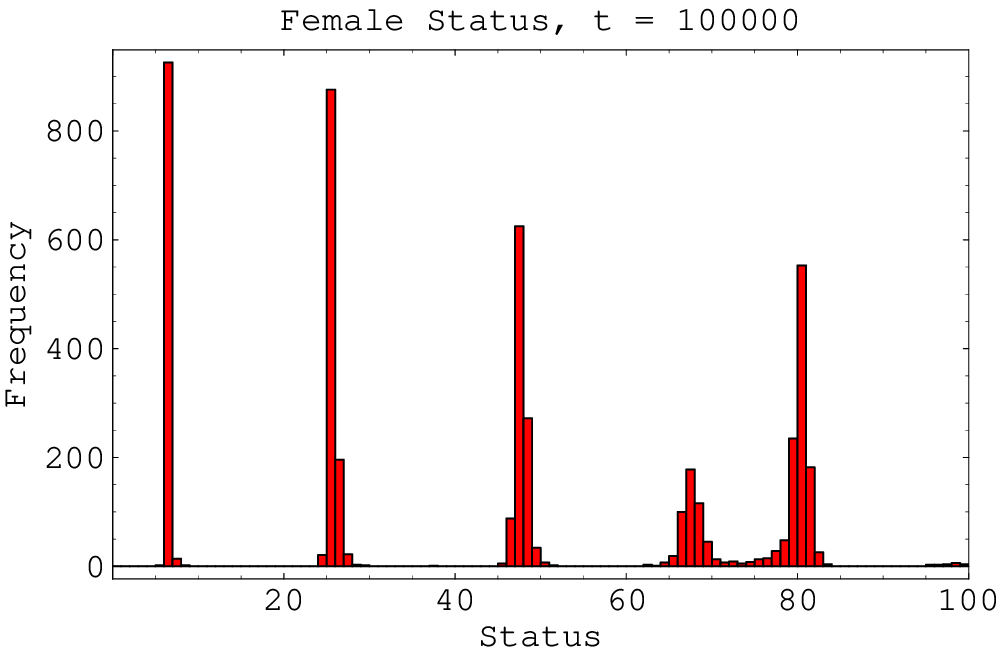}%
	\label{self-inherited-noachieved.1.fHistogram.100000}%
}%
\\
\subfigure[Male status histo\-gram, for one run]{%
	\includegraphics[width=1.5in]{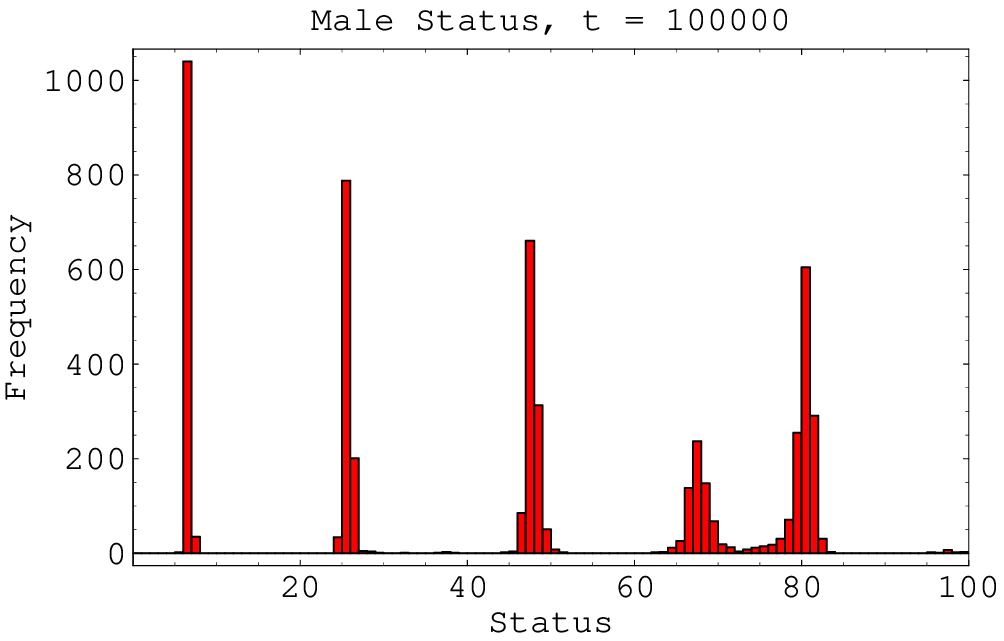}%
	\label{self-inherited-noachieved.1.mHistogram.100000}%
}%
}%
\end{center}
\caption{Marriage frequency, hypergamy $h,$ and status histograms at
$t = 100,000$ marriages, for the model $\mathbf{M}_{6}$ from
Section~\ref{sec:inherited}, using the interval around self strategy
$\mathbf{S}_{3}.$ Children inherited the average of their parents'
statuses, and agents did not achieve any status over their lifetimes.}
\label{fig:self-inherited-noachieved.1.100000}%
\end{figure}

Each of the previous computer models $\mathbf{M}_{1}, \mathbf{M}_{2},$
and $\mathbf{M}_{3}$ was modified to use this form of inheritance,
with children's sex determined randomly, producing a new set of models
$\mathbf{M}_{4}, \mathbf{M}_{5},$ and $\mathbf{M}_{6}.$ The results
are shown in
Figures~\ref{fig:rational-inherited-noachieved.1.10000}--\ref{fig:rational-inherited-noachieved.1.100000}
for model $\mathbf{M}_{4}$ using the rational strategy
$\mathbf{S}_{1}$ from Section~\ref{sec:rational},
Figures~\ref{fig:learning-inherited-noachieved.1.10000}--\ref{fig:learning-inherited-noachieved.1.100000}
for model $\mathbf{M}_{5}$ using the learning algorithm
$\mathbf{S}_{2}$ from Section~\ref{sec:learning}, and
Figures~\ref{fig:self-inherited-noachieved.1.10000}--\ref{fig:self-inherited-noachieved.1.100000}
for model $\mathbf{M}_{6}$ using the interval strategy
$\mathbf{S}_{3}$ from Section~\ref{sec:self}. Note that the hypergamy
metric $h$ breaks down as a proxy for measuring class endogamy when
the peaks in the status histogram become very narrow, as shown in
Figure~\ref{fig:rational-inherited-noachieved.1.100000}.  This is
because $h = 0$ at many points, or is undefined, reducing the size of
the jumps from low $h$ to high $h,$ despite the fact that the degree
of class endogamy is very high.

Under all three strategies, this form of inherited status has the
effect of converting a uniform status distribution into a multimodal one,
with peaks and valleys in the status histogram. Over long time
periods, gaps in the status distribution appear, in which there are
very few agents having a given status level, or none at all. 

The multimodality and gaps in the status distribution are caused by a
kind of regression towards the mean: A child's expected status lies in
the middle of an endogamous class. Thus, the class boundaries move
inwards over time.  Even under strategy $\mathbf{S}_{3}$ in model
$\mathbf{M}_{6},$ this effect alone is strong enough to produce class
endogamy: Gaps appear in the status distribution after about $30,000$
marriages, reducing agents' marriage options outside of the local peak
in the status histogram. This seems to be driven by the fact that the
highest- and lowest-status individuals are constrained to find spouses
within intervals $\epsilon$ lower and higher than them,
respectively. These constraints produce sufficient class boundaries
for the regression effect to produce gaps in the status
histogram. This effect seems to be very similar to the segregation
phenomenon investigated by Weisbuch and
coworkers~\cite{weisbuch:etal:2002}, in their simplest model. Although
they do not provide an explanation for this phenomenon, they do show
that the number of clusters formed is generally equal to
$\lfloor 100/(2 (\epsilon + 1)) \rfloor$ (translated into the parameters used
here). This prediction seems to be borne out in model $\mathbf{M}_{6},$
since the number of classes formed by the end of a run tends to be
$\lfloor 100/(2 (9 + 1)) \rfloor = 5,$ as can be seen in
Figure~\ref{fig:self-inherited-noachieved.1.marriages.100000}. This
model may also have some similarities with Axelrod's culture diffusion
model~\cite{axelrod:1997b}, in which a lattice of villages coalesces
into a few discrete ``cultures''.

In general, averaging inheritance increases the degree of class
endogamy under all three strategies. Hence, this type of inheritance is
another force that can produce class endogamy. (If children inherited
a weighted average of their parents' statuses, this force would still
drive the model towards class endogamy, though less strongly.)

% rational-inherited-noachieved
%

%

% learning-inherited-noachieved
%

%

% self-inherited-noachieved
%

\subsection{Achieved Status} \label{sec:achieved}

\begin{figure}
\begin{center}
\subfigure[Marriage frequency, \newline for one run]{%
	\includegraphics[width=1.5in]{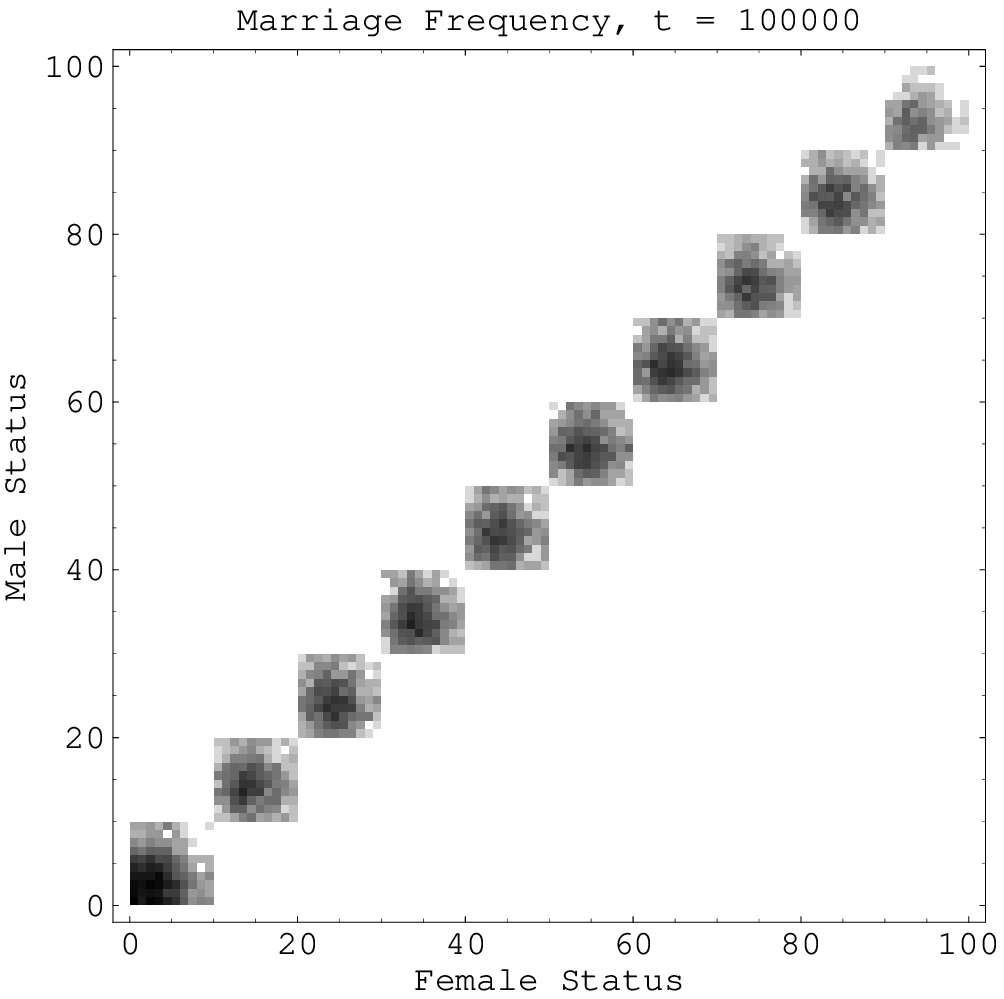}%
	\label{fig:rational-inherited-achieved.1.marriages.100000}%
}%
\goodgap
\parbox{1.5in}{%
\subfigure[Hypergamy $h,$ for one run]{%
	\includegraphics[width=1.5in]{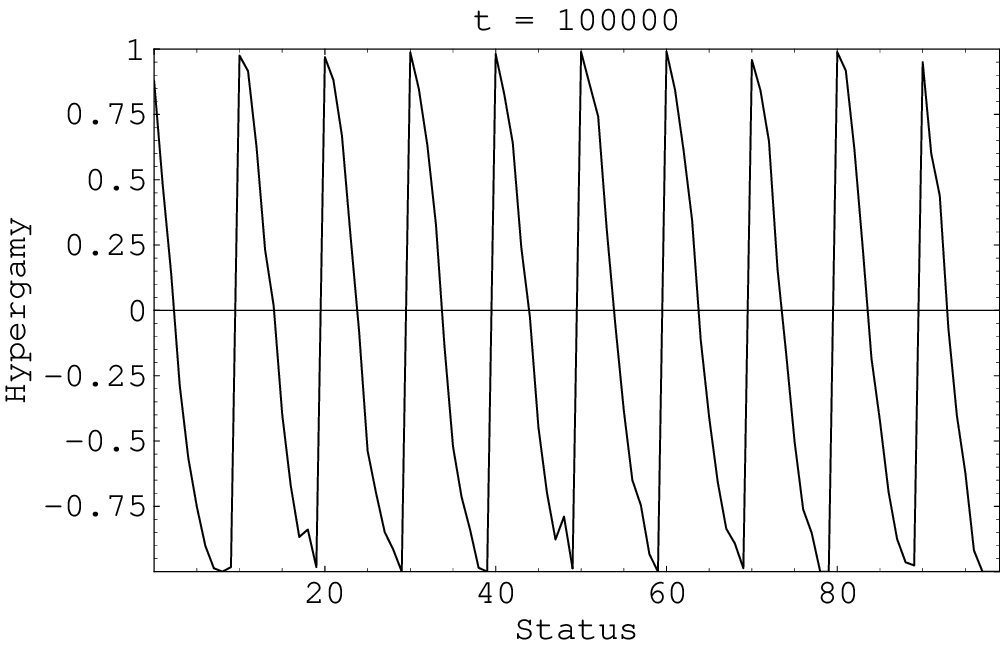}%
	\label{fig:rational-inherited-achieved.1.hypergamy.100000}%
}%
\\
\subfigure[Hypergamy $h,$ for 50 runs]{%
	\includegraphics[width=1.5in]{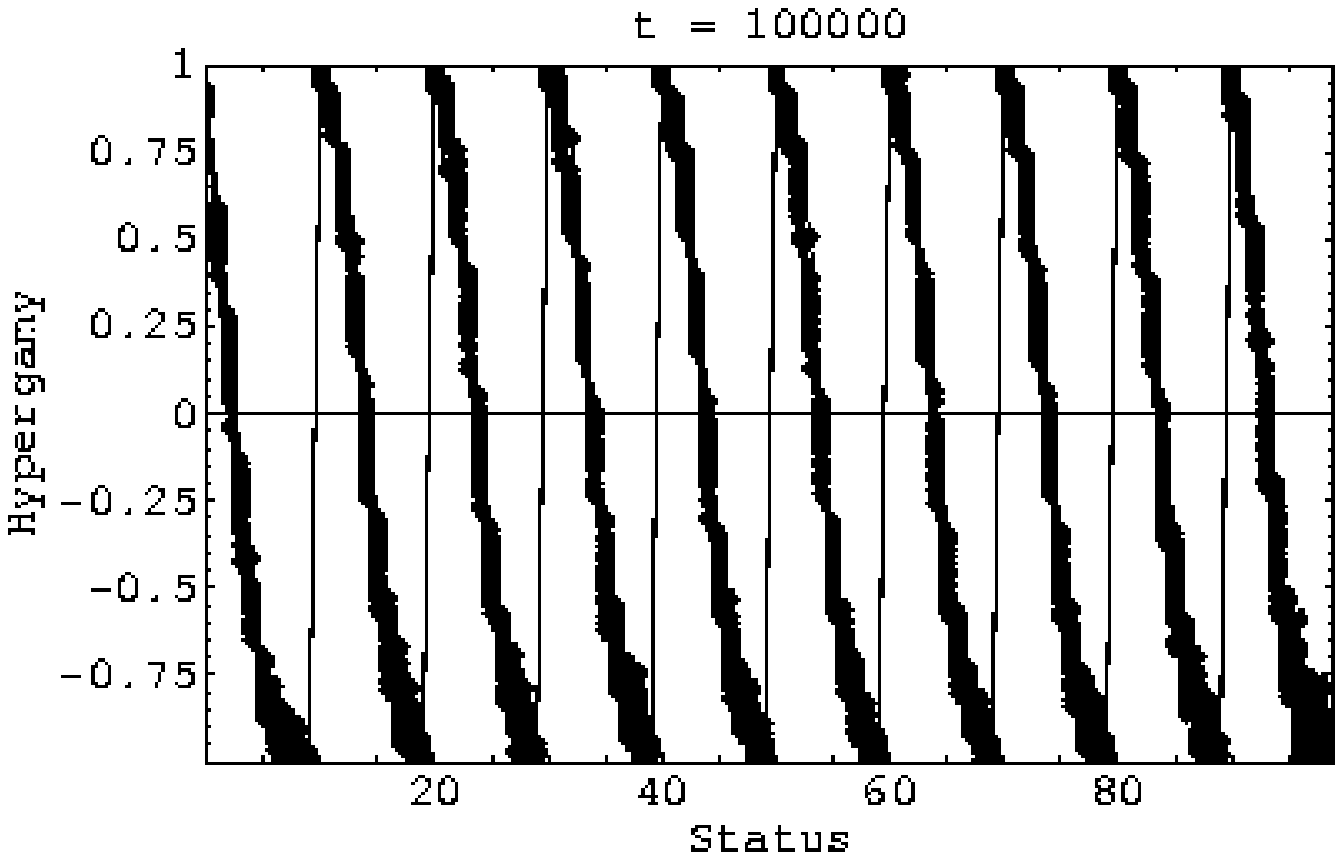}%
	\label{fig:rational-inherited-achieved.multiHypergamy.100000}%
}%
}%
\goodgap
\parbox{1.5in}{%
\subfigure[Female status histo\-gram, for one run]{%
	\includegraphics[width=1.5in]{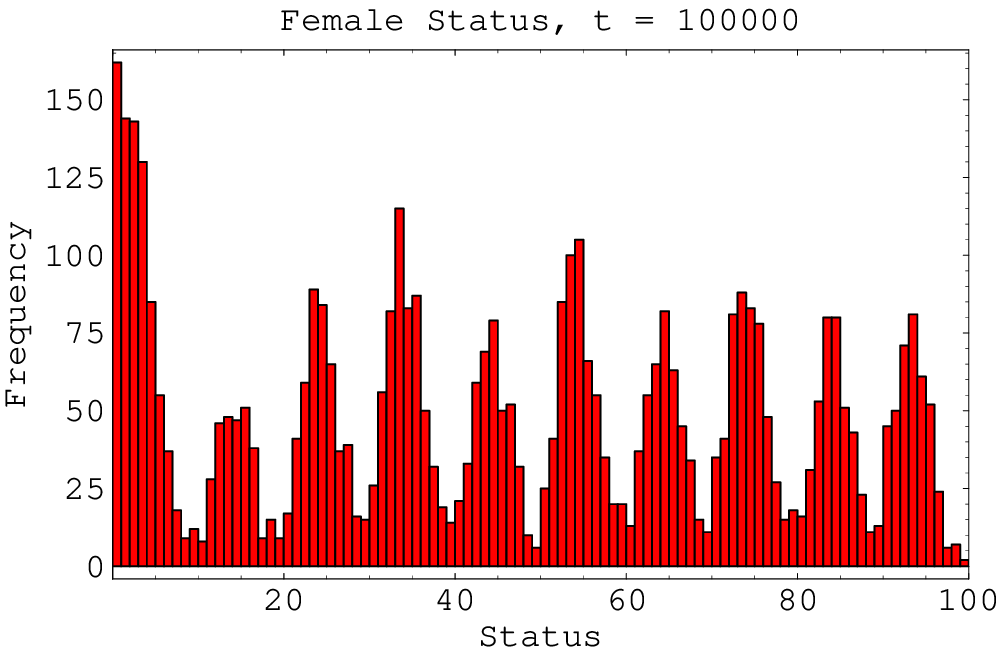}%
	\label{rational-inherited-achieved.1.fHistogram.100000}%
}%
\\
\subfigure[Male status histo\-gram, for one run]{%
	\includegraphics[width=1.5in]{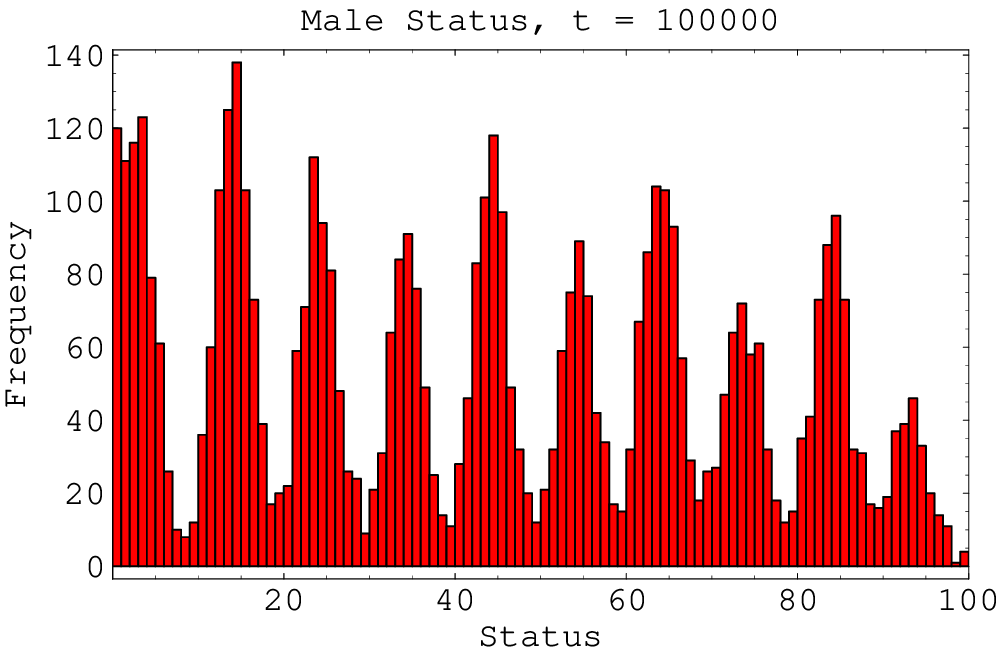}%
	\label{rational-inherited-achieved.1.mHistogram.100000}%
}%
}%
\end{center}
\caption{Marriage frequency, hypergamy $h,$ and status histograms at
$t = 100,000$ marriages, for the model $\mathbf{M}_{7}$ from
Section~\ref{sec:achieved}, using the rational strategy
$\mathbf{S}_{1}.$ Children inherited the average of their parents'
statuses, and agents achieved status over their lifetimes, according
to a Gaussian distribution with $\mu = 0$ and $\sigma = 2.0$.}
\label{fig:rational-inherited-achieved.1.100000}%
\end{figure}

\begin{figure}
\begin{center}
\subfigure[Marriage frequency, \newline for one run]{%
	\includegraphics[width=1.5in]{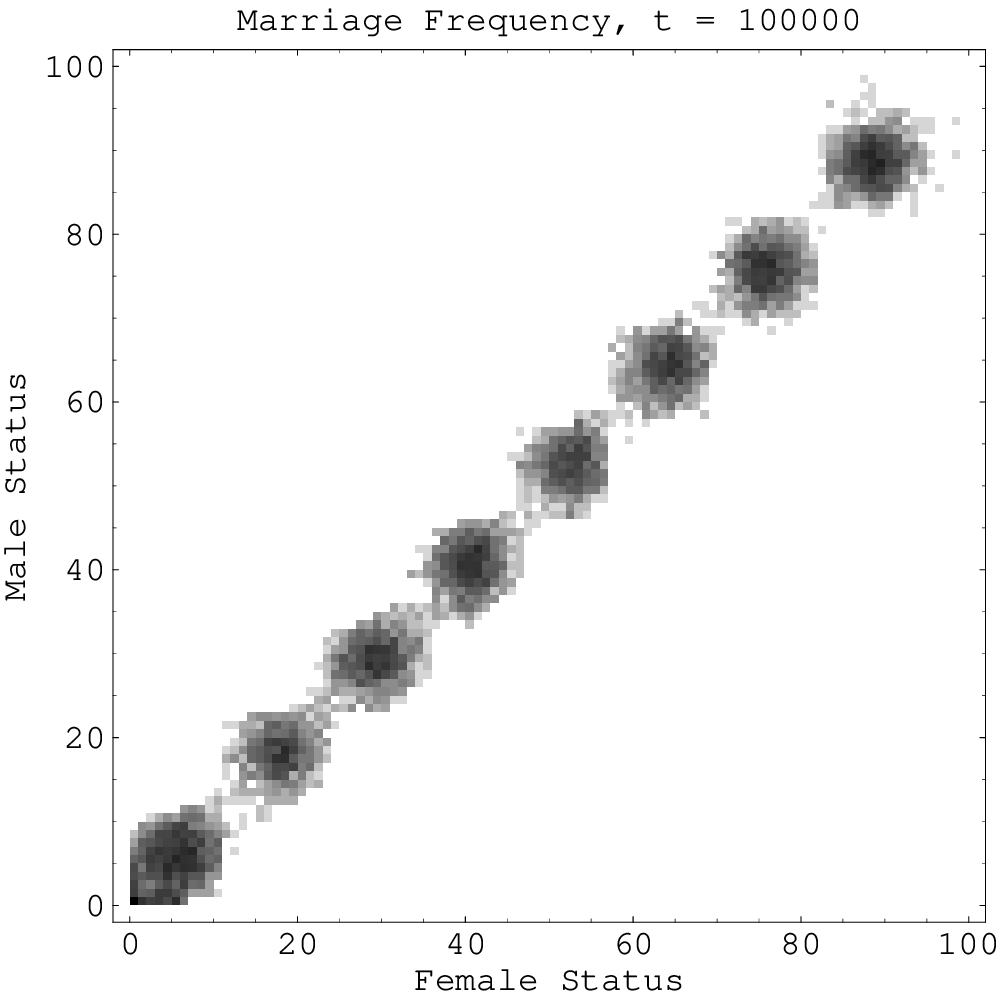}%
	\label{fig:learning-inherited-achieved.1.marriages.100000}%
}%
\goodgap
\parbox{1.5in}{%
\subfigure[Hypergamy $h,$ for one run]{%
	\includegraphics[width=1.5in]{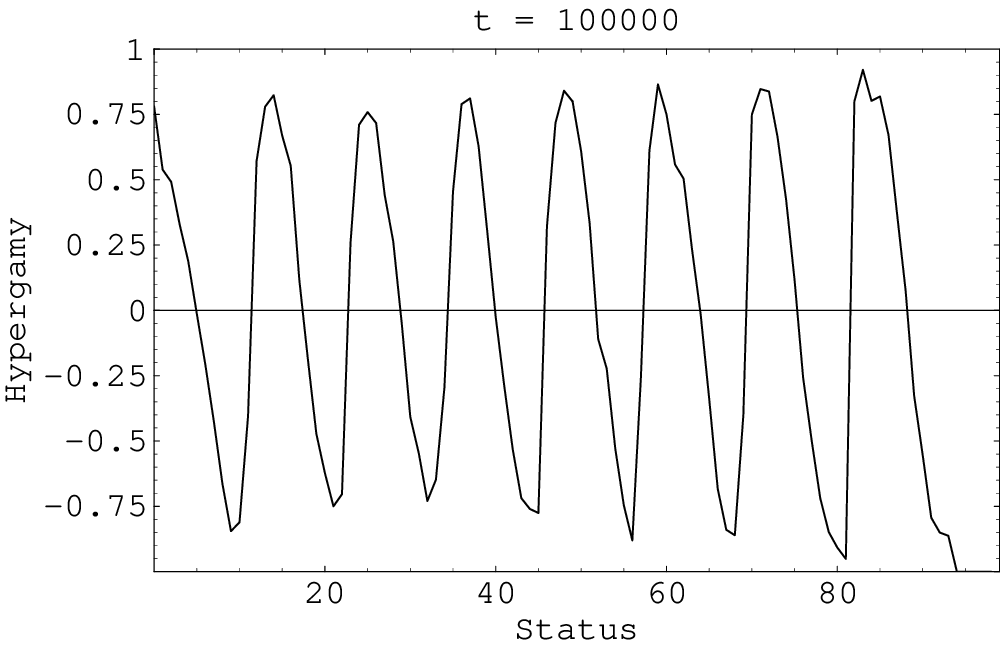}%
	\label{fig:learning-inherited-achieved.1.hypergamy.100000}%
}%
\\
\subfigure[Hypergamy $h,$ for 50 runs]{%
	\includegraphics[width=1.5in]{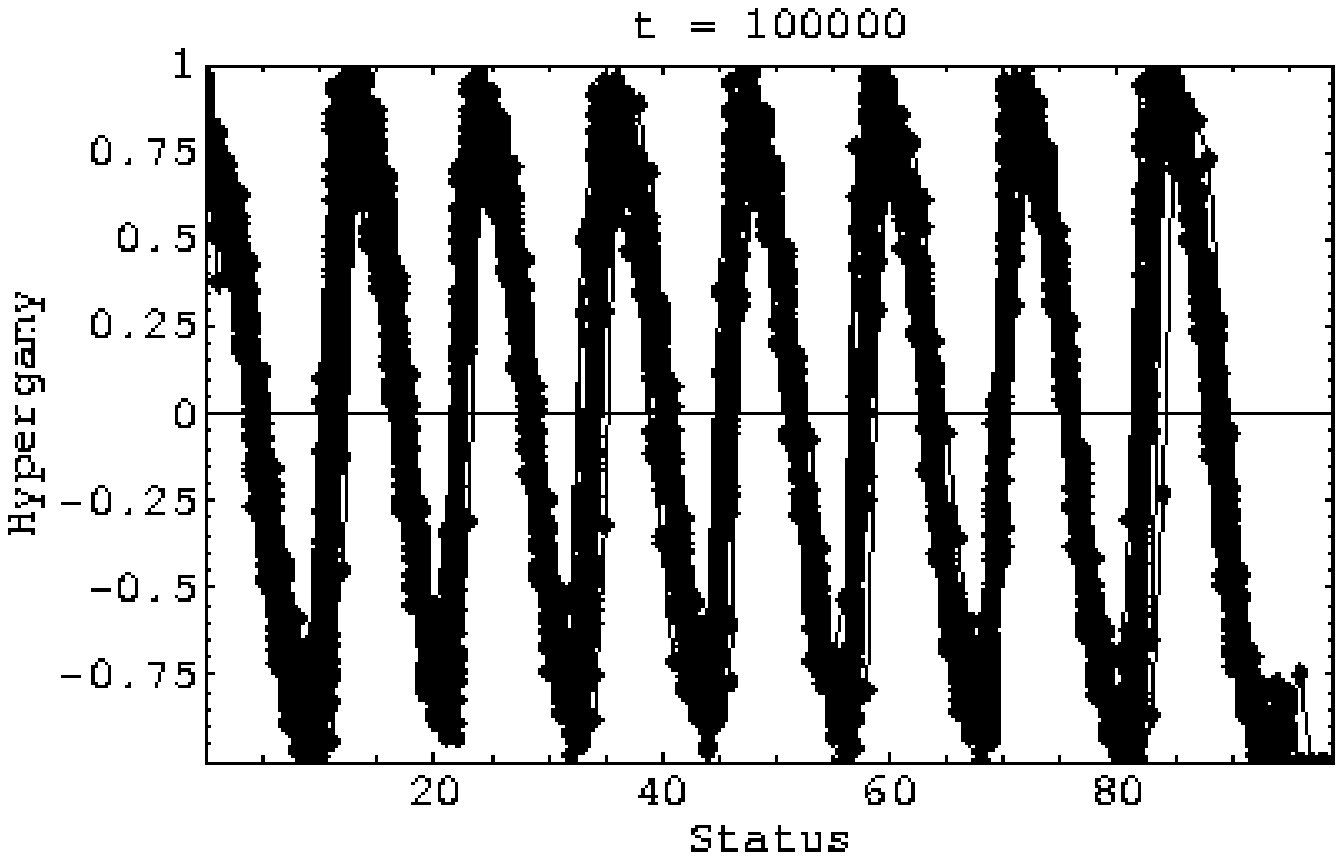}%
	\label{fig:learning-inherited-achieved.multiHypergamy.100000}%
}%
}%
\goodgap
\parbox{1.5in}{%
\subfigure[Female status histo\-gram, for one run]{%
	\includegraphics[width=1.5in]{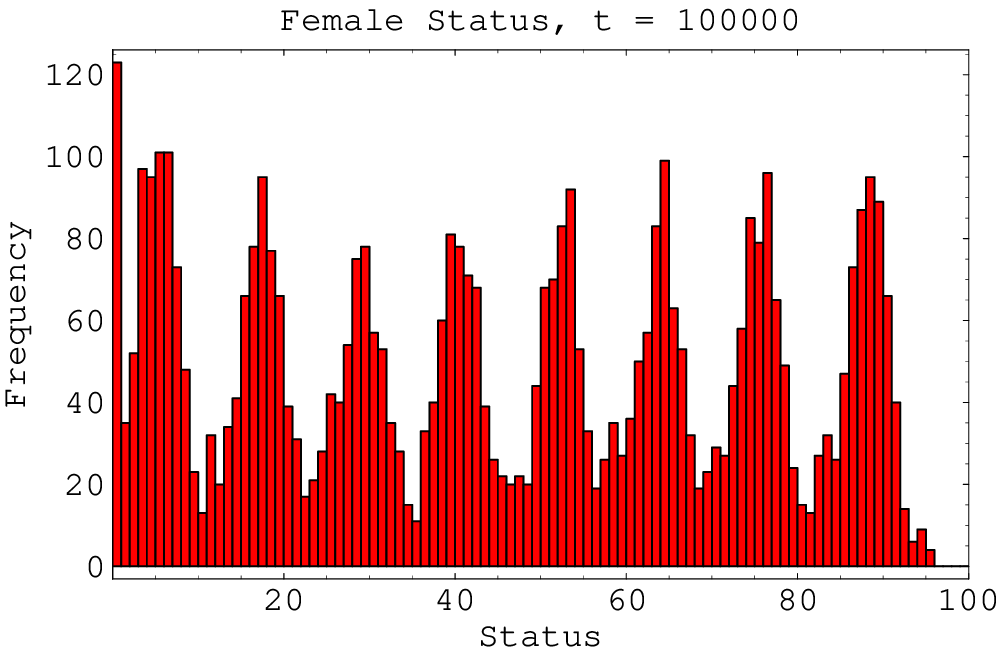}%
	\label{learning-inherited-achieved.1.fHistogram.100000}%
}%
\\
\subfigure[Male status histo\-gram, for one run]{%
	\includegraphics[width=1.5in]{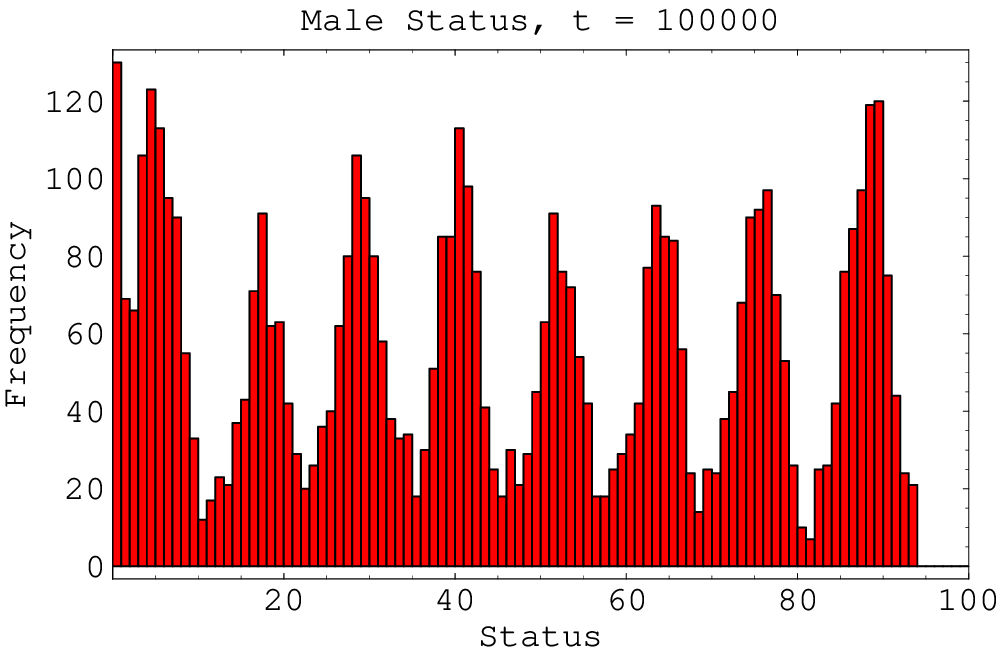}%
	\label{learning-inherited-achieved.1.mHistogram.100000}%
}%
}%
\end{center}
\caption{Marriage frequency, hypergamy $h,$ and status histograms at
$t = 100,000$ marriages, for the model $\mathbf{M}_{8}$ from
Section~\ref{sec:achieved}, using the learning algorithm
$\mathbf{S}_{2}.$ Children inherited the average of their parents'
statuses, and agents achieved status over their lifetimes, according
to a Gaussian distribution with $\mu = 0$ and $\sigma = 2.0$.}
\label{fig:learning-inherited-achieved.1.100000}%
\end{figure}

\begin{figure}
\begin{center}
\subfigure[Marriage frequency, \newline for one run]{%
	\includegraphics[width=1.5in]{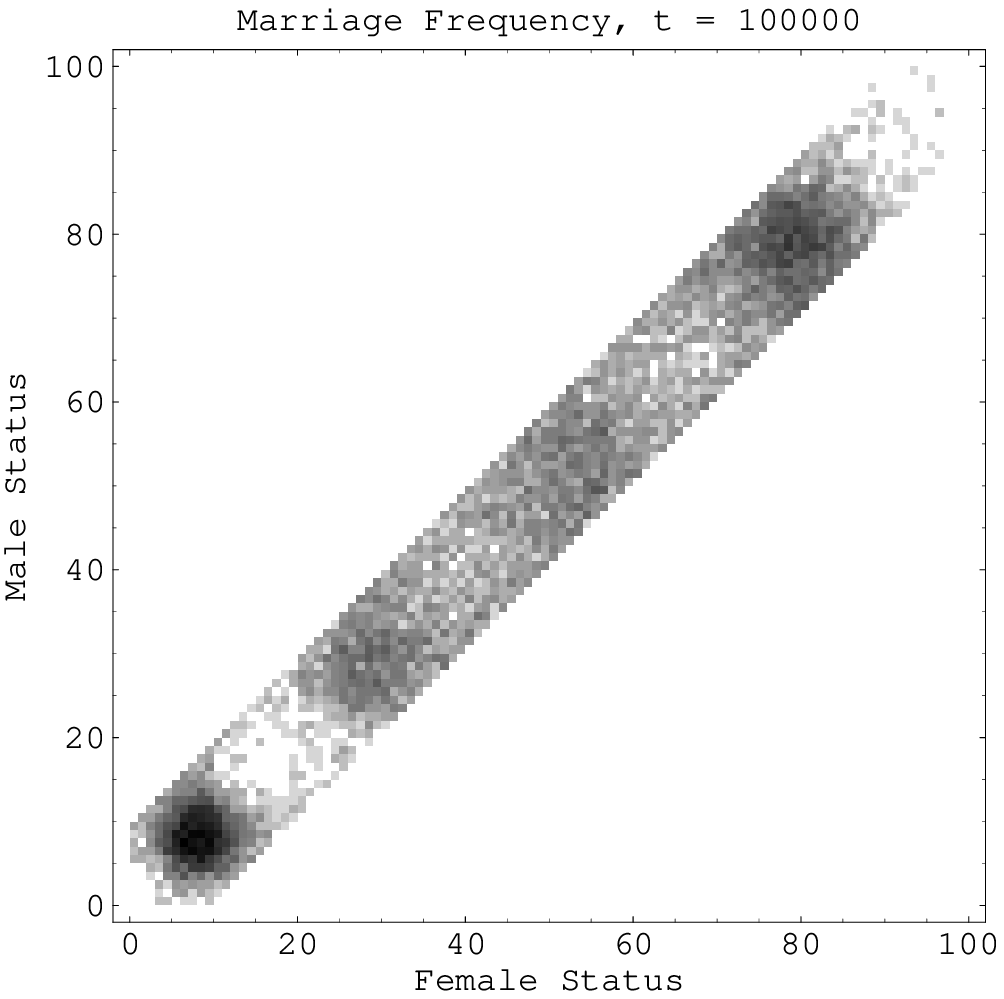}%
	\label{fig:self-inherited-achieved.1.marriages.100000}%
}%
\goodgap
\parbox{1.5in}{%
\subfigure[Hypergamy $h,$ for one run]{%
	\includegraphics[width=1.5in]{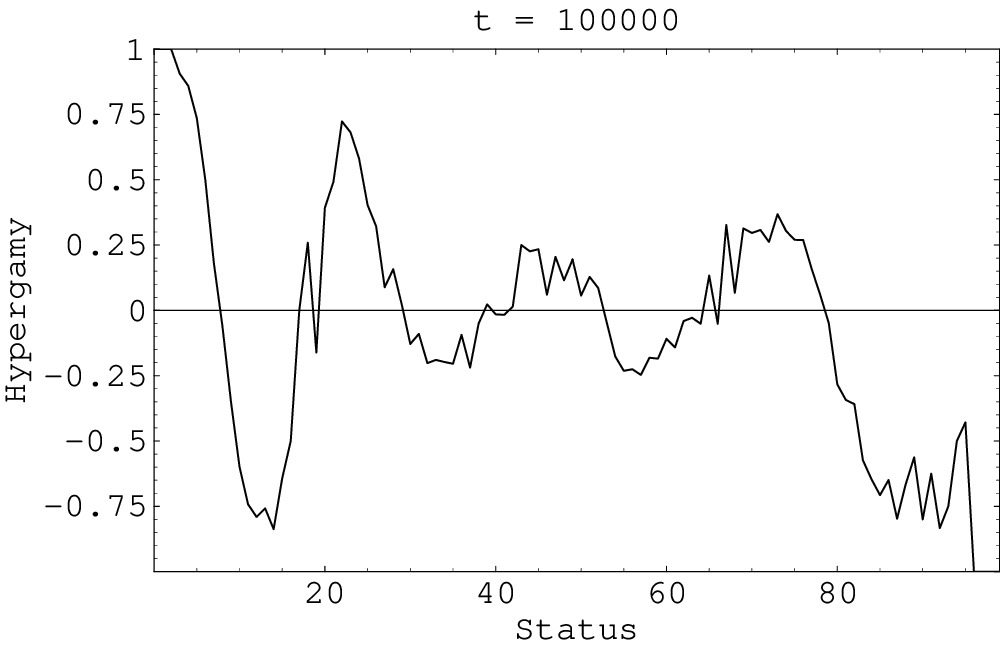}%
	\label{fig:self-inherited-achieved.1.hypergamy.100000}%
}%
\\
\subfigure[Hypergamy $h,$ for 50 runs]{%
	\includegraphics[width=1.5in]{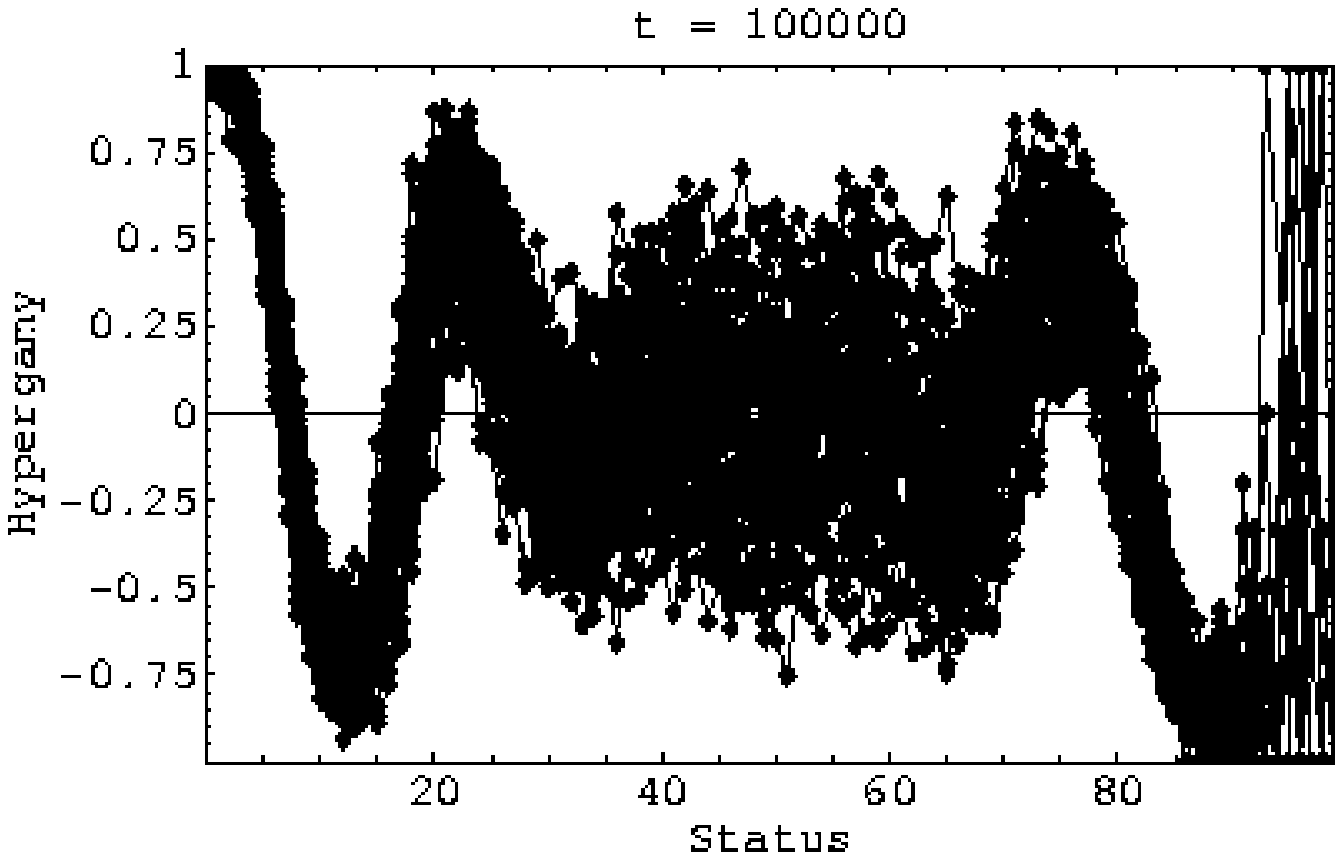}%
	\label{fig:self-inherited-achieved.multiHypergamy.100000}%
}%
}%
\goodgap
\parbox{1.5in}{%
\subfigure[Female status histo\-gram, for one run]{%
	\includegraphics[width=1.5in]{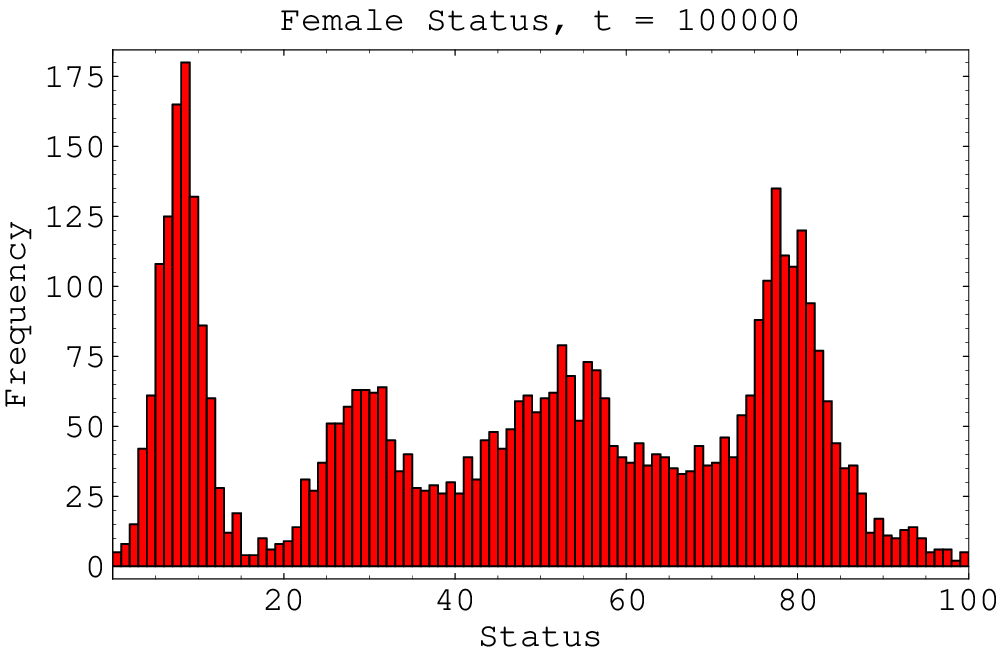}%
	\label{self-inherited-achieved.1.fHistogram.100000}%
}%
\\
\subfigure[Male status histo\-gram, for one run]{%
	\includegraphics[width=1.5in]{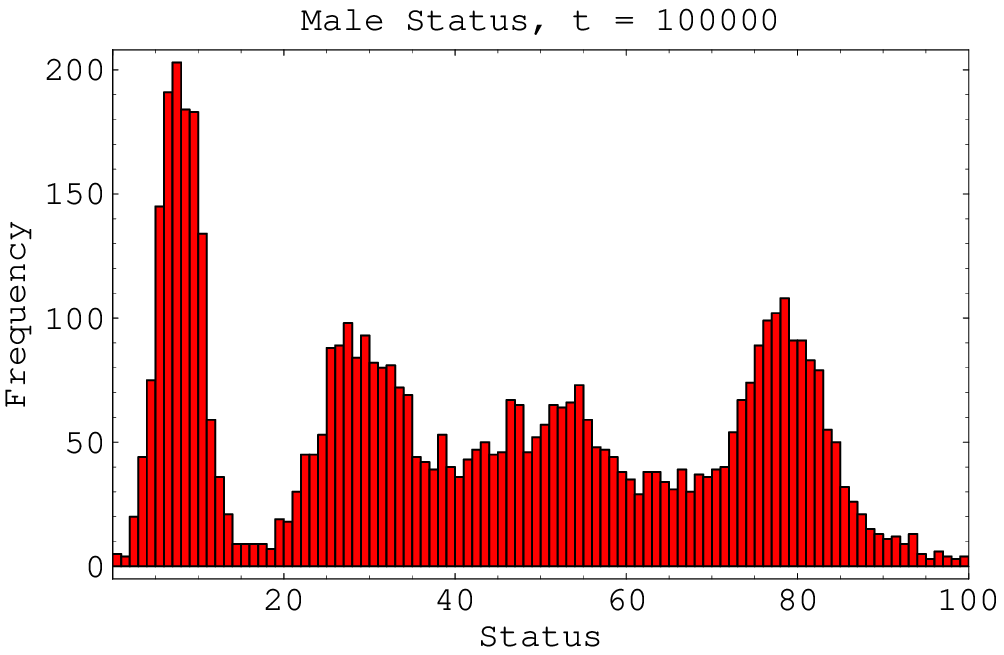}%
	\label{self-inherited-achieved.1.mHistogram.100000}%
}%
}%
\end{center}
\caption{Marriage frequency, hypergamy $h,$ and status histograms at
$t = 100,000$ marriages, for the model $\mathbf{M}_{9}$ from
Section~\ref{sec:achieved}, using the interval around self strategy
$\mathbf{S}_{3}.$ Children inherited the average of their parents'
statuses, and agents achieved status over their lifetimes, according
to a Gaussian distribution with $\mu = 0$ and $\sigma = 2.0$.}
\label{fig:self-inherited-achieved.1.100000}%
\end{figure}

In the previous models, agents inherited status at birth but did not
achieve or lose any status over their lifetimes. What happens if
achieved status is added? Models $\mathbf{M}_{4}, \mathbf{M}_{5},$ and
$\mathbf{M}_{6}$ from Section~\ref{sec:inherited} were modified so
that agents achieved or lost an amount of status $s_{\mathrm{a}}$
drawn from a Gaussian distribution with mean $\mu = 0$ and standard
deviation $\sigma = 2.0,$ producing models $\mathbf{M}_{7},
\mathbf{M}_{8},$ and $\mathbf{M}_{9}.$ An agent's overall status was
then simply the sum of its inherited and achieved statuses: $s =
s_{\mathrm{i}} + s_{\mathrm{a}}.$ Just as in
Section~\ref{sec:inherited}, both children inherited the average of
their parents' overall statuses, and children's sex was determined
randomly. The results at $t = 100,000$ marriages are shown in
Figure~\ref{fig:rational-inherited-achieved.1.100000} for model
$\mathbf{M}_{7}$ using the rational strategy $\mathbf{S}_{1}$ from
Section~\ref{sec:rational},
Figure~\ref{fig:learning-inherited-achieved.1.100000} for model
$\mathbf{M}_{8}$ using the learning algorithm $\mathbf{S}_{2}$ from
Section~\ref{sec:learning}, and
Figure~\ref{fig:self-inherited-achieved.1.100000} for model
$\mathbf{M}_{9}$ using the interval strategy $\mathbf{S}_{3}$ from
Section~\ref{sec:self}.

As can be seen, this type of achieved status acts as a type of noise,
smoothing out the status distribution and filling in the gaps caused
by inheritance. Gaps still occur in the status histogram if the spread
$\sigma$ of achieved status is small relative to the class
size. However, the peaks do not narrow past a point determined by
$\sigma,$ and if $\sigma$ is large enough, the status distribution
remains smooth.  Thus, this force acts in opposition to the
inheritance force from Section~\ref{sec:inherited}.

% rational-inherited-achieved
%
%

% learning-inherited-achieved
%
%

% self-inherited-achieved
%

\section{Conclusion} \label{sec:conclusion}

This paper has presented a variety of agent-based mathematical and
computer models that can produce class endogamy, based on Marcus and
Flannery's verbal model~\cite{marcus:flannery:1996}. One force that
can produce class endogamy occurs if agents are only willing to marry
suitors having status no less than some fixed value below the status
of their highest-status suitor.  This result was proved mathematically
in Section~\ref{sec:proof} by generalizing and simplifying Burdett and
Coles's model~\cite{burdett:coles:1997, burdett:coles:2001}, and
verified with a computer model in
Section~\ref{sec:rational}. Moreover, the simulation results in
Section~\ref{sec:learning} showed that agents can easily learn the
status of their highest-status suitor, reducing the amount of
rationality required for class endogamy to be produced.
Section~\ref{sec:inherited} demonstrated that another such force
results if children inherit the average of their parents'
statuses. This force can produce multimodality and gaps in the status
distribution. If this effect is strong enough, class endogamy results
simply due to a scarcity of potential mates outside of the local peak
in the status histogram. In fact, actual gaps can appear in the status
histogram. In contrast, Section~\ref{sec:achieved} showed that status
achieved over an agent's lifetime can be viewed as noise, analogous to
mutation in biological evolution, counteracting the inheritance force
and having the effect of smoothing out the status distribution. I
propose that class endogamy may have been produced by the interaction
of forces such as these, along with other factors such as
ideology. (This is analogous to the manner in which biological
evolution is the product of multiple forces including natural
selection, crossover, mutation, and genetic
drift~\cite{futuyma:1998}.) In sum, these models have shown that
simple, plausible heuristic mechanisms can in theory produce class
endogamy. How stupid can agents be and still form social strata?
Perhaps rather stupid, indeed. Whether these models have any bearing
on how class endogamy actually evolved in any particular human society
must of course be tested against the ethnographic, ethnohistorical,
archaeological, and psychological data.

One question raised by these models is, if these mechanisms can
reliably and quickly produce class endogamy, why didn't class endogamy
and social stratification emerge immediately after ranked societies
developed? In other words, why do chiefdoms exist at all in the
historical and archaeological record? One possible answer is that
complex chiefdoms seem to have been unstable, with recurring cycles of
aggregation into complex chiefdoms followed by collapse into smaller
and simpler chiefdoms~\cite{wright:1994}. Chiefdoms, without social
stratification, may simply have been a more stable organizational
form. (However, even chiefdoms seem to have been somewhat unstable,
sometimes collapsing into unranked societies~\cite{leach:1954}.)
Another answer may be that newly-developed chiefdoms were relatively
small and egalitarian, and became larger and less egalitarian only
over time. If the status difference between the highest- and
lowest-status members of a society was relatively small, and the
selection of eligible mates was small as well, then individuals may
have been less selective, reducing the impact of the forces described
here. Finally, there was likely resistance to stratification that took
time to overcome.

Burdett and Coles~\cite{burdett:coles:1997, burdett:coles:2001} argue
that their model applies to contemporary Western societies. I am
dubious about applying these models to the present day. First, the
models assume that men and women choose whether to marry based on a
single factor: status. In contrast, there seem to be multiple factors
upon which men and women base their decisions in modern societies,
such as personality, common interests, intelligence, education,
wealth, and physical attractiveness, as well as social status. While
married couples seem to be positively correlated for many such
factors~\cite{burdett:coles:1997}, correlation alone does not imply
endogamy, as was shown in Section~\ref{sec:self}. Furthermore, these
models assume that men and women each have a single overall
desirability metric for the opposite sex. However, in contemporary
societies, different people often weight these factors very
differently: Two men can easily disagree as to how desirable a woman
is, and vice versa. In general, it is unclear whether endogamous
classes form by these mechanisms in modern societies. In contrast,
religions and ethnicities may form endogamous groups; however, these
are relatively discrete groups to begin with, in contrast to the
continuous status distributions considered here. The same is true of
education, since it is usually achieved in discrete amounts and
classified into categories such as ``MBA from an Ivy League
university''.

\section{Future Work} \label{sec:future-work}

The most obvious extension of this paper would be to test whether the
mechanisms proposed here were in fact present in any real societies,
using ethnographic, ethnohistorical, and archaeological data. It is
also important to test whether the type of class endogamy produced by
these models is consistent with that in real societies. One possible
source of empirical data is the Human Relations Area Files
(HRAF)~\cite{hraf}.  It might also be possible to devise psychological
tests to examine whether the rational strategy $\mathbf{S}_{1}$ from
Section~\ref{sec:rational} is realistic within some particular
contemporary society.

Another immediate extension to the model would be to add randomness or
noise to the agents' decision making under the various strategies, in
order to test how robust the emergence of class endogamy is.  This
noise would represent such factors as chance events, imperfect
information, and love. (In contrast, in contemporary Western
societies, the effects of social status could be considered as noise
in relation to the dominant effects of love and attraction.)  It would
also be useful to define a class endogamy metric, so that simulation
results could be compared objectively. One such metric could be the
average distance between local minima and maxima on the hypergamy
plots, but this breaks down as a proxy for measuring class endogamy if
the peaks in the status histogram are very narrow, as can be seen in
Figure~\ref{fig:rational-inherited-noachieved.1.100000}.
Finally, Miller's Active Nonlinear Tests (ANTs)~\cite{miller:1996}
could be used to search for regions of parameter space where these
models break down.
 
Another way to make the model more realistic would be to add death and
reproductive rates that vary by agent age~\cite{weiss:1973}, and
perhaps some form of discounting. 
Multidimensional status could be added as well, with agents weighing
such factors as social status, wealth, education, and attractiveness
to produce an overall status or desirability score. It seems likely
that this would be an obstacle to class formation, like noise, since
it would increase the likelihood for agents who would otherwise be in
separate classes, based on social status alone, to marry.

More aspects of the models could be analyzed mathematically, for
instance the rate of regression towards the mean when inherited status
is averaged, the magnitude of the smoothing caused by achieved status,
and the conditions under which the learning algorithm $\mathbf{S}_{2}$
is effective.

Henry Wright has proposed another verbal model of the evolution of
class endogamy:
\begin{quote}
Neither the difficult question of why some societies ascribe the right
to make community-wide decisions to office-holders drawn from a
limited social sub-group, nor the question of why some networks of
simple ascriptively ranked societies develop social classes are
crucial to this essay. Regarding the latter question, it suffices to
suggest that --- if productive systems can sustain a continuity of the
social network in time (and many cannot) --- with time, intermarriages
and disputes among the ranking families will disperse claimants to
office. Many individuals may compete for offices with which few will
have any local connection. Indeed, the ranking or noble class as a
whole can be expected to oppose any local interests. Thus, the
development within a network of chiefly polities of a class competing
for positions, but opposing others outside the class, may be simply
explained.~\cite[pp. 70--71]{wright:1994}
\end{quote}
It would be interesting to implement this model on a computer, to
verify whether such distributed intermarriage is indeed another
mechanism for producing class endogamy. Such a model would be more
complicated than the ones presented in this paper, however. The set of
agents eligible for the office of chief would have to be explicitly
defined, as well as the effect of office holding on status, the
network of chiefdoms, and the conditions under which two agents from
separate chiefdoms would marry.

Another mechanism might result from the interaction between hypogamous
marriage and the averaging of inherited status. As a thought
experiment, consider the following situation: A paramount chief, with
status $99,$ arranges a hypogamous marriage between one of his
daughters, also with status $99,$ and a subchief in an outlying
village, with status $49.$ The inhabitants of the paramount's seat
have status ranging from $0$ to $99,$ while those in the subchief's
village have status from $0$ to $49.$ If children inherit the average
of their parents' statuses, then the offspring of this marriage would
have a status of $(99 + 49) / 2 = 79$ at birth, far outranking both
their father and all of the other local villagers. Such offspring
would thus naturally look outside of their own village for potential
mates, leading to the kind of noble class that Wright refers to. This
model, while merely a caricature, demonstrates another way in which
gaps in a chiefdom's status distribution could have resulted, in
addition to the inheritance force from Section~\ref{sec:inherited}.

One last extension might be to keep track of the agents'
lineages over time and examine what effect the mechanisms presented
here had on their genealogical trees.

Finally, similar models have been used to study the causes of
sympatric speciation~\cite{hilscher:2004}. It would be interesting to
investigate whether there are analogies, at an abstract mathematical
level, between class endogamy and biological speciation.

\section*{Acknowledgments}

I would like to thank Eric Rupley and Henry Wright for many fruitful
discussions. This paper would not exist without the inspiration of
Joyce Marcus and Kent Flannery's verbal model, as well as Henry
Wright's.  I have also had many helpful conversations with Troy
Tassier, and I am especially indebted to him for pointing out Burdett
and Coles's papers to me. Any misinterpretations of those models or of
the archaeological, anthropological, and economics literature are my
own, of course. I am also grateful to the members of the University of
Michigan Royal Road Group for their comments and support.  Finally,
this work benefited from being presented to the UM Complex Systems
Advanced Academic Workshop, as well as to the SwarmFest 2004 workshop.

\bibliographystyle{plain} 
\bibliography{endogamy} % bib filename without .bib 
\end{document}